\documentclass[12pt]{article}%
\usepackage[nosort]{cite}
\usepackage{graphicx}
\usepackage{multicol}
\usepackage{amsfonts}
\usepackage{amssymb}
\usepackage{amsmath}
\usepackage{heck}
\usepackage{ytableau}
\usepackage{afterpage}
\usepackage{setspace}
\usepackage{verbatim}
\usepackage{color}
\usepackage{longtable}
\usepackage{float}
\usepackage{subcaption}
\usepackage{epsfig}
\usepackage{epstopdf}
\usepackage{adjustbox}
\usepackage{tikz}
\usepackage[margin=1in]{geometry}
\usepackage{titletoc}
\usepackage{hyperref}%
\usepackage{mathrsfs}
\usepackage{dsfont}
\usepackage{algorithm}
\usepackage{algpseudocode}
\usepackage{tikz}
\usepackage{pgfplots}
\usepackage{algorithmicx, algpseudocode,algorithm}
\usepackage{caption}
\usepackage{float}
\numberwithin{equation}{section}

\usepackage{tikz}
\usetikzlibrary{shapes.geometric, decorations.markings}

\newsavebox{\mysavebox}

\hypersetup{colorlinks,citecolor=black,filecolor=black,linkcolor=black,urlcolor=black}

\usetikzlibrary{decorations.markings}
\usetikzlibrary{decorations.markings}
\usetikzlibrary{hobby}
\usepackage{tikz}
\usetikzlibrary{arrows}
\usetikzlibrary{arrows.meta}
\usetikzlibrary{arrows,decorations.pathmorphing}
\usetikzlibrary{shapes.geometric,calc,arrows, positioning,shapes.misc,decorations.markings}
\tikzset{
  big arrow/.style={
    decoration={markings,mark=at position 1 with {\arrow[scale=2,#1]{>}}},
    postaction={decorate},
    shorten >=0.4pt},
  big arrow/.default=black}

\pgfdeclarelayer{edgelayer}
\pgfdeclarelayer{nodelayer}
\pgfsetlayers{edgelayer,nodelayer,main}
\pgfplotsset{compat=1.16}
\tikzstyle{none}=[inner sep=0pt]

\tikzstyle{NodeCross}=[draw, shape=circle, cross out, inner sep=0pt, minimum size=6pt,line width=0.25mm]
\tikzstyle{Circle}=[draw, shape=circle, black, fill=black, inner sep=0pt, minimum size=6pt]
\tikzstyle{circle}=[draw, shape=circle, black, fill=black, inner sep=0pt, minimum size=16pt]
\tikzstyle{Star}=[draw, shape=star, fill=black, star points=8, inner sep=0pt, minimum size=8pt]
\tikzstyle{CircleRed}=[draw, shape=circle, black, fill=red, inner sep=0pt, minimum size=6pt]
\tikzstyle{StarP}=[draw={rgb,255: red,128; green,0; blue,128}, shape=star, fill={rgb,256: red,128; green,0; blue,128}, star points=8, inner sep=0pt, minimum size=12pt]
\tikzstyle{ShadedCircRed}=[draw=red, shape=circle, fill={rgb, 255: red,255; green,114; blue, 118}, inner sep=0pt, minimum size=80pt, line width=0.5mm, fill opacity=0.2]
\tikzstyle{ShadedCircRed2}=[draw=red, shape=circle, fill={rgb, 255: red,255; green,114; blue, 118}, inner sep=0pt, minimum size=10pt]
\tikzstyle{ShadedCircRed3}=[draw=black, shape=rectangle, fill={rgb, 255: red,255; green,114; blue, 118}, inner sep=0pt, minimum size=113pt, line width=0.25mm]
\tikzstyle{ShadedCirc}=[draw=red, shape=circle, fill=white, inner sep=0pt, minimum size=45pt,  fill opacity=1.0,  line width=0.5mm]
\tikzstyle{CircleBlue}=[draw, shape=circle, fill=blue, inner sep=0pt, minimum size=6pt]
\tikzstyle{BigCirclePurple}=[draw, shape=circle, fill={rgb,255: red,191; green,0; blue,191}, inner sep=0pt, minimum size=12pt]
\tikzstyle{CirclePurple}=[draw, shape=circle, fill={rgb,255: red,191; green,0; blue,191}, inner sep=0pt, minimum size=5pt]
\tikzstyle{EmptyCircle}=[draw, shape=circle, inner sep=0pt, minimum size=4pt]
\tikzstyle{GreenCircle}=[draw, shape=circle,  fill={rgb,255: red,80; green,200; blue,120}, inner sep=0pt, minimum size=8pt]
\tikzstyle{BrownCircle}=[draw, shape=circle,  fill={rgb,255: red,210; green,105; blue,30}, inner sep=0pt, minimum size=8pt]
\tikzstyle{CirclePurpleSmall}=[draw, shape=circle, fill={rgb,255: red,191; green,0; blue,191}, inner sep=0pt, minimum size=4pt]
\tikzstyle{BigCircleGreen}=[draw, shape=circle, fill={rgb,255: red,0; green,191; blue,0}, inner sep=0pt, minimum size=12pt]
\tikzstyle{BigCircleBlue}=[draw, shape=circle, fill={rgb,255: red,0; green,0; blue,191}, inner sep=0pt, minimum size=12pt]
\tikzstyle{BigCircleRed}=[draw, shape=circle, fill={rgb,255: red,191; green,0; blue,0}, inner sep=0pt, minimum size=12pt]
\tikzstyle{BrownCircleSmall}=[draw, shape=circle,  fill={rgb,255: red,210; green,105; blue,30}, inner sep=0pt, minimum size=6pt]
\tikzstyle{SmallCircleBrown}=[draw, shape=circle,  fill={rgb,255: red,210; green,105; blue,30}, inner sep=0pt, minimum size=5pt]
\tikzstyle{SmallCircleRed}=[draw, shape=circle, fill={rgb,255: red,191; green,0; blue,0}, inner sep=0pt, minimum size=6pt]
\tikzstyle{DashedLine}=[-, densely dashed, line width=0.25mm]
\tikzstyle{DottedLine}=[-, dotted, line width=0.25mm]
\tikzstyle{ThickLine}=[-, line width=0.25mm]
\tikzstyle{ArrowLineRight}=[-, -{Stealth[scale=1.25]}, line width=0.25mm, scale=5]
\tikzstyle{ArrowLineRed}=[-, draw={rgb,255: red,191; green,0; blue,0}, -{Stealth[scale=1.75]}, line width=0.1mm, scale=5]
\tikzstyle{RedLine}=[-, draw={rgb,255: red,191; green,0; blue,0}, fill=none, line width=0.5mm]
\tikzstyle{DashedLineThin}=[-, densely dashed, line width=0.125mm, fill=none, draw=black]
\tikzstyle{DottedRed}=[-, dotted, draw={rgb,255: red,191; green,0; blue,0}, fill=none, line width=0.25mm]
\tikzstyle{DashedRed}=[-, densely dashed, draw={rgb,255: red,191; green,0; blue,0}, fill=none, line width=0.25mm]
\tikzstyle{BlueLine}=[-, draw={rgb,255: red,0; green,0; blue,191}, fill=none, line width=0.5mm]
\tikzstyle{ArrowLineBlue}=[-, draw={rgb,255: red,0; green,0; blue,191}, -{Stealth[scale=1.75]}, line width=0.1mm, scale=5]
\tikzstyle{GreenDoubleArrow}=[<->, draw={rgb,155: red,0; green,255; blue,0},  line width= 0.5mm, scale=5]
\tikzstyle{RedDoubleArrow}=[<->, draw={rgb,255: red,255; green,0; blue,0},  line width= 0.5mm, scale=5]
\tikzstyle{BlueDottedLight}=[-, dotted, draw={rgb,255: red,0; green,0; blue,191}, fill=none, line width=0.3mm]
\tikzstyle{BrownLine}=[-, draw={rgb,255: red,210; green,105; blue,30}, fill=none, line width=0.5mm]
\tikzstyle{DottedRed}=[-, dotted, draw={rgb,255: red,191; green,0; blue,0}, fill=none, dotted, line width=0.5mm]
\tikzstyle{DottedPurple}=[-, dotted, draw={rgb,255: red,191; green,0; blue,191}, fill=none, dotted, line width=0.5mm]
\tikzstyle{BlueDottedLight}=[-, dotted, draw={rgb,255: red,0; green,0; blue,191}, fill=none, line width=0.5mm]
\tikzstyle{ArrowLinePurple}=[-, draw={rgb,255: red,191; green,0; blue,191}, -{Stealth[scale=1.75]}, line width=0.5mm, scale=5]
\tikzstyle{DashedLineGreen}=[-, densely dashed, draw={rgb,255: red,74; green,103; blue,65}, line width=0.25mm]
\tikzstyle{LineGreen}=[-, draw={rgb,255: red, 74; green,200; blue,65}, line width=0.5mm]
\tikzstyle{ArrowLineGreen}=[-, draw={rgb,255: red,0; green,191; blue,0}, -{Stealth[scale=1.75]}, line width=0.5mm, scale=5]
\tikzstyle{GreenLine}=[-, draw={rgb,255: red,0; green,191; blue,0}, fill=none, line width=0.5mm]
\tikzstyle{PurpleLine}=[-, draw={rgb,255: red,191; green,0; blue,191}, fill=none, line width=0.5mm]
\tikzstyle{PPurpleLine}=[-, draw={rgb,255: red,191; green,0; blue,191}, fill=none, line width=2.5mm]
\tikzstyle{DPurpleLine}=[-, dotted, draw={rgb,255: red,191; green,0; blue,191}, fill=none, line width=0.5mm]
\tikzstyle{SBrownLine}=[-, draw={rgb,255: red,191; green,0; blue,191}, fill=none, opacity=0.35, line width=2.5mm]
\tikzstyle{DottedBlue}=[-, dotted, draw=blue, fill=none, dotted, line width=0.5mm]
\tikzstyle{DashedPurpleLine}=[-, densely dashed, draw={rgb,255: red,191; green,0; blue,191}, fill=none, line width=0.5mm]
\tikzstyle{SmallCircleBlue}=[draw, shape=circle, fill=blue, inner sep=0pt, minimum size=5pt]
\tikzstyle{SmallCirclePurple}=[draw, shape=circle, fill={rgb,255: red,191; green,0; blue,191}, inner sep=0pt, minimum size=5pt]
\tikzset{snake it/.style={decorate, decoration=snake}}

\tikzset{
dashstar/.style={
 dash pattern=on 5pt off 5pt,
 postaction={
  decorate,
  decoration={
   markings,
   mark=between positions 9pt and 1 step 10pt with {
     \node[color=red] {*};
   }
  }
 }
},
dashstarstar/.style={
 dash pattern=on 5pt off 10pt,
 postaction={
   decorate,
   decoration={
     markings,
     mark=between positions 10pt and 1
          step 15pt
           with {
            \node at (-2pt,0pt) {\pgfuseplotmark{star}};
            \node at (2pt,0pt) {\pgfuseplotmark{star}};
           }
   }
 }
}
}

\begin{document}

\date{May 2026}

\title{Quiver Approach to Symmetry Theories}

\institution{PENN}{\centerline{$^{1}$Department of Physics and Astronomy, University of Pennsylvania, Philadelphia, PA 19104, USA}}
\institution{PENNmath}{\centerline{$^{2}$Department of Mathematics, University of Pennsylvania, Philadelphia, PA 19104, USA}}
\institution{Maribor}{\centerline{$^{3}$Center for Applied Mathematics and Theoretical Physics, University of Maribor, Maribor, Slovenia}}

\authors{
Vivek Chakrabhavi\worksat{\PENN}\footnote{e-mail: \texttt{vivekcm@sas.upenn.edu}},
Mirjam Cveti\v{c}\worksat{\PENN,\PENNmath,\Maribor}\footnote{e-mail: \texttt{cvetic@physics.upenn.edu}}, \\ \vspace{4mm}
Jonathan J. Heckman\worksat{\PENN,\PENNmath}\footnote{e-mail: \texttt{jheckman@sas.upenn.edu}}, and Shani Meynet\worksat{\PENN}\footnote{e-mail: \texttt{smeynet@sas.upenn.edu}}
}

\abstract{
Global symmetry anomalies of a quantum field theory (QFT) can be packaged as specific couplings of a higher-dimensional symmetry theory (SymTh).
In this work we show that for 5D superconformal field theories (SCFTs) engineered from M-theory backgrounds $X$ a Calabi-Yau cone, this data can be extracted from the path algebra of branes probing $X$. This provides a complementary algebraic approach compared with more geometric computations based on the explicit calculation of triple intersection numbers in a resolved geometry and / or $\eta$-invariants extracted from the boundary geometry $\partial X$. Our method applies in situations where the counterpart geometric computation is either unknown or combinatorially unwieldy. We illustrate with several toric threefold examples, including orbifolds $\mathbb{C}^{3} / \Gamma$ and more general non-orbifold Calabi-Yau cones of Sasaki-Einstein five-manifolds.
}

\maketitle

\enlargethispage{\baselineskip}

\setcounter{tocdepth}{3}

\tableofcontents

\newpage

\section{Introduction} \label{sec:Intro}

Symmetries provide important constraints / selection rules in a wide variety of quantum systems. Recently, the notion of
a global symmetry in quantum field theory (QFT) has revealed a remarkably rich topological structure \cite{Gaiotto:2014kfa}.\footnote{See e.g., \cite{Sharpe:2015mja, McGreevy:2022oyu,
Freed:2022iao,
Gomes:2023ahz,
Brennan:2023mmt,
Bhardwaj:2023kri,
Shao:2023gho,
Costa:2024wks,
Kaidi:2026urc} for some recent reviews.} For a $D$-dimensional QFT$_D$, this symmetry data can in turn be packaged in terms of a higher-dimensional symmetry theory (SymTh) / symmetry topological field theory (SymTFT), where the original QFT$_{D}$ is decompressed on an extra-dimensional interval with a relative QFT$_D$ (in the sense of \cite{Freed:2012bs}) localized at one end, and its global form (specifying the precise global symmetry and operator content) is specified at the other end.\footnote{See \cite{Freed:2012bs, Reshetikhin:1991tc, Turaev1992StateSI, Barrett:1993ab, Witten:1998wy, Fuchs:2002cm, Kirillov:2010nh, Kapustin:2010if, Kitaev_2012, Fuchs:2012dt, Freed:2018cec, Freed:2022qnc, Heckman:2017uxe, Gaiotto:2020iye, Apruzzi:2021nmk, Kaidi:2022cpf, Brennan:2024fgj, Argurio:2024oym, DelZotto:2024tae, Apruzzi:2024htg, Antinucci:2022vyk, Lawrie:2023tdz, Baume:2023kkf, Yu:2023nyn, Antinucci:2024zjp, Heckman:2024zdo, Franco:2024mxa, Gagliano:2024off, Cvetic:2024dzu, Cordova:2024iti, Bonetti:2024cjk, Bhardwaj:2024igy, Jia:2025jmn, Yu:2024jtk, Apruzzi:2023uma, Apruzzi:2025mdl, Heckman:2024oot, Heckman:2025lmw, Apruzzi:2025hvs, Torres:2025jcb, Bergman:2026lnz, Xue:2026hji, Jia:2026tsl, Jia:2026vcr, Cvetic:2025kdn, DeMarco:2025pza, Cummings:2026giw, Heckman:2026beg, Yu:2026gdf} for a partial list of references.}

Quite remarkably, many aspects of these deep topological structures naturally lift to the extra-dimensional geometry of string / M- / F-theory backgrounds. For example, in an 11D M-theory background of the form:
\begin{equation}
M_D \times X_{11 - D},
\end{equation}
one engineers a QFT$_D$ on $D$-dimensional Minkowski space $M_D$ by specifying a local singular geometry on $X$. A particularly well-studied situation is that where $X = \mathrm{Cone}(Y)$, where $Y$ can be viewed as $\partial X$, the boundary of the cone ``at infinity''. In this setup, the local degrees of freedom of the QFT originate from the tip of the singular cone, and heavy defects arise from branes stretching out to the boundary. Symmetry operators come from branes which link / intersect with these defects.\footnote{The picture has been developed over several years, see e.g., \cite{DelZotto:2015isa, Albertini:2020mdx, Morrison:2020ool} for relations to defects and \cite{Apruzzi:2022rei, GarciaEtxebarria:2022vzq, Heckman:2022muc, Heckman:2022xgu, Cvetic:2023plv, Bergman:2024aly} for the origin of symmetry operators.} Finite higher-form symmetries involve linking in the extra-dimensional geometry, while continuous higher-form symmetries involve intersections in the extra-dimensional geometry. It is also worth noting that the ``branes'' in question need not be BPS $p$-branes of string / M-theory. They can also include isometries and automorphisms of the internal geometry, as in \cite{Cvetic:2025kdn, Bah:2025vfu, Bah:2026gcf} 

In many cases of interest where the QFT is most simply \textit{defined} by the singular geometry, this approach provides one of the only quantitative ways to access its symmetry data. For example, higher-form symmetries of these backgrounds are directly
packaged in terms of the ``defect group'' (introduced in \cite{DelZotto:2015isa}), as computed by quotients of the form $H_{\ast}(X, \partial X) / H_{\ast}(X)$, which can in turn often be recast purely in terms of the boundary homology $H_{\ast}(\partial X)$. In this context, global forms of the absolute QFT$_{D}$ correspond to self-consistent boundary conditions for bulk fields of the M-theory background geometry. A change of boundary conditions in the SymTh / SymTFT corresponds to changing these boundary conditions at infinity, i.e., by gauging some of the generalized global symmetries. Obstructions to this operation are captured by anomalies of the QFT$_D$, as encoded in ``interaction terms'' of the bulk SymTh / SymTFT.

In many cases of interest, the anomalies of the boundary QFT$_D$ can also be extracted directly from geometry. In M-theory, one can start with the 11D topological terms $C_3 \wedge G_4 \wedge G_4$ and $C_{3} \wedge X_{8}(R)$, and a suitable dimensional reduction on $\partial X$ then results in a general procedure for extracting some part of the SymTh / SymTFT interaction terms, and the corresponding anomalies of the QFT$_{D}$.
In practice, this amounts to resolving the singularities of $X$, and then extracting the intersection numbers for cycles in the resolved geometry. Insofar as the anomalies of interest are independent of the choice of resolution, this provides a well-defined procedure for extracting the defining data of the SymTh / SymTFT engineered by $X$. In practice carrying out such calculations can quickly become unwieldy.

There are also more conceptual limitations to using purely intersection-theoretic techniques to extract these anomalies. For example, in many cases of interest the QFT$_D$ also enjoys a non-abelian flavor symmetry, as engineered by singularities of $X$ which extend out to $\partial X$. The resolution procedure also breaks these flavor symmetries, making it more subtle to extract this data. One way to bypass these concerns is to instead work directly with the singular geometry, focusing entirely on the geometry ``at infinity,'' namely $\partial X$. This is the strategy adopted in \cite{Cvetic:2025lat}, where it was shown that for 5D superconformal field theories (SCFTs) engineered by M-theory on orbifolds $\mathbb{C}^{3} / \Gamma$, suitable $\eta$-invariants and their refinements can be used to extract both the $p$-form symmetries, as well as refinements involving the flavor symmetries and their embedding in higher-group symmetries of the 5D SCFT. This procedure was explicitly carried out in \cite{Cvetic:2025lat} for boundary geometries given by generalized lens spaces $S^{5} / \Gamma$ with $\Gamma$ a finite abelian group.

But there are also many local geometries where extracting the necessary $\eta$-invariants is either difficult to extract, or simply unknown. This includes many cases of interest, including 5D SCFTs engineered by toric Calabi-Yau threefolds which are not orbifolds. In such situations, it would clearly be desirable to have a complementary method which computes the relevant anomaly data directly from the singular space $X$.

In this work we present a uniform procedure for extracting both the generalized symmetries, as well as anomalies directly from the singular space $X$. We focus on the case of 5D SCFTs engineered by singular local threefolds $X$, though we expect the method to carry through for more general backgrounds engineered in string theory. The main idea underlying our approach is to focus on the structure of branes probing the local singular geometry. For example, in a 5D SCFT engineered from M-theory on $X$, the reduction on a circle leads to a 4D KK theory. The corresponding KK momenta are interpreted as D0-brane particles probing the singularity, realizing a quiver quantum mechanics which directly tracks the electric and magnetic BPS defects of the 5D theory and their reduction to 4D. As shown in \cite{DelZotto:2022fnw}, the defect group (which encodes the candidate electric one-form and magnetic two-form symmetries) of the 5D SCFT is encoded in the adjacency matrix of this quiver simply because this information is tantamount to computing the Dirac pairing of the 4D KK theory (see \cite{DelZotto:2022ras}). One of our aims in this work will be to show how additional refined quiver data encodes the anomalies of the 5D SCFT, and its IIA counterpart (after dimensional reduction).

The main idea underlying our approach is to reinterpret the $\eta$-invariants of $X$ in terms of distinct paths modulo F-term relations of the corresponding quiver gauge theory. This geometric correspondence has been most fully explored in IIB backgrounds with D3-branes probing $X$, and this is related (after compactifying on a $T^3$ and T-dualizing) to IIA with D0-branes probing the same geometry, so we can freely interchange the two setups insofar as we are primarily interested in extracting the relevant geometric data of $X$.

At the level of extracting the relevant geometric data directly from our probe theory, we can equally well work either with type IIA and D0-branes probing $X$, or with type IIB and D3-branes probing the same geometry. In particular, we show that the $\eta$-invariant / anomaly coefficient of the SymTh / SymTFT is extracted from a matrix-valued quiver Hilbert series $H_{ij}(t)$ \cite{Eager:2010yu, Martelli:2006yb} (see also \cite{10.1093/qmath/han016}) constructed from independent (up to F-term relations) paths $i \rightarrow j$ weighted by $t^{\Delta_{ij}}$ with $\Delta_{ij}$ the scaling dimension of the corresponding chiral mesonic operator associated with the path.

Our main proposal is that the $\eta$-invariant, and hence the anomaly of the one-form symmetry, is computed by a residue of the difference between the twisted and untwisted Hilbert series:
\begin{equation}
\alpha = \frac{1}{2}\eta= \operatorname*{Res}_{t=1} \left[ \frac{H_{\mathrm{tw}}(t)-H_{\mathrm{untw}}(t)}{1-t}\right] \qquad \mathrm{mod}\,1,
\label{eq:IntroEtaFromHilbert}
\end{equation}
where $H_{\mathrm{tw}} = H_{0j}$ counts paths from $i = 0$ to some other node $j \neq i$, and $H_{\mathrm{untw}} = H_{00}$ counts paths from the node $i = 0$ associated with the trivial sheaf $\mathcal{O}_X$ back to itself.\footnote{In many cases the quantum symmetry / automorphism of the quiver allows us to instead use another node besides $i = 0$.}

This perspective has several advantages. First, it bypasses the need to resolve the singular geometry. The computation is performed directly in the singular theory and applies equally well to isolated and non-isolated singularities. Additionally, this procedure extends to examples where a direct geometric computation of the $\eta$-invariant is not presently available. In these cases, one should view the results of this paper as quiver based predictions for the $\eta$-invariant. Finally, it gives a concrete field-theoretic interpretation of the geometric $\eta$-invariant.

The treatment of non-isolated singularities requires one additional refinement. When the orbifold action has fixed loci on the boundary $Y$, additional flavor sectors are supported on lower-dimensional strata. These flavor symmetries can mix with the one-form symmetry through a 2-group structure. We find that this data is captured by tracking the higher-order poles of the Laurent expansion:
\begin{equation}
\alpha^{\mathrm{ref}}= \frac{1}{2}\eta^{\mathrm{ref}} = {\operatorname*{Res}}^{(2)}_{t=1} \left[ \frac{H_{\mathrm{tw}}(t)-H_{\mathrm{untw}}(t)}{1-t} \right] + \operatorname*{Res}_{t=1} \left[ \frac{H_{\mathrm{tw}}(t)-H_{\mathrm{untw}}(t)}{1-t} \right] \qquad \mathrm{mod}\,1,
\label{eq:IntroRefinedEta}
\end{equation}
where Res$^{(2)}$ captures data from the $(1-t)^{-2}$ term in the Laurent expansion of the Hilbert series. This gives the quiver realization of the refined $\eta$-invariant appropriate to stratified singularities.

We illustrate the proposal in several classes of examples. For abelian orbifolds $X=\mathbb{C}^3/\mathbb{Z}_N$, the quiver computation reproduces the known defect groups and twisted $\eta$-invariants of the generalized lens-space link $S^5/\mathbb{Z}_N$. We then study examples with non-isolated singularities, where the quiver detects additional contributions associated with lower-dimensional fixed loci and reproduces the expected refined anomaly data. We also apply the same method to non-orbifold toric singularities such as the cones over $Y^{p,q}$ \cite{Gauntlett:2004yd}. In these cases the relevant quiver and its Hilbert series are known, while a direct geometric computation of the corresponding $\eta$-invariant is generally unavailable.\footnote{For some recent progress in extracting $\eta$-invariants in related geometries, see reference \cite{savale2026etainvariantcirclebundle}.} The quiver therefore provides a practical way to predict the anomaly data of the associated symmetry theory. 
Additional examples including product orbifolds with product groups $\Gamma = \mathbb{Z}_N \times \mathbb{M}$ and discrete torsion (in the case of related IIA backgrounds) are deferred to an Appendix.

The rest of this paper is organized as follows. In section \ref{sec:REVIEW} we review some aspects of 
5D SCFTs and their symmetry theories. In sections \ref{sec:QUIVERS} and \ref{sec:Refinements} we present our proposal for extracting all of the relevant symmetry and anomaly data for these theories directly from the brane probe quiver gauge theory. In section \ref{sec:EXAMPLES} we present a variety of examples. In section \ref{sec:CONC} we present our conclusions and directions for future research. In Appendix \ref{app:ToricGeometry} we compare our $\eta$-invariants to known anomaly coefficients computed via toric geometry. We defer details on the quiver adjacency matrices to Appendices \ref{app:QuiverAdjacency} and \ref{app:QuantumSymmetry}. Appendix \ref{app:DISCTORSION} provides a treatment of IIA orbifolds with discrete torsion.

\section{5D SCFTs and Symmetry Theories} \label{sec:REVIEW}

In this section we review some aspects of 5D SCFTs engineered by geometry, as well as their corresponding SymTh / SymTFTs.\footnote{See \cite{Seiberg:1996bd, Douglas:1996xp, Intriligator:1997pq} for some early constructions of 5D SCFTs as well as \cite{Argyres:2022mnu} for a recent brief review.}
We shall be interested in 5D SCFTs engineered from M-theory on the background $\mathbb{R}^{4,1} \times X$ with $X$ a local Calabi-Yau threefold given by a singular cone of the form $X = \mathrm{Cone}(Y)$ with $Y = \partial X$. The local degrees of freedom of the 5D SCFT are localized at the singular tip of the cone, and additional flavor symmetry factors will be engineered by allowing additional singularities to emanate out from the tip to the boundary $\partial X$. We realize an SCFT when the Calabi-Yau cone has collapsing divisors, i.e., we have effective strings which have vanishing tension as we approach the singular point in moduli space \cite{Seiberg:1996bd}. Heavy defects arise from branes which stretch from the singularity out to the boundary $\partial X$, and these can be partially screened by the dynamical states obtained from branes wrapped on collapsing cycles of the conical geometry. Extracting the overall higher-form symmetry depends to some extent on whether the $\partial X = Y$ geometry also contains singularities, i.e., flavor symmetries as well. Nevertheless, there is by now an algorithmic procedure to read this data directly from the geometry, see in particular \cite{Cvetic:2022imb, DelZotto:2022joo, Acharya:2023bth, Cvetic:2024dzu}.

Our aim in subsequent sections will be to extract the relevant symmetry data for these QFTs directly from quiver gauge theory data. We begin by discussing how to read off generalized global symmetries from such backgrounds, and how this is packaged in terms of the defect group. We then turn to a discussion of anomalies and the structure of the SymTh / SymTFT.

\subsection{Higher-Form Symmetries}

The geometric data of heavy defects and their screening can be read off from the BPS quiver of the 4D KK theory obtained from reducing the 5D SCFT on a circle. Indeed, as found in \cite{DelZotto:2022fnw}, the spectrum of unscreened electric and magnetic line operators is economically captured in terms of the adjacency matrix of the D0-brane quiver quantum mechanics obtained from brane probes of the tip of the conical geometry $X$.\footnote{We review the procedure for extracting the quiver gauge theory from a brane probe in Appendix \ref{app:QuiverAdjacency}.}
On general grounds, this is to be expected because the quiver gauge theory directly packages the basis of fractional branes / collapsed branes in terms of stable quiver representations of the Calabi-Yau (see \cite{Aspinwall:2004jr} for a review). As found in \cite{DelZotto:2022fnw, DelZotto:2022ras, Braeger:2024jcj, Braeger:2025rov}, the corresponding Dirac pairing for the 4D KK theory is given by:
\begin{equation}
B = A - A^{T},
\end{equation}
where $A$ is the directed adjacency matrix of the probe brane theory. In the 4D KK theory, there are therefore unscreened electric and magnetic one-form symmetries and these are obtained directly from \cite{DelZotto:2022fnw}. Lifting back to 5D, we have:\footnote{This formula also works for type IIA backgrounds of the form $\mathbb{R}^{3,1} \times X$ with discrete torsion, as well as non-supersymmetric orbifolds,
where one uses the adjacency matrix for fermionic degrees of freedom in the quiver \cite{Braeger:2024jcj, Braeger:2025rov}.}
\begin{equation}
\mathrm{Tor}(\mathrm{Coker}(B)) = \mathbb{D} = \mathbb{D}^{(1)}_{\mathrm{el}} \oplus \mathbb{D}^{(2)}_{\mathrm{mag}}.
\end{equation}
We remark that additional data concerning the flavor symmetry algebra of the 5D SCFT and accompanying 4D KK theory can also be extracted directly from $A$, a feature we return to in section \ref{sec:QUIVERS}. The candidate electric one-form symmetries of the defect group directly lift to one-form symmetries of the 5D theory, and the magnetic one-form symmetries lift to magnetic two-form symmetries of the 5D theory. The absolute form of the theory is specified by a (non-anomalous) choice of polarization. For the most part, we shall implicitly specialize to the electric polarization since there can be obstructions to other polarizations.

\subsection{SymTh / SymTFT, and Anomalies}

The global symmetry data of a $D$-dimensional QFT$_{D}$ can conveniently be packaged in terms of a higher-dimensional SymTh / SymTFT. In the SymTFT framework, the higher-dimensional bulk is a TFT, but more generally, a SymTh can support gapless free fields. Both situations naturally arise in stringy constructions, and so we shall aim to keep our discussion general enough to cover both sorts of possibilities. We comment that typically, this bulk system is taken to reside in a $(D+1)$-dimensional space by extending the original spacetime $M_{D}$ by an interval $M_D \times I$, with one end supporting a relative QFT, i.e., a physical boundary condition, and the other supporting a topological or free boundary condition, i.e., a specification of the absolute form of the QFT$_D$. In some situations it is convenient to continue decompressing the symmetry data, eventually filtering it to a fully gapped bulk, with different gapless modes localized on edges and corners of this higher-dimensional system \cite{Cvetic:2024dzu}.

Given a quantum field theory in $D$ spacetime dimensions with global symmetry $G$, one may couple the theory to background gauge fields $B$ for $G$ and consider its partition function
\begin{equation}
Z_D[B].
\end{equation}
In the presence of an anomaly, this partition function is not invariant under background gauge transformations, but instead transforms by a phase which we schematically write as:\footnote{We use additive notation for gauge transformations.}
\begin{equation}
Z_D[B + \delta B] = Z_D[B] \exp\left(2\pi i \int_{M_D} \omega_D(B,\delta B)\right),
\end{equation}
where $\omega_D$ is a local functional characterizing the anomaly.

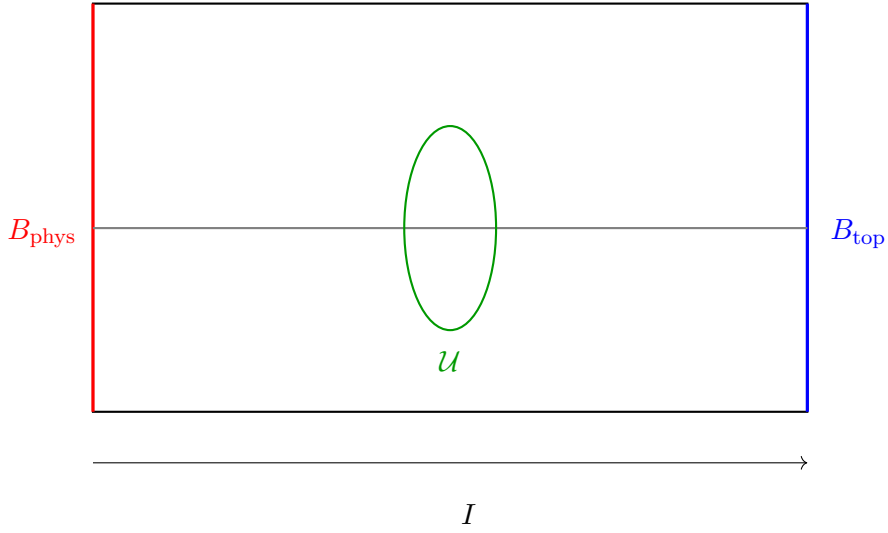
\begin{figure}[t]
\centering
\begin{tikzpicture}[scale=1.35, every node/.style={font=\small}]
\begin{scope}[rotate=270, shift={( -3,-2)}, shift={(2,-3)}]
    \draw[thick] (0,0) rectangle (4,7);

    \draw[very thick, blue] (0,7) -- (4,7);
    \draw[very thick, red] (0,0) -- (4,0);

    \node[blue, above] at (2.5,7.5) {$B_{\mathrm{top}}$};
    \node[red, below] at (2,-0.5) {$B_{\mathrm{phys}}$};

    \draw[->] (4.5,0) -- (4.5,7);
    \node[right] at (5,3.5) {$I$};

    \draw[thick, gray] (2.2,7) -- (2.2,0);

    \draw[thick, green!60!black] (1.2,3.5) arc[start angle=180,end angle=540,x radius=1.0,y radius=0.45];
    \node[green!60!black] at (3.5,3.5) {$\mathcal{U}$};
\end{scope}
\end{tikzpicture}
\caption{The anomaly inflow / SymTFT slab picture. The $D$-dimensional theory is realized as a physical boundary condition $B_{\mathrm{phys}}$ of a $(D+1)$-dimensional bulk SymTFT, while $B_{\mathrm{top}}$ denotes a topological boundary condition. Extended defects stretch across the slab, and topological symmetry operators (e.g., $\mathcal{U}$ in the figure) act on them by linking.}
\label{fig:SymTFTslab}
\end{figure}

This anomalous variation can be canceled by introducing a $(D+1)$-dimensional invertible topological theory whose action $S_{D+1}[B]$ satisfies
\begin{equation}
\delta S_{D+1}[B] = - \int_{M_D} \omega_D(B,\delta B).
\end{equation}
The combined system,
\begin{equation}
Z_{\mathrm{tot}} = Z_D[B] \,
\exp\left(2\pi i \int_{M_{D+1}} S_{D+1}[B]\right),
\end{equation}
is gauge invariant. The $(D+1)$-dimensional topological theory is the SymTh / SymTFT, and encodes the anomaly via inflow (see figure \ref{fig:SymTFTslab}).

In the case of immediate interest where we have a 5D SCFT and a discrete one-form symmetry, the background field is a torsion valued two-form gauge field $B_2$, and the anomaly is captured by topological couplings constructed from $B_2$ and characteristic classes of spacetime. There can in principle also be various zero-form symmetries of the 5D SCFT, and these are characterized by additional bulk one-form gauge fields.

In this work the relevant terms of the SymTh / SymTFT split up as a sum of generalized BF terms quadratic in the fields, and anomaly / interaction terms involving more than two fields:
\begin{equation}
S_{\mathrm{SymTh / SymTFT}} = S_{\mathrm{BF}} + S_\mathrm{anom}.
\end{equation}
The BF terms track the candidate $\mathbb{Z}_{K}$ electric one-form and magnetic two-form symmetries of the system, which we present as $\mathbb{Z}_{K}$ valued cocycles. In this normalization, the BF term takes the form:
\begin{equation}
S_{\mathrm{BF}} = \frac{2 \pi}{K} \int_{6D} B_2 \cup \delta C_3,
\end{equation}
where $B_2$ is the electric one-form potential and $C_3$ is its magnetic counterpart. The anomaly / interaction terms involve terms of the form $B_2 \cup B_2 \cup B_2$, as well as possible mixing terms with the continuous non-abelian symmetries obtained from non-isolated singularities:
\begin{align}\label{eq:SYMTHANOM}
S_{\mathrm{SymTH}}^{\mathrm{anom}} = 2\pi &\int_{6D} \left[ \beta  B_2 \cup B_2 \cup B_2 + \gamma B_2 \cup \frac{p_1}{4} \right] \nonumber \\
+ &\int_{6D} \left[ \frac{1}{2(2\pi)^2}\sum_{k} \delta_k\, B_2 \cup \mathrm{Tr}\left(F^{(k)}_2 F^{(k)}_2\right) + \frac{1}{6(2\pi)^3}\sum_{k} \epsilon_k\, \mathrm{Tr}\left(F^{(k)}_2 F^{(k)}_2 F^{(k)}_2\right)\right]\,,
\end{align}
Here, the first line constitute contributions involving possible anomalies of the one-form symmetry and a mixed anomaly with gravity.
The second line involves contributions with continuous non-abelian flavor symmetries with curvature / field strength $F^{(k)}$. The normalization for the field strength terms is chosen so that the corresponding Chern classes are integer normalized. Observe that a shift in the background value of $B_2$ can be compensated by a shift in the spin connection / possible flavor symmetries. As such, we shall seek to extract the combinations:
\begin{align}
\alpha &\equiv \beta + \gamma \qquad \mathrm{mod}\;1 \\
\alpha^{\mathrm{ref}} &\equiv \beta + \gamma + \underset{k}{\sum} \delta_k + \underset{k}{\sum} \epsilon_k \,\,\, \mathrm{mod} \, 1,
\end{align}
where the ``ref'' refers to a refined combination which also detects contributions from possible flavor symmetries. In general, $\alpha$ can take fractional values which are more refined than the order of elements in the 1-form symmetry group itself. This is because it is measured according to elements in $H^{6}(B^2 G^{(1)}, U(1))$ (see \cite{Kapustin:2013uxa}) with $G^{(1)}$ the 1-form symmetry group.

Another comment is that even with imperfect data concerning the full structure of the non-abelian flavor symmetry group, we can still reliably extract the structure of its center, and in particular possible anomalies which involve this data. This information is enough to completely fix the coefficients $\delta_k$ and $\epsilon_k$ and as such we can extract an unambiguous answer even without knowing the full flavor group.\footnote{For example, in the 5D $T_3$ theory, the geometry detects an $\mathfrak{su}(3)^3$ flavor symmetry algebra which is known to enhance to $\mathfrak{e}_6$ \cite{Benini:2009gi}. The structure of the center-flavor / one-form symmetry mixing is nevertheless fully controlled by the considerations we present.}

In the specific context of 5D SCFTs, there are a few different strategies one can adopt to extract these interaction terms of the SymTh / SymTFT. First of all, when a gauge theory phase is available, one can hope to extract the relevant higher-form symmetries and anomalies through a direct gauge theory computation \cite{Gukov:2020btk, Apruzzi:2021vcu}. One concern with this approach is that it can in principle miss non-perturbative / accidental enhancements in the symmetry structure which are present directly at the conformal fixed point. A related geometric strategy is to choose a resolution of $X$ and then extract the coefficients of the bulk SymTh / SymTFT by dimensional reduction of the M-theory topological terms $C_3 \wedge G_4 \wedge G_4$ and $C_3 \wedge X_8(R)$ on the internal linking geometry $\partial X$. In practical terms, the calculation of anomaly coefficients then reduces to the computation of certain triple intersections of divisors modulo compactly supported ones (see Appendix \ref{app:ToricGeometry} for further details). This approach is subject to the same concern that certain intrinsic properties of the SCFT may be destroyed by working on the Coulomb branch, but one can explicitly check by working in different chambers of moduli space that the results are independent of a chosen chamber \cite{Apruzzi:2021nmk}. An additional worry, however, is that the resolution of singularities can modify / destroy the non-abelian symmetries engineered by the geometry, though one can often piece together complementary information to extract the necessary data of the SymTh / SymTFT.

An alternative strategy is to work directly with the singular geometry $X$, since this is the method for constructing the 5D SCFT in the first place. This approach was developed in \cite{Cvetic:2025lat}, where it was shown that the relevant coefficients of the SymTh / SymTFT are encoded in twisted $\eta$-invariants of $\partial X$.

To explain the appearance of $\eta$-invariants, let $X$ be an auxiliary bulk with boundary $Y=\partial X$, and let $L'\to X$ be a line bundle whose restriction to the boundary is $L=L'|_{\partial X}$. The Atiyah-Patodi-Singer theorem (see \cite{Atiyah:1975jf, Atiyah:1976jg, Atiyah:1976qjr}) gives the index formula for the Dirac operator $\slashed D$ twisted by $L'$. Schematically,
\begin{align}\label{eq:dirac_index}
\mathrm{Index}(\slashed{D},L',X)&=\int_X \mathrm{ch}(L')\,\hat A(TX) -\frac{1}{2}\left(\eta^{\slashed{D}}_{L}(Y)+h^{\slashed{D}}_L(Y)\right), 
\end{align}
Thus the bulk topological terms and the boundary spectral asymmetry are not independent, but rather are tied together by index theory. In particular, once the bulk integral is written in terms of characteristic classes, the boundary correction is precisely the corresponding APS $\eta$-invariant. For the Dirac operator, these are
\begin{equation}
\eta^{\slashed{D}}_{L}(\partial X) =\lim_{\epsilon\to 0^+}\sum_k e^{-\epsilon |\lambda_k|}\,\mathrm{sign}(\lambda_k),\qquad h^{\slashed{D}}_{L}(\partial X)= \dim \ker(\slashed{D}_{\partial X},L),
\label{eq:APSBoundaryDirac}
\end{equation}
where $\lambda_k$ are the eigenvalues of the induced tangential Dirac operator on $\partial X$ twisted by $L$.\footnote{Of course the boundary is not a complex manifold, but one can nevertheless speak of a $\overline{\partial}$-type operator, as constructed from the bulk.}

Writing $c_1(L')=T_2$, one can expand \ref{eq:dirac_index} up to the degree six term to get 
\begin{equation}
\alpha=\frac{1}{6}\int_X T_2 \left( T_2^2-\frac{p_1(TX)}{4} \right)\qquad \mathrm{mod}\;1.
\label{eq:AlphaTodd}
\end{equation}
Applying the APS theorem to the twisted and untwisted $\slashed{D}$-operators and subtracting the untwisted contribution
removes the bundle-independent bulk term. One then obtains
\begin{equation}
\alpha_L= \frac{1}{2}\Bigl(\eta^{\slashed{D}}_{L}(Y)-\eta^{\slashed{D}}_{\varnothing}(Y)\Bigr)+ \frac{1}{2}\Bigl(h^{\slashed{D}}_{L}(Y)-h^{\slashed{D}}_{\varnothing}(Y)\Bigr) \qquad \mathrm{mod}\;1.
\label{eq:AlphaRelativeGeneral}
\end{equation}
For the positively curved links relevant in this paper, the boundary kernels vanish, so this simplifies to
\begin{equation}
\alpha_L = \frac{1}{2}\Bigl(\eta^{\slashed{D}}_{L}(Y)-\eta^{\slashed{D}}_{\varnothing}(Y)\Bigr)\qquad \mathrm{mod}\;1.
\label{eq:AlphaRelativeEta}
\end{equation}
The main conceptual point is that the physically relevant anomaly coefficient is a relative $\eta$-invariant, namely the twisted boundary contribution minus the untwisted one.\footnote{See \cite{Cvetic:2025lat} for the full derivation.}

We now turn to the explicit computation of these $\eta$-invariants for geometries of the form $X=\mathbb C^n/\Gamma$ whose boundary is $Y=S^{2n-1}/\Gamma$. Let $\Gamma\subset U(n)$ be a finite subgroup admitting a lift $\widetilde\epsilon:\Gamma\to \widetilde U(n)$, and let $(\chi,V_\chi)$ be a representation of $\Gamma$. Degeratu’s theorem expresses the twisted $\eta$-invariant of the orbifold $S^{2n-1}/\Gamma$ in representation-theoretic form \cite{10.1093/qmath/han016}.\footnote{See also \cite{Bar1996}.} For the Dirac operator twisted by the orbibundle associated with $\chi$, one has
\begin{equation}
\eta_\chi=2(-1)^n \frac{1}{|\Gamma|}\sum_{\genfrac{}{}{0pt}{}{\gamma\in \Gamma}{\gamma\neq 1}} \frac{\mathrm{Tr}\left(\widetilde\epsilon(\gamma),D_{1/2}\right)\chi(\gamma)}{\det(1-\gamma)}.
\label{eq:DegeratuDirac}
\end{equation}
Equivalently, if $M_\chi(t)$ denotes the twisted Molien series, which counts the number of $\Gamma$-invariant holomorphic functions of the Calabi-Yau, then
\begin{equation}
\eta_\chi = 2(-1)^n \,\underset{t=1}{\mathrm{Res}}\, \frac{M_{\chi\otimes D_{1/2}}(t)}{1-t}.
\label{eq:DegeratuMolienDirac}
\end{equation}
When $\Gamma\subset SU(n)$, there is a canonical lift and the half-determinant representation becomes trivial, so the formula simplifies accordingly. In the notation of \cite{Cvetic:2025lat}, one has
\begin{equation}
\frac{1}{2}\eta^{\slashed{D}}_\chi\left(S^{2n-1}/\Gamma\right)= \frac{(-1)^n}{|\Gamma|}\sum_{\gamma\in \Gamma_{\mathrm{FF}}} \frac{\chi(\gamma)}{\det(1-\gamma)},
\label{eq:dbarEtaRepresentation}
\end{equation}
where $\Gamma_{\mathrm{FF}}\subset \Gamma$ is the subset of elements acting
fixed-point freely on the sphere. The twisted Molien’s series is given as follows
\begin{equation}
M_\chi(t)=\frac{1}{|\Gamma|}\sum_{\gamma\in\Gamma}\frac{\chi(\gamma)}{\det(1-t\gamma)}.
\label{eq:twisted-Molien}
\end{equation}
Thus, the expression for the $\eta^{\slashed{D}}$-invariant can be rewritten as
\begin{equation}
\frac{1}{2}\eta^{\slashed{D}}_\chi\!\left(S^{2n-1}/\Gamma\right)= (-1)^n\, \underset{t=1}{\mathrm{Res}}\,\frac{M_\chi(t)}{1-t}.
\label{eq:dbar-eta-Molien}
\end{equation}
This is the form that will be most useful later, since it is precisely the bridge between boundary $\eta$-invariants and the quiver computation.

For orbifolds of the form
\begin{equation}
X=\mathbb C^3/\mathbb Z_N(m_1,m_2,m_3),\qquad Y=S^5/\mathbb Z_N(m_1,m_2,m_3),
\end{equation}
with $\omega=e^{2\pi i/N}$ and generator acting as $(z_1,z_2,z_3)\mapsto (\omega^{m_1}z_1,\omega^{m_2}z_2,\omega^{m_3}z_3)$, the above formulas simplify considerably. If $L_Q$ is the line bundle associated with the one-dimensional representation of charge $Q$, then
\begin{equation}
\frac{1}{2}\eta^{\slashed{D}}_{L_Q}(Y) = -\frac{1}{N}\sum_{k=1}^{N-1} \frac{\omega^{kQ}} {(1-\omega^{k m_1})(1-\omega^{k m_2})(1-\omega^{k m_3})}.
\label{eq:dbarEtaCyclic}
\end{equation}

The reason the relative $D$-invariant is the object we actually want is now clear. The physical anomaly is not the absolute $\eta$-invariant of some chosen twist, but the change in the boundary APS correction relative to the untwisted sector. Equivalently,
\begin{equation}
\alpha_L \sim \frac{1}{2}\Bigl(\eta^{\slashed{D}}_{\mathrm{twisted}} - \eta^{\slashed{D}}_{\mathrm{untwisted}}\Bigr)\qquad \mathrm{mod}\;1,
\label{eq:AlphaRelativeEtaShort}
\end{equation}
and it is this relative quantity which will later be reproduced by the quiver based computation. Put differently, the untwisted subtraction removes the universal background contribution and isolates the symmetry-sensitive part of the spectral asymmetry. This is exactly the combination that matches the one-form anomaly coefficient in the SymTFT.

We conclude this section noting that a vanishing $\alpha_L$ coefficient is not enough to conclude that the one-form symmetry is free of anomalies. Indeed, $\alpha_L=\beta_L+\gamma_L$ and it might be zero if $\beta_L=-\gamma_L\neq0$. In order to extract the coefficients $\beta_L$ and $\gamma_L$, one can compute $\frac{1}{2}\eta^{\slashed{D}}_{L_Q}(Y)$ and $\frac{1}{2}\eta^D_{L_{2Q}}(Y)$ as in \cite{Cvetic:2025lat}, leading to
\begin{equation}
\beta_L = \frac{1}{12}\eta^{\slashed{D}}_{L_{2Q}}(Y) - \frac{1}{6}\eta^{\slashed{D}}_{L_Q}(Y) \, , \quad \gamma_L = \frac{-1}{12}\eta^{\slashed{D}}_{L_{2Q}}(Y) + \frac{2}{3}\eta^{\slashed{D}}_{L_Q}(Y) \, .
\end{equation}
Only when both coefficients are zero, the one-form symmetry is free of anomalies and can be gauged.

\subsection{Illustrative Example: \texorpdfstring{$\mathbb C^3/\mathbb Z_3$}{C3/Z3}} \label{sec:IllustrativeExamplePart1}

We now illustrate the general discussion in the simplest isolated example,
\begin{equation}
X=\mathbb C^3/\mathbb Z_3(1,1,1),\qquad Y=\partial X=S^5/\mathbb Z_3.
\end{equation}
Let $g$ denote the generator of $\mathbb Z_3$, acting on the coordinates of $\mathbb C^3$ as
\begin{equation}
g:(z_1,z_2,z_3)\mapsto (\omega z_1,\omega z_2,\omega z_3), \qquad \omega=e^{2\pi i/3}.
\end{equation}
Since the orbifold group is abelian and acts fixed point free, the defect group contains an electric $\mathbb Z_3$ factor, and hence the 5D SCFT carries a $\mathbb Z_3$ one-form symmetry. Geometrically, the charged line defect is realized by an M2-brane wrapping a non-compact relative two-cycle, while the dual topological operator is realized by an M5-brane wrapping a boundary three-cycle in $Y$ that links with it.

Our goal is to determine the anomaly coefficient
\begin{equation}
\alpha=\beta+\gamma \qquad \mathrm{mod}\; 1
\end{equation}
for this $\mathbb Z_3$ one-form symmetry. As explained in the previous subsection, $\alpha$ is computed by a relative $\eta^{\slashed{D}}$-invariant, or equivalently by the difference between the twisted and untwisted Molien series. More precisely, we have
\begin{equation}
\alpha = \frac{1}{2}\left( \eta^{\slashed{D}}_{\mathrm{tw}}(Y) - \eta^{\slashed{D}}_{\mathrm{untw}}(Y) \right) = \underset{t=1}{\mathrm{Res}}\,\frac{M_{\mathrm{tw}}(t)-M_{\mathrm{untw}}(t)}{1-t}.
\label{eq:AlphaRelativeResidueExample}
\end{equation}

We begin with the untwisted Molien series, corresponding to the trivial character of $\mathbb Z_3$:
\begin{equation}
M_{\mathrm{untw}}(t)=\frac{1}{3}\sum_{k=0}^{2}\frac{1}{(1-\omega^k t)^3} = \frac{1+7t^3+t^6}{(1-t^3)^3}.
\label{eq:MolienUntwistedZ3Closed}
\end{equation}

Next we turn to the nontrivial twist. Let $\chi_\omega$ be the $\mathbb{Z}_3$ character with $\chi_\omega(g)=\omega$. The corresponding twisted Molien series is
\begin{equation}
M_{\mathrm{tw}}(t) = \frac{1}{3}\sum_{k=0}^{2}\frac{\chi_\omega(g^k)^{-1}}{(1-\omega^k t)^3}=\frac{1}{3}\sum_{k=0}^{2} \frac{\omega^{-k}}{(1-\omega^k t)^3} =\frac{3(t+2t^4)}{(1-t^3)^3}.
\label{eq:MolienTwistedZ3Closed}
\end{equation}
This is the Molien series associated with the nontrivial $\mathbb Z_3$ twist relevant for the symmetry defect.

We now take the relative residue. Using
\eqref{eq:MolienUntwistedZ3Closed} and
\eqref{eq:MolienTwistedZ3Closed}, we find
\begin{equation}
\alpha= \underset{t=1}{\mathrm{Res}}\,\frac{M_{\mathrm{tw}}(t)-M_{\mathrm{untw}}(t)}{1-t}=-\frac{1}{9}\qquad \mathrm{mod}\;1.
\label{eq:AlphaZ3Example}
\end{equation}
Equivalently,
\begin{equation}
\frac{1}{2}\left(\eta^{\slashed{D}}_{\mathrm{tw}}(Y)-\eta^{\slashed{D}}_{\mathrm{untw}}(Y)\right)=-\frac{1}{9}\qquad \mathrm{mod}\;1.
\end{equation}
Thus the anomaly theory for the $\mathbb Z_3$ one-form symmetry contains the
topological coupling
\begin{equation}
S_{\mathrm{SymTFT}}^{\mathrm{anom}}\;=\; 2\pi\left(  \int_{6D} \beta B_2\cup B_2\cup B_2+ \gamma \int_{6D} B_2\cup \frac{p_1}{4} \right),
\end{equation}
in the sense that the combination of coefficients obeys:
\begin{equation}
\alpha=\beta+\gamma=-\frac{1}{9}\qquad \mathrm{mod}\;1.
\end{equation}
The individual contributions in this case are $\beta = 1/18$ and $\gamma = -1/6$ (see Appendix \ref{app:ToricGeometry} for details).

This example already exhibits the main theme of the paper. The one-form symmetry is determined by the topology of the link $Y=S^5/\mathbb Z_3$, while its anomaly is extracted from a relative $\eta$-invariant on the same boundary geometry. In the next section we will recover this same quantity by a much simpler calculation using generating functions associated with the brane-probe theory.

Let us note that although $B_2$ is a $\mathbb{Z}_3$ valued flat gerbe, the $\mathbb{Z}_9$ valued anomaly is consistent with the topological classification of finite group anomalies. Indeed, following \cite{Kapustin:2013uxa}, we expect this anomaly to be captured by a class in $H^6(B^2\mathbb{Z}_3,U(1)) \cong \mathbb{Z}_9$. (This is true for the $B_2^3$ term, while the latter should be associated to a cobordism class, being a mixed anomaly between the center symmetry and gravity.)

\section{Quiver Approach to SymTh / SymTFTs} \label{sec:QUIVERS}

In the previous section we illustrated a geometric approach to extracting the SymTh / SymTFT for QFTs engineered from singular string / M-theory / F-theory backgrounds. We also saw that there are limitations to a purely geometric approach to extracting the relevant interaction terms of the SymTh / SymTFT. Motivated by these considerations, our aim in this section will be to extract the relevant data directly from the quiver gauge theory defined by a brane probe of the singular geometry.

Our general expectation is that the relevant data of the SymTh / SymTFT is encoded in the defining data of a quiver gauge theory, in particular the gauge group, matter content and interaction terms of the theory. In this section we demonstrate that this expectation is borne out. We shall develop an algorithm for counting paths in the quiver modulo F-term relations, as specified by the superpotential of the 4D $\mathcal{N} = 1$ theory obtained from D3-branes probing $X$, as well as its descendant as obtained from compactification on a $T^3$, i.e., the theory of D0-branes probing $X$. For each such path, the geometric content of holomorphic volumes in $X$ is encoded (in the 4D treatment) in terms of the scaling dimension / R-charge assignments (as extracted from $a$-maximization \cite{Intriligator:2003jj} in the 4D $\mathcal{N} = 1$ SCFT for bifundamental fields in the quiver).\footnote{Strictly speaking, we ought to only discuss scaling dimension / R-charge assignments for gauge invariant operators of the 4D SCFT, but for holomorphic quantities there is no ambiguity in making such assignments to individual bifundamental fields. Another comment is that although the gravity dual of $a$-maximization is volume minimization \cite{Martelli:2006yb}, this only involves the metric in a rather indirect way; only algebraic numbers actually appear, retaining the topological / number theoretic structure expected in the corresponding $\eta$-invariants and anomalies.}

This sort of counting problem was actually considered in \cite{Eager:2010yu, Eager:2012hx}, where an explicit correspondence between geometric quantities in the local Calabi-Yau $X$ and paths of the quiver gauge theory obtained from $N$ D3-branes probing $X$ was established. We shall be interested in weighted sums of closed paths $i \rightarrow i$, as associated to gauge invariant products of bifundamentals around the quiver, as well as weighted sums of open paths $i \rightarrow j$ for $i \neq j$, as associated to gauge non-invariant products of bifundamentals around the quiver. These appear to define sensitive quantities in the stringy setup and we will also propose a purely quiver theoretic interpretation of this data as well.

With these motivations in mind, our first aim will be to simply state the algorithm, and how it works, initially deferring a physical interpretation. With this in place, we then explain why this is a sensitive quantity to compute from the perspective of the quiver gauge theory. We then turn to some examples illustrating how our quantity matches to geometry, and goes beyond results attainable through current geometric methods.

\subsection{Quiver Hilbert Series and \texorpdfstring{$\eta$}{eta}-Invariants}

In this section we present our general algorithm for reading off the twisted $\eta$-invariants of $X$ directly
from a quiver Hilbert series. The main computational tool that will be used in the remainder of this paper is a matrix-valued Hilbert series derived from the brane-probe of $X$. The basic point is that this object provides a quiver-theoretic refinement of the Molien series discussed in the previous section. On the geometric side, the Molien series counts holomorphic functions on the Calabi-Yau cone $X$. On the gauge theory side, these same holomorphic functions are identified with chiral operators in the worldvolume theory of D3-branes probing $X$ constructed from the bifundamentals. Passing from the Molien series to the quiver Hilbert series is simply a shift in what is being counted: from geometric quantities to gauge theory quantities.

More concretely, let $X$ be a local Calabi-Yau threefold singularity. A brane probing this 
singularity then specifies a quiver gauge theory. The corresponding path algebra, modulo F-term relations, is the noncommutative algebra governing the holomorphic sector of the theory. Following \cite{Eager:2010yu} (see also \cite{Martelli:2006yb}), the Hilbert series is naturally promoted from a single generating function to a matrix $H(t)$, whose $(i,j)$-entry counts F-term equivalence classes of paths from node $i$ to node $j$, weighted by their total scaling dimension:
\begin{equation}
H_{ij}(t) \equiv \underset{(i \rightarrow j) / \mathrm{F-terms}}{\sum} t^{\Delta_{ij}},
\end{equation}
i.e., $t$ is a fugacity, and $\Delta_{ij} = 3 R_{ij} / 2$ is the scaling dimension, and $R_{ij}$ is the R-charge of the scalar operator associated with the chiral mesonic operator obtained by taking a product of bifundamentals starting at node $i$ and ending at node $j$. In this sense, the quiver Hilbert series is strictly richer than the ordinary Molien series: the diagonal entries count loops based at a given node, while the off-diagonal entries count open paths interpolating between distinct fractional branes.

Formally speaking, then, we can always construct a corresponding Hilbert matrix which counts paths $i \rightarrow j$ (up to relations) in the quiver. When $i = j$ we have a closed loop as obtained by taking products of bifundamental fields around the quiver. This defines a gauge invariant operator of the quiver gauge theory, and we can count these using standard techniques such as the superconformal index of the associated 4D $\mathcal{N} = 1$ SCFT obtained from a stack of D3-branes probing $X$.

On the other hand, we shall also be interested in counting paths $i \rightarrow j$ where $i \neq j$. Taking a product of bifundamentals in this case produces local operators of the 4D gauge theory $\mathcal{O}_{i \rightarrow j}(x)$ which are manifestly not gauge invariant. Our motivation for considering such paths stems from the geometric computation of twisted $\eta$-invariants for orbifolds $\mathbb{C}^3 / \Gamma$ performed in \cite{10.1093/qmath/han016}, where an additional weighting by a representation of $\Gamma$ figures in the construction of the $\eta$-invariant.
We give a physical interpretation of this counting problem shortly, but for now we observe that it is formally well-defined to construct the matrix Hilbert series in this way.

For Calabi-Yau three-fold quivers, the matrix Hilbert series takes a particularly simple closed form. The main input is the adjacency matrix $A_{ij}$ and the IR scaling dimension $\Delta_{ij}$ associated to each bifundamental scalar of the quiver. Introduce the fugacity weighted matrix $A(t)$ with entries:
\begin{equation}
A_{ij}(t) = A_{ij} t^{\Delta_{ij}} = A_{ij} t^{3R_{ij} / 2} \qquad \mathrm{(no\;summation)}.
\label{eq:WeightedAdjacency}
\end{equation}
The R-charge which appears here is that of the IR R-symmetry; it is fixed by $a$-maximization in the corresponding 4D $\mathcal{N} = 1$ SCFT \cite{Intriligator:2003jj}, which corresponds to volume minimization in the Sasaki-Einstein manifold $\partial X$ \cite{Martelli:2006yb}. In our conventions the R-charge of the scalar component of a free chiral multiplet is $2/3$ (i.e., scaling dimension $1$). Note that in spite of having such fractional powers, the Hilbert series still organizes according to integer powers. In what is to follow, we will refer to $A(t)$ as the weighted adjacency matrix and $A$ simply as the adjacency matrix.

As a matrix equation, the relevant matrix Hilbert series can then be extracted from the results of \cite{Eager:2010yu, bocklandt2006}:
\begin{equation}
H(t)=\Bigl(\mathbf 1-A(t)+t^3 A^{T}(t^{-1})-t^3 \mathbf 1\Bigr)^{-1},
\label{eq:GeneralQuiverHilbert}
\end{equation}
where $\mathbf 1$ is the identity matrix. This is the natural CY$_3$ specialization of the matrix Hilbert series formula derived from the projective resolution of a graded Calabi-Yau algebra.\footnote{Observe that although equation (\ref{eq:GeneralQuiverHilbert}) appears to explicitly depend only on the weighted adjacency matrix, the F-term relations (and thus the path algebra) is implicitly enforced via the Calabi-Yau condition, i.e., that the moduli space for the quiver gauge theory reconstructs motion of a probe brane on a singular Calabi-Yau threefold.} Physically, the first term counts the trivial path, the term $A(t)$ counts single-arrow insertions, the transpose term incorporates the F-term relations implied by the superpotential, and the final subtraction reflects the Calabi-Yau threefold condition. Equivalently, the matrix inverse resums all allowed quiver paths modulo the relations of the chiral ring.

\paragraph{Example: Orbifolds}

To illustrate, consider the special case of an orbifold $X=\mathbb C^3/\Gamma$ where the general expression for the matrix Hilbert series greatly simplifies. Let $\{R_i\}$ denote the irreducible representations of $\Gamma$, and let $R_{\mathrm{pt}}$ be the point group representation by which $\Gamma$ acts on $\mathbb{C}^3$. One obtains $A_{ij}$, the McKay adjacency matrix by computing
\begin{equation}
R_{\mathrm{pt}} \otimes R_i=\bigoplus_j A_{ij}\, R_j .
\end{equation}
From, this we can compute the weighted adjacency matrix as
\begin{equation}
    A(t) = A_{ij}t,
    \label{eq:C3WeightedAdjacency}
\end{equation}
as every bifundamental field carries a R-charge of $R=2/3$ in this case.
Then, the matrix Hilbert series is given by
\begin{equation}
H^{(\mathrm{orb})}(t)=\left(\mathbf 1-At+ A^{T}t^{2}-t^3 \mathbf 1\right)^{-1}.
\label{eq:EagerMatrixHilbert}
\end{equation}

A general comment here is that for orbifolds, all the R-charge assignments are equal. This 
does not hold for more general quiver gauge theories, e.g., the $Y^{p,q}$ examples we consider later. In this case, the structure of the weighted adjacency matrix will necessarily be more intricate.

The $(i,j)$-entry $H_{ij}(t)$ counts paths from node $i$ to node $j$, or equivalently open strings / operators transforming from the $i$-th fractional brane sector to the $j$-th one. The diagonal entries $H_{ii}(t)$ count loops based at node $i$. We single out $i = 0$ for the distinguished trivial sheaf $\mathcal{O}_X$, which for orbifolds $\mathbb{C}^3 / \Gamma$ is associated with the trivial 1D representation. $H_{00}(t)$ corresponds to the ordinary Hilbert series of the orbifold.

This last point is the direct bridge back to the Molien series. Recall that in the previous section we presented the ordinary Molien series for $\Gamma\subset SU(3)$:
\begin{equation}
M(t)=\frac{1}{|\Gamma|}\sum_{g\in \Gamma}\frac{1}{\det(1-t R(g))},
\end{equation}
which counts $\Gamma$-invariant holomorphic functions on $\mathbb C^3$. Since holomorphic functions on the cone correspond to mesonic chiral operators in the probe theory, this is equally the generating function for gauge-invariant mesonic operators in the quiver. In the Hilbert matrix formalism, the Molien series is refined to
\begin{equation}
H_{ij}(t)=\frac{1}{|\Gamma|}\sum_{g\in\Gamma}\frac{\chi_i(g)\chi_j(g^{-1})}{\det(1-t R(g))},
\label{eq:MatrixMolienAgain}
\end{equation}
where $\chi_i$ is the character of the irreducible representation $R_i$. This is precisely the matrix-valued Molien series appearing earlier. The ordinary untwisted Molien series is recovered as the $(0,0)$-component,
\begin{equation}
\label{eq:H00Molien}
H_{00}(t)=\frac{1}{|\Gamma|}\sum_{g\in\Gamma}\frac{1}{\det(1-t R(g))}=M(t),
\end{equation}
Observe that this choice is not completely unique, automorphisms of the quiver permute this to other choices. Summarizing, the usual Molien series is simply the diagonal trivial-representation entry of the full matrix Hilbert series.

The off-diagonal entries of the matrix Hilbert series are also important. For $i\neq j$, the function $H_{ij}(t)$ counts operators running from node $i$ to node $j$. These are not gauge-invariant loops at a single node, but rather open-string sectors between distinct fractional branes. They can be dressed by additional F1-strings which stretch to infinity, and thus can be counted as well (see subsection \ref{ssec:INTERP}).
From the representation-theoretic viewpoint, they are the Molien series refined by the insertion of nontrivial characters. In other words, the matrix Hilbert series organizes both the untwisted and twisted sectors simultaneously:
\begin{equation}
\text{untwisted sector}\longleftrightarrow H_{ii}(t),\qquad \text{refined/twisted sector}\longleftrightarrow H_{ij}(t)\quad (i\neq j).
\end{equation}
In particular, the off-diagonal entries reproduce exactly the twisted Molien series introduced previously, with the character insertions implementing the twist. This is the precise sense in which the matrix Hilbert series contains more information than the ordinary Molien series: it packages the entire family of twisted and untwisted generating functions.

It is useful to spell this out in representation-theoretic language. The coefficient of $t^n$ in $H_{ij}(t)$ may be written as
\begin{equation}
H_{ij}(t)=\sum_{n=0}^{\infty}\dim \operatorname{Hom}_{\Gamma}\left(R_j, \operatorname{Sym}^n(\mathbb C^3)\otimes R_i \right)t^n .
\end{equation}
Thus, $H_{ij}(t)$ counts degree-$n$ monomials in the orbifold coordinates, but with prescribed source and target representation labels. The diagonal case $i=j$ reduces to the usual invariant counting, while $i\neq j$ records how the polynomial ring decomposes into nontrivial $\Gamma$-representations. This is exactly what one should expect from the quiver: the matrix index remembers not only how many operators exist, but also where in the quiver they begin and end.

\subsection{Anomalies and \texorpdfstring{$\eta$}{eta}-Invariants from the Quiver Hilbert Series} \label{sec:RevisitingEta}

We now return to the main physical quantity of interest in this paper: the anomaly of the discrete one-form symmetry of the 5D SCFT engineered by M-theory on a Calabi-Yau threefold $X$. As reviewed in the previous section, the corresponding anomaly theory in six dimensions contains two terms,
\begin{equation}
S_{\mathrm{SymTFT}}^{\mathrm{anom}}=2\pi\left( \beta \int B_2 \cup B_2 \cup B_2+\gamma \int B_2 \cup \frac{p_1}{4} \right),
\end{equation}
where $B_2$ is the background two-form gauge field for the one-form symmetry. The first term is the cubic self-anomaly, while the second is the mixed gravitational anomaly. For the purposes of the present analysis, the combination that is directly accessible from spectral data is
\begin{equation}
\alpha=\beta+\gamma \qquad \mathrm{mod}\;1.
\end{equation}

In the previous section, this quantity was expressed geometrically in terms of a relative $\eta$-invariant on the Sasaki-Einstein link $Y=\partial X$. More precisely, if $L$ denotes the flat bundle associated to the relevant symmetry defect, then one has
\begin{equation}
\alpha_L=\frac{1}{2}\Bigl(\eta^{\slashed{D}}_{L}(Y)- \eta^{\slashed{D}}_{\varnothing}(Y) \Bigr)\qquad \mathrm{mod}\;1.
\label{eq:AlphaRelativeEtaQuiver}
\end{equation}
Equivalently, in the Molien-series language this same quantity is given by the residue:
\begin{equation}
\alpha_L= \underset{t=1}{\mathrm{Res}} \frac{M_{\mathrm{tw}}(t)-M_{\mathrm{untw}}(t)}{1-t}\qquad\textrm{mod}\,1,
\label{eq:AlphaMolienResidue}
\end{equation}
where $M_{\mathrm{untw}}(t)$ is the untwisted Molien series and $M_{\mathrm{tw}}(t)$ is the twisted Molien series associated to the relevant one-form symmetry sector.

The point of this subsection is to rewrite this formula entirely in terms of the matrix Hilbert series $H_{ij}(t)$. In what follows, we refer to $i = j = 0$ as the path connecting the trivial sheaf $\mathcal{O}_{X}$ back to itself, i.e., it is a particular choice of fractional brane.
In particular, the diagonal entry $H_{00}(t)$ reproduces the untwisted Molien series, while the off-diagonal entries $H_{ij}(t)$ reproduce the twisted sectors. Thus the residue formula above admits a direct quiver-theoretic reformulation:
\begin{equation}
M_{\mathrm{untw}}(t)=H_{00}(t),\qquad M_{\mathrm{tw}}(t)=H_{0i}(t)\qquad (i\neq 0).
\end{equation}
Substituting this into \eqref{eq:AlphaMolienResidue}, we obtain
\begin{equation}
\alpha_i=\underset{t=1}{\mathrm{Res}} \frac{H_{0i}(t)-H_{00}(t)}{1-t}\qquad\textrm{mod}\,1, \qquad i\neq 0.
\label{eq:AlphaFromQuiverGeneral}
\end{equation}
The particular choice of $i$ corresponds to choosing a particular generator for the one-form symmetry. We make the physically motivated choice to twist by $H_{01}$ to remain sensitive to refinements coming from 2-group data. This is explained further in section \ref{sec:Refinements} and Appendix \ref{app:QuantumSymmetry}.

This is the basic quiver formula that we will use in the remainder of the paper. It expresses the physically relevant anomaly coefficient of the one-form symmetry of the 5D SCFT directly in terms of the matrix Hilbert series of the four-dimensional brane-probe quiver.

From the viewpoint of the previous subsection, the use of the matrix Hilbert series is fairly natural. The matrix already packages both untwisted and twisted sectors into a single object:
\begin{equation}
H_{00}(t) \text{ counts gauge-invariant operators}, \quad H_{0i}(t) \text{ counts operators in the ith twisted sector}.
\end{equation}
Taking their difference therefore produces the direct quiver analog of the relative Molien series used in the geometric computation. The residue at $t=1$ then isolates precisely the part of the asymptotic expansion that computes the relative $\eta$-invariant.

It is useful to write this in a slightly more parallel way with the previous section. There, the twisted Molien series satisfied
\begin{equation}
\frac{1}{2}\eta^{\slashed{D}}_{\chi} \left(S^{5}/\Gamma\right) =
 \underset{t=1}{\mathrm{Res}} \frac{M_{\chi}(t)}{1-t},
\end{equation}
and the physical anomaly coefficient was obtained only after subtracting the untwisted sector. 
In quiver language, the same logic yields
\begin{equation}
\frac{1}{2}\Bigl( \eta^{\slashed{D}}_{i} - \eta^{\slashed{D}}_{0} \Bigr) = \underset{t=1}{\mathrm{Res}} \frac{H_{0i}(t)-H_{00}(t)}{1-t},
\label{eq:EtaDifferenceFromHij}
\end{equation}
so that
\begin{equation}
\alpha_i = \frac{1}{2} \Bigl( \eta^{\slashed{D}}_{i}- \eta^{\slashed{D}}_{0} \Bigr) = \underset{t=1}{\mathrm{Res}}\frac{H_{0i}(t)-H_{00}(t)}{1-t} \qquad \mathrm{mod}\,1.
\end{equation}
Thus the quiver Hilbert series reproduces both the counting of operators as well as the same relative spectral invariant that governs the anomaly of the one-form symmetry in the 5D SCFT.

This reformulation is powerful because it replaces a geometric computation on the link $Y=\partial X$ by a purely quiver-theoretic calculation. Once the matrix Hilbert series is known, the anomaly coefficient follows immediately by taking a residue. In particular, this gives a practical route to examples for which a direct geometric treatment is cumbersome.

A few comments are in order. First, the subtraction by $H_{00}(t)$ is essential: it removes the universal untwisted contribution and isolates the symmetry-sensitive part of the spectral asymmetry, exactly as in the geometric APS analysis. Second, the choice of the $i$-th entry $H_{0i}(t)$ has a precise meaning. Any non-diagonal component probes a nontrivial twisted sector, and hence may be used to extract the the various $\eta$-invariant for different choices of the flat bundle $L$.

\subsection{Physical Interpretation} \label{ssec:INTERP}

We have now presented a self-contained algorithm for extracting the anomalies of the
SymTh / SymTFT directly from the matrix Hilbert series of a quiver gauge theory. On the other hand, this
prescription requires us to count paths associated with operators which are not gauge invariant in the quiver. Our aim
in this section will be to provide a physical interpretation, explaining why the counting problem
is still physically meaningful.

\begin{figure}[t!]
\centering
\makebox[\textwidth][c]{%
\begin{tikzpicture}[
    >=Stealth,
    gauge/.style={circle,draw,thick,minimum size=12mm,inner sep=1pt,fill=white},
    flavor/.style={rectangle,draw,thick,minimum size=10mm,inner sep=1pt,fill=white},
    bifund/.style={-{Stealth[length=2.7mm,width=2.0mm]},thick},
    flavarrow/.style={-{Stealth[length=2.2mm,width=1.7mm]},thick},
    every node/.style={font=\small}
]

\def\R{3.2}
\def\F{5.4}

\node[gauge] (g0) at (150:\R) {$G_{i-3}$};
\node[gauge] (g1) at (100:\R) {$G_{i-2}$};
\node[gauge] (g2) at ( 50:\R) {$G_{i-1}$};
\node[gauge] (g3) at (  0:\R) {$G_i$};
\node[gauge] (g4) at (-50:\R) {$G_{i+1}$};
\node[gauge] (g5) at (-100:\R) {$G_{i+2}$};
\node[gauge] (g6) at (-150:\R) {$G_{i+3}$};

\node[flavor] (f0) at (150:\F) {$F_{i-3}$};
\node[flavor] (f1) at (100:\F) {$F_{i-2}$};
\node[flavor] (f2) at ( 50:\F) {$F_{i-1}$};
\node[flavor] (f3) at (  0:\F) {$F_i$};
\node[flavor] (f4) at (-50:\F) {$F_{i+1}$};
\node[flavor] (f5) at (-100:\F) {$F_{i+2}$};
\node[flavor] (f6) at (-150:\F) {$F_{i+3}$};

\draw[bifund] (g0) to[bend left=16] (g1);
\draw[bifund] (g1) to[bend left=16] (g2);
\draw[bifund] (g2) to[bend left=16] (g3);
\draw[bifund] (g3) to[bend left=16] (g4);
\draw[bifund] (g4) to[bend left=16] (g5);
\draw[bifund] (g5) to[bend left=16] (g6);

\node[font=\Large,fill=white,inner sep=1pt,rotate=90] at (180:\R) {$\cdots$};

\foreach \g/\f in {g0/f0,g1/f1,g2/f2,g4/f4,g5/f5,g6/f6}{
    \draw[flavarrow] (\f) to[bend left=13] (\g);
    \draw[flavarrow] (\g) to[bend left=13] (\f);
}

\draw[flavarrow] (f3) to[bend left=13] (g3);
\draw[flavarrow] (g3) to[bend left=13] (f3);

\node[fill=white,inner sep=1.3pt]
    at ($(g3)!0.52!(f3)+(0,0.58)$) {$Z_i$};

\node[fill=white,inner sep=1.3pt]
    at ($(g3)!0.52!(f3)+(0,-0.58)$) {$Z_i^{c}$};

\end{tikzpicture}%
}

\caption{To account for strings which stretch off to $\partial X$ in the geometry,
one can enlarge the original brane probe quiver by adding vectorlike pairs
$Z_{i} \oplus Z_{i}^{c}$ to each quiver node. Adding a mass deformation
$m_{i} Z_{i} Z_{i}^{c}$ returns the system to the original quiver. In this
enlarged system, gauge non-invariant operators $\mathcal{O}_{i \rightarrow j}$
can be dressed by the additional $Z$ and $Z^c$'s to construct gauge invariant
operators. Physically, this corresponds to F1 strings which stretch out to
infinity, but are connected in the interior by $\mathcal{O}_{i \rightarrow j}$.}
\label{fig:quiver_with_flavor}
\end{figure}
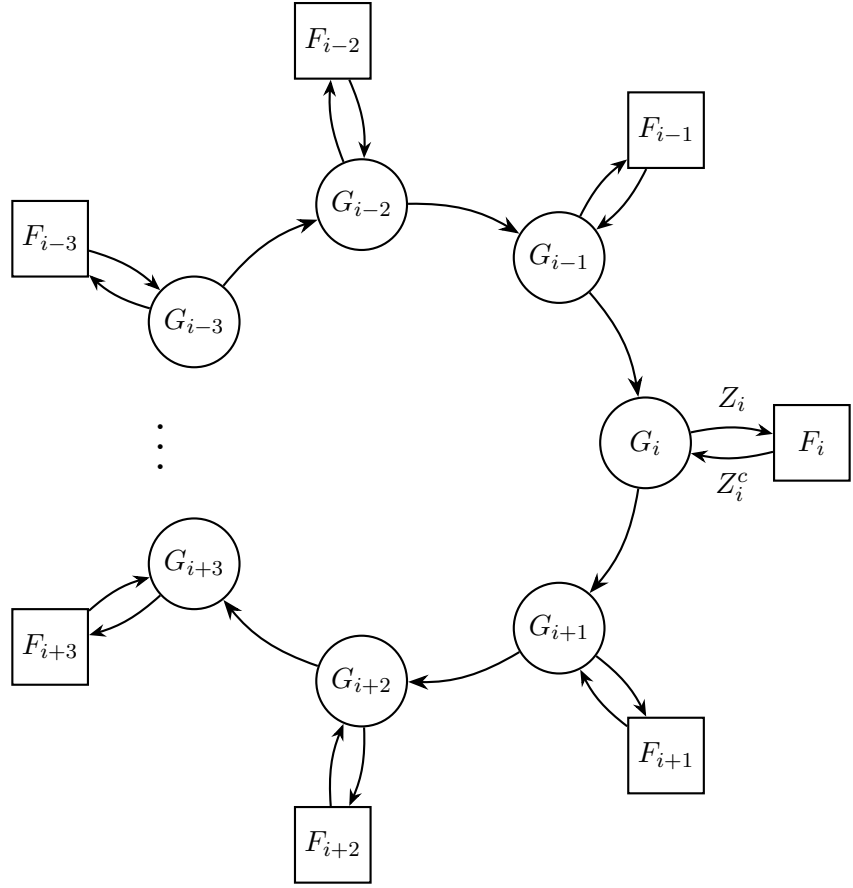

To frame the discussion to follow, observe that from a bottom up perspective we are building operators $\mathcal{O}_{i \rightarrow j}$ by taking a product of bifundamentals as we stretch from node $i$ to node $j$. This operator is manifestly not gauge invariant, but we can remedy this by considering the closely related quiver obtained by adding a vector-like pair $Z_{i} \oplus Z^{c}_{i}$ attached to each quiver node. Here, $Z_i$ is in the fundamental of $G_i$ and $Z^{c}_i$ is in the conjugate representation. Observe that after adding this set of fields, we can build gauge invariant operators $Z^{c}_i \mathcal{O}_{i \rightarrow j} Z_j$. Of course, adding such fields ``by hand'' modifies the IR behavior of the quiver gauge theory, but we can also add a mass term $m_{i} Z_i Z^{c}_i$ to the superpotential which removes these fields as well. From this point of view, the counting problem we are considering naturally embeds in a ``bigger'' system, as depicted in figure \ref{fig:quiver_with_flavor}.

This prescription is also natural from the top down point of view. In terms of the extra-dimensional geometry, it is natural to interpret
the operator $\mathcal{O}_{i \rightarrow j}$ as a string which stretches between the fractional branes concentrated near the tip of the singular cone. After adding a mass term, an operator such as $Z_i$ (or $Z^{c}_{i}$) amounts to a long string stretching from the tip of the cone ``out to infinity.'' This is a long F1-string attached to a specific fractional brane. To make this a pointlike object in the 4D spacetime, the F1-string also needs to wrap a (typically torsional) 1-cycle of $\partial X$. Note also that the use of a stretched F1-string makes sense in both the D3-brane and D0-brane probes of the geometry $X$. In the case of orbifolds $X = \mathbb{C}^{3}/ \Gamma$, the twisted $\eta$-invariants of \cite{10.1093/qmath/han016} explicitly reference a choice of orbifold group generator: this is the same data specified by the wrapped F1-string ``at infinity.'' From the perspective of the quiver gauge theory, this amounts to enlarging our Hilbert space to allow these additional heavy objects / states.

\subsection{Illustrative Example: \texorpdfstring{$\mathbb C^3/\mathbb Z_3$}{C3/Z3}} \label{sec:IllustrativeExamplePart2}

We now revisit the example of the orbifold singularity
\begin{equation}
X=\mathbb C^3/\mathbb Z_3
\end{equation}
from the quiver perspective. The goal is to show explicitly that the brane-probe quiver reproduces both the higher-form symmetry structure and the anomaly data of the corresponding 5D SCFT.

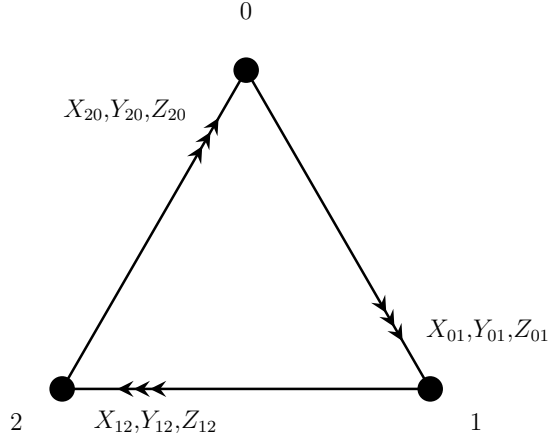
\begin{figure}
    \usetikzlibrary{arrows}
    \centering
    \scalebox{0.8}{
    \begin{tikzpicture}
        \node[draw=none, minimum size=7cm,regular polygon,regular polygon sides=3] (a) {};

        \foreach \x in {1,2,3}\fill (a.corner \x) circle[radius=6pt, fill = none];

        \begin{scope}[very thick,decoration={markings, mark=at position 0.85 with {\arrow[scale = 1.5, >=stealth]{>>>}}}]
        \draw[postaction = {decorate}] (a.corner 1) to (a.corner 3);
        \draw[postaction = {decorate}] (a.corner 3) to (a.corner 2);
        \draw[postaction = {decorate}] (a.corner 2) to (a.corner 1);
        \end{scope}
        \node at (0,4.5) {$0$};
        \node at (3.8,-2.3) {$1$};
        \node at (-3.8,-2.3) {$2$};
        \node at (-2,2.85) {$X_{20}$,$Y_{20}$,$Z_{20}$};
        \node at (4,-0.8) {$X_{01}$,$Y_{01}$,$Z_{01}$};
        \node at (-1.5,-2.3) {$X_{12}$,$Y_{12}$,$Z_{12}$};
    \end{tikzpicture}
    }
    \caption{Quiver for D-brane probe of $\mathbb{C}^3/\mathbb{Z}_3$. The quiver has three nodes corresponding to three irreducible representations of $\mathbb Z_3$ and bifundamental matter connecting each node.}
    \label{fig:Z3Example}
\end{figure}

The quiver for $\mathbb C^3/\mathbb Z_3$ has three nodes, with three arrows connecting each node to the next in cyclic order. See figure \ref{fig:Z3Example}. Its adjacency matrix is therefore
\begin{equation}
A=
\begin{pmatrix}
0 & 3 & 0 \\
0 & 0 & 3 \\
3 & 0 & 0
\end{pmatrix}.
\end{equation}
From this we obtain the Dirac pairing
\begin{equation}
B=A-A^T=
\begin{pmatrix}
0 & 3 & -3 \\
-3 & 0 & 3 \\
3 & -3 & 0
\end{pmatrix}.
\end{equation}
The higher-form symmetry is determined by the torsion subgroup of the cokernel of this pairing \cite{DelZotto:2022fnw}. In the present case one finds
\begin{equation}
\mathrm{Tor}\,\mathrm{Coker}(B)\simeq \mathbb Z_3\oplus \mathbb Z_3.
\end{equation}
This reproduces the expected defect group of the 5D SCFT: one $\mathbb Z_3$ factor corresponds to the electric one-form symmetry, while the other corresponds to the magnetic two-form symmetry. In the present subsection, we focus on the anomaly of the electric $\mathbb Z_3$ one-form symmetry.

To extract the anomaly coefficient, we now compute the quiver Hilbert series. Substituting the adjacency matrix $A$ into the matrix Hilbert series formula for orbifolds \eqref{eq:EagerMatrixHilbert} gives:
\begin{equation}
H(t)=\left(\mathbf 1-tA+t^2 A^T-t^3\mathbf 1\right)^{-1},
\end{equation}
one obtains
\begin{equation}
H(t)=
\begin{pmatrix}
\displaystyle -\frac{1+7t^3+t^6}{(-1+t^3)^3}
&
\displaystyle -\frac{3(t+2t^4)}{(-1+t^3)^3}
&
\displaystyle -\frac{3t^2(2+t^3)}{(-1+t^3)^3}
\\[1.2em]
\displaystyle -\frac{3t^2(2+t^3)}{(-1+t^3)^3}
&
\displaystyle -\frac{1+7t^3+t^6}{(-1+t^3)^3}
&
\displaystyle -\frac{3(t+2t^4)}{(-1+t^3)^3}
\\[1.2em]
\displaystyle -\frac{3(t+2t^4)}{(-1+t^3)^3}
&
\displaystyle -\frac{3t^2(2+t^3)}{(-1+t^3)^3}
&
\displaystyle -\frac{1+7t^3+t^6}{(-1+t^3)^3}
\end{pmatrix}.
\end{equation}
The important point is that this matrix reproduces exactly the twisted and untwisted Molien data discussed in the previous section. In particular, the diagonal component
\begin{equation}
H_{00}(t)= -\frac{1+7t^3+t^6}{(-1+t^3)^3}
\end{equation}
is precisely the ordinary untwisted Molien series for $\mathbb C^3/\mathbb Z_3$ (see \eqref{eq:MolienUntwistedZ3Closed}), while the off-diagonal component
\begin{equation}
H_{01}(t)= -\frac{3(t+2t^4)}{(-1+t^3)^3}
\end{equation}
is precisely the twisted Molien series associated to twisting by the character $\chi_\omega(g)=\omega$ (see \eqref{eq:MolienTwistedZ3Closed}). This is exactly the match anticipated in the general discussion above: the matrix Hilbert series packages both the untwisted and twisted sectors into a single quiver-theoretic object.\footnote{In this case, the quantum symmetry of the quiver is cyclic permutation of all the nodes, thus any starting node $0$ and any ending node $i$ can be used to compute $\alpha_i$. For $\mathbb{C}^3$ orbifolds with non-isolated singularities, this is no longer the case.}

The anomaly coefficient is then obtained from the same residue formula as before, now written directly in quiver language:
\begin{equation}
\alpha=\underset{t=1}{\mathrm{Res}}\frac{H_{01}(t)-H_{00}(t)}{1-t}\qquad\textrm{mod}\,1.
\end{equation}
Evaluating this residue gives
\begin{equation}
\alpha=-\frac{1}{9}\qquad\textrm{mod}\,1,
\end{equation}
in perfect agreement with the result obtained previously from the geometric analysis of the relative $\eta$-invariant.

At this point, we slow down to understand what each component of the Hilbert series is counting. We first look at
\begin{equation}
H_{00}(t)= -\frac{1+7t^3+t^6}{(-1+t^3)^3}
\end{equation}
which counts the number of $\mathbb{Z}_3$-invariant functions in the Calabi-Yau. To see this, we can expand $H_{00}$ around $t=0$ to get
\begin{equation}
    H_{00} \simeq 1+10t^3+\dots,
\end{equation}
which tells us that there are ten $\mathbb Z_3$-invariant polynomials of degree three. These are given by $f(X,Y,Z)=X^3,\,X^2Y,\,X^2Z,\,XY^2,\,XZ^2,\,XYZ,\,Y^3,\,Y^2Z,\,YZ^2,\,\textrm{and}\,Z^3$. On the quiver side, this should be matched with ten gauge invariant operators / paths made up of three chiral fields that complete a loop from node zero to itself. These operators are $X_{01}X_{12}X_{20},\,X_{01}X_{12}Y_{20},\,X_{01}X_{12}Z_{20},\,X_{01}Y_{12}Y_{20},\,X_{01}Z_{12}Z_{20},\,X_{01}Y_{12}Z_{20},\,Y_{01}Y_{12}Y_{20},\,Y_{01}Y_{12}Z_{20}$,\\ $Y_{01}Z_{12}Z_{20}, \textrm{ and } Z_{01}Z_{12}Z_{20}$. All other operators are related to one of these ten via F-term relations given by the superpotential for this theory:
\begin{equation}
    W= \text{Tr} \left[X_{01}Y_{12}Z_{20}-X_{01}Z_{12}Y_{20} + X_{12}Y_{20}Z_{01}-X_{12}Z_{20}Y_{01} + X_{20}Y_{01}Z_{12}-X_{20}Z_{01}Y_{12}\right].
\end{equation}
Thus, we see that we can exchange counting $\mathbb Z_3$-invariant holomorphic functions for mesonic chiral operators in the quiver.

We can similarly look at the twisted Hilbert series
\begin{equation}
H_{01}(t)= -\frac{3(t+2t^4)}{(-1+t^3)^3}.
\end{equation}
Expanding this term around $t=0$ gives
\begin{equation}
    H_{01} \simeq 3t+15t^4+\dots,
\end{equation}
which tells us that there are three operators / paths between node zero and node one. These are simply $X_{01}$ $Y_{01}$, and $Z_{01}$. Thus, we once again see the expected match between geometry and field theory.

Thus, in this simple example, the quiver perspective reproduces the full structure of interest. The antisymmetrized adjacency matrix correctly captures the higher-form symmetry group, while the matrix Hilbert series reproduces the twisted and untwisted Molien series and hence the relative $\eta$-invariant governing the anomaly of the electric one-form symmetry. This provides a concrete check of the general formalism and illustrates how the anomaly data of the 5D SCFT may be extracted directly from quiver data (see section \ref{sec:EXAMPLES} for further examples).

\section{Flavor Refinements} \label{sec:Refinements}

In the discussion so far, we have focused only on the discrete one-form symmetry of the theory. More generally, however, the theories of interest can also carry a zero-form flavor symmetry. This happens when the orbifold action contains group elements that generate fixed points on the boundary $\partial X$. This flavor symmetry can mix non-trivially with the one-form symmetry through a 2-group structure.\footnote{See \cite{Cordova:2018cvg, Benini:2018reh, Cvetic:2022imb, DelZotto:2022joo, Kapustin:2013uxa, Lee:2021crt}.} As a result, the anomaly theory contains additional terms beyond those associated purely with the one-form symmetry. We give a general discussion in this section, deferring explicit worked examples to section \ref{sec:EXAMPLES}.

The zero-form symmetries we focus on are those which are locally presented as ADE singularities in the geometry $X$, i.e., they are locally of the form $\mathbb{C}^2 / \Gamma_{\mathrm{ADE}}$ for $\Gamma_{\mathrm{ADE}}$ a finite subgroup of $SU(2)$. These naturally arise when $X$ has a non-isolated singularity, as for example occurs in many $\mathbb{C}^3 / \Gamma_{SU(3)}$ orbifolds. Observe that this also means $\partial X$ will be singular, and there is a refined $\eta$-invariant calculation on $\partial X$ which takes such singularities into account \cite{Cvetic:2025lat}. 

From the viewpoint of the five-dimensional system these additional singular loci furnish a flavor symmetry with Lie algebra:
\begin{equation}
\mathfrak f \equiv \bigoplus_i \mathfrak g^{(i)}_{\mathrm{ADE}}:
\label{eq:flavor-lie-algebra}
\end{equation}
A general comment here is that while the enhancement to a non-abelian flavor symmetry algebra can be difficult to explicitly track, many aspects of this also directly follow from the quiver obtained from a brane probe of the theory. Indeed, returning to the anti-symmetrized adjacency matrix $B = A - A^{T}$, we can compute $\mathrm{Coker}(B)$:
\begin{equation}
\mathrm{Coker}{B} = \underset{i}{\bigoplus}\; \mathbb{Z} / p^{e_i} \mathbb{Z} \;\oplus\; \mathbb{Z}^{f+1},
\end{equation}
where $f$ is the rank of the flavor symmetry algebra.\footnote{The additional free generator in $\mathrm{Coker}(B)$ corresponds to the KK charge of the 4D KK theory.}
The torsional piece directly extracts the defect group for the 5D SCFT \cite{DelZotto:2022fnw}, while the free part determines the rank of the flavor symmetry algebra. Moreover, there is a natural intersection pairing on the free part $\mathbb{Z}^{f}$, and this is just the Cartan matrix of the corresponding Lie algebra in line (\ref{eq:flavor-lie-algebra}). We will track this structure in examples in section \ref{sec:EXAMPLES}.

As a general comment, we note that quiver based techniques most directly provide access to the rank of the flavor group, as well as the global form of the center. This serves to constrain candidate global forms for the flavor symmetry group, but does not by itself directly tell us the full form of the zero-form flavor symmetry or the 2-group. Nevertheless, the anomaly terms of the SymTh / SymTFT can still be extracted since this data primarily involves mixing between the center of the flavor symmetry and the one-form symmetries.

Assuming we have extracted a candidate zero-form symmetry, one can then ask whether this entwines with the one-form symmetry
the corresponding global symmetry data, as organized by the long exact sequence for a 2-group \cite{Bhardwaj:2021wif,Lee:2021crt}:
\begin{equation}
1\longrightarrow A \longrightarrow \widetilde A \longrightarrow \widetilde F \longrightarrow F \longrightarrow 1,
\label{eq:2group-sequence-refined}
\end{equation}
where $\widetilde{F} = \prod_i \widetilde{G}_{\mathrm{ADE}}^{(i)}$ is the ``naive'' product of simply connected ADE Lie groups, each with Lie algebra $\mathfrak g_{\mathrm{ADE}}^{(i)}$. This is corrected to $F$, the actual flavor symmetry group of the 5D theory. Likewise, $\widetilde A$ is the naive one-form symmetry obtained before accounting for the fixed-locus sectors, while $A$ is the actual one-form symmetry of the 5D SCFT after quotienting by the subgroup generated by the fixed-point data. In this sense, the appearance of non-isolated ADE singularities simultaneously reduces the discrete one-form symmetry and introduces new continuous flavor symmetry factors.

Returning to the anomaly terms of the SymTh / SymTFT of equation (\ref{eq:SYMTHANOM}):
\begin{align}
S_{\mathrm{SymTH}}^{\mathrm{anom}} = 2\pi & \int_{6D} \left[ \beta B_2 \cup B_2 \cup B_2 + \gamma B_2 \cup \frac{p_1}{4} \right] \\
+ & \int_{6D} \left[ \frac{1}{2(2\pi)^2}\sum_{k} \delta_k\, B_2 \cup \mathrm{Tr}\left(F^{(k)}_2 F^{(k)}_2\right) + \frac{1}{6(2\pi)^3}\sum_{k} \epsilon_k\, \mathrm{Tr}\left(F^{(k)}_2 F^{(k)}_2 F^{(k)}_2\right)\right]\,,
\end{align}
the coefficients $\delta_k$ and $\epsilon_k$ are determined by a refined $\eta$-invariant. More precisely, the combination detected by the refined $\eta$-invariant is:
\begin{equation}
\frac{1}{2}\eta^{\mathrm{ref}}=\alpha^{\mathrm{ref}} = \beta+\gamma+\sum_k\delta_k+\sum_k\epsilon_k\qquad\textrm{mod}\,1.
\label{eq:RefinedEtaDefinition}
\end{equation}
This refined quantity is already encoded in the same matrix Hilbert series we have been using. In the isolated case, the relative quantity
\begin{equation}
\frac{H_{0i}(t)-H_{00}(t)}{1-t}
\end{equation}
had only a simple pole at $t=1$, and its residue reproduced the ordinary combination
\begin{equation}
\alpha=\beta+\gamma\qquad\textrm{mod}\,1.
\end{equation}
For non-isolated orbifolds, the singular structure is richer, and the expansion around $t=1$ develops higher-order poles:
\begin{equation}
\frac{H_{0i}(t)-H_{00}(t)}{1-t}= \frac{a_i}{(1-t)^2} +\frac{b_i}{1-t} +\text{regular}.
\label{eq:HigherPoleExpansion}
\end{equation}

In this treatment, singularities concentrated on lower-dimensional subspaces simply correspond to higher-order poles in the geometry. This is to be expected since moving a brane probe away from the tip of the cone $X$ but still along an ADE singularity results in a non-trivial quiver with accompanying matrix Hilbert series.

Motivated by this structure, we define the refined $\eta$-invariant extracted from the quiver by summing the coefficients of all singular terms in the expansion about $t=1$. In the present cyclic examples this amounts to
\begin{equation}
\alpha = \frac{1}{2}\eta^{\mathrm{ref}}_i= {\operatorname*{Res}}^{(2)}_{t=1} \left( \frac{H_{0i}(t)-H_{00}(t)}{1-t} \right) + \operatorname*{Res}_{t=1}\left( \frac{H_{0i}(t)-H_{00}(t)}{1-t} \right)\qquad\textrm{mod}\,1,
\label{eq:RefinedEtaFromPoles}
\end{equation}
where $\operatorname*{Res}^{(2)}_{t=1}$ denotes the coefficient of $(1-t)^{-2}$. Equivalently, in terms of the Laurent expansion \eqref{eq:HigherPoleExpansion},
\begin{equation}
\alpha = \frac{1}{2}\eta^{\mathrm{ref}}_i=a_i+b_i\qquad\textrm{mod}\,1 \, ,
\label{eq:RefinedEtaab}
\end{equation}
where again the index $i$ is chosen according to the the action of the quantum symmetry described in \ref{app:QuantumSymmetry}.

This provides the quiver realization of the refinement: the full singular part of the expansion, rather than only the simple pole, contributes to the anomaly data.

It is useful to summarize this geometrically. The refined quantity is naturally organized as a sum over singular strata of the boundary. Denoting by $\Gamma_{\mathrm{fix}}\subset \Gamma$ the subset of orbifold elements generating fixed loci, we may write schematically
\begin{equation}
\alpha = \frac{1}{2}\eta^{\mathrm{ref}} \sim \sum_{\gamma\in\Gamma_{\mathrm{fix}}} \frac{1}{2}\eta\bigl(Y_\gamma/\Gamma\bigr)\qquad\textrm{mod}\,1,
\label{eq:RefinedEtaStrata}
\end{equation}
where $Y_\gamma$ denotes the contribution from the stratum fixed by $\gamma$. Depending on the element $\gamma$, this contribution may come either from the full five-dimensional boundary sector $S^5/\Gamma$ or from one of the three-dimensional links $S^3/\mathbb{Z}_{g_k}$ associated with the codimension-four $A$-type loci. For isolated singularities only the simple pole survives, reproducing the anomaly combination $\beta+\gamma$. For non-isolated singularities, higher-order poles appear and encode the additional $\delta_k$ and $\epsilon_k$.

Both $\eta^{\mathrm{ref}}$ and $\eta^{\mathrm{unref}}$ can be computed from the quiver by picking the $i$ index according to the full quantum group or the subgroup associated to the one-form symmetry, as discussed in Appendix \ref{app:QuantumSymmetry}. The two invariants have different physical meanings: the unrefined one is sensitive only to the self-anomaly and gravitational mixed anomaly of the one-form symmetry. The refined one instead measures data associated with flavor symmetry sectors and possible 2-group structures.

\section{Examples} \label{sec:EXAMPLES}

In this section we illustrate the general discussion of the previous sections with some examples. We show how to extract the $\eta$-invariant(s) and the corresponding symmetry theory structures in a wide range of geometries. We primarily focus on abelian orbifolds of the form $\mathbb{C}^3 / \mathbb{Z}_N$, as well as Calabi-Yau cones over the Sasaki-Einstein $Y^{p,q}$ geometries. Additional examples, including the orbifolds $\mathbb{C}^3 / \mathbb{Z}_M \times \mathbb{Z}_N$ and their IIA generalizations with discrete torsion are deferred to Appendix \ref{app:DISCTORSION}.

We summarize our results in table \ref{tab:example_summary}. In Appendix \ref{app:DISCTORSION} we discuss some related examples given by IIA on orbifolds with discrete torsion.

\begin{table}[t]
\centering
\begin{tabular}{|c|c|c|c|}
\hline
& & &\\
Geometry $X$ & $\mathbb{D} = \mathbb{D}^{(1)}_{\mathrm{el}} \oplus \mathbb{D}^{(2)}_{\mathrm{mag}}$ & Unrefined $\eta$-invariant & Refined $\eta$-invariant \\
& & &\\
\hline
& & &\\

$\mathbb{C}^3/\mathbb{Z}_5(1,1,3)$
& $\mathbb{Z}_5 \oplus \mathbb{Z}_5$
& $\frac{1}{2}\eta^{\mathrm{unref}}=\frac{1}{5}\ \mathrm{mod}\,1$
& ---  
\\[1em]

$\mathbb{C}^3/\mathbb{Z}_6(1,1,4)$
& $\mathbb{Z}_3 \oplus \mathbb{Z}_3$
& $\frac{1}{2}\eta^{\mathrm{unref}}=\frac{1}{9}\ \mathrm{mod}\,1$
& $\frac{1}{2}\eta^{\mathrm{ref}}=\frac{2}{9}\ \mathrm{mod}\,1$ \\[1em]

$\mathbb{C}^3/\mathbb{Z}_9(1,2,6)$
& $\mathbb{Z}_3 \oplus \mathbb{Z}_3$
& $\frac{1}{2}\eta^{\mathrm{unref}}=0\ \mathrm{mod}\,1$
& $\frac{1}{2}\eta^{\mathrm{ref}}=\frac{1}{9}\ \mathrm{mod}\,1$ \\[1em]

$\mathbb{C}^3/\mathbb{Z}_8(2,1,5)\times \mathbb{Z}_2(1,0,1)$ with DT
& $\mathbb{Z}_2 \oplus \mathbb{Z}_2$
& $\frac{1}{2}\eta^{\mathrm{unref}}=0\ \mathrm{mod}\,1$
& $\frac{1}{2}\eta^{\mathrm{ref}}=\frac{1}{8}\ \mathrm{mod}\,1$ \\[1em]

$\mathrm{Cone}(Y^{4,2})$
& $\mathbb{Z}_2 \oplus \mathbb{Z}_2$
& $\frac{1}{2}\eta^{\mathrm{unref}}=0\ \mathrm{mod}\,1$
& --- 
\\[1em]

$\mathrm{Cone}(Y^{6,3})$
& $\mathbb{Z}_3 \oplus \mathbb{Z}_3$
& $\frac{1}{2}\eta^{\mathrm{unref}}=\frac{2}{9}\ \mathrm{mod}\,1$
& --- 
\\

& & &\\
\hline
\end{tabular}
\caption{Summary of the defect group, $\mathbb{D}$, unrefined/refined $\eta$-invariants for the examples discussed in this section. Here DT stands for discrete torsion (see Appendix \ref{app:DISCTORSION}). The $\eta^{\textrm{ref}}$ column is empty for theories with no refinement.}
\label{tab:example_summary}
\end{table}

\subsection{Abelian Orbifolds of \texorpdfstring{$\mathbb C^3$}{C3}} \label{sec:C3Orbifolds}

In this section, we compute the (refined) $\eta$-invariant for various abelian orbifolds of the form $\mathbb{C}^3/\mathbb{Z}_N$.

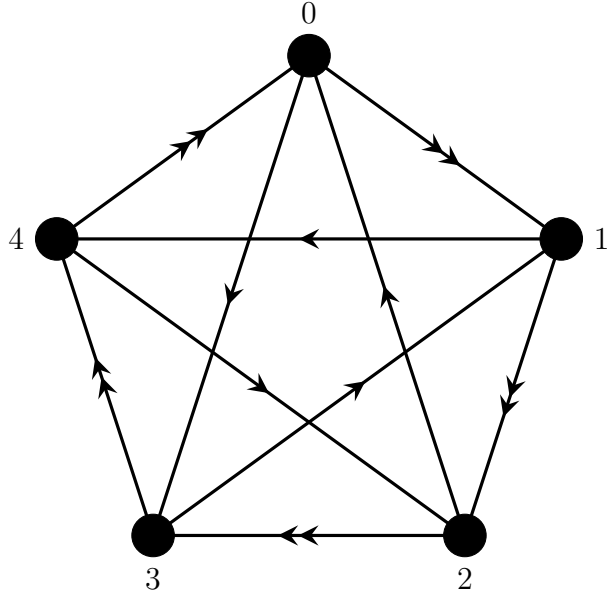
\begin{figure}
    \centering
    \begin{tikzpicture}

    \node[draw=none, minimum size=7cm, regular polygon, regular polygon sides=5] (a) {};

    \foreach \corner/\labelnum/\pos in {
        1/0/above,
        2/4/left,
        3/3/below,
        4/2/below,
        5/1/right
    }
      \node[draw, circle, inner sep=2pt, fill=black, label=\pos:$\labelnum$] at (a.corner \corner) {};

    \begin{scope}[very thick, decoration={
        markings,
        mark=at position 0.6 with {\arrow[scale = 1.5, >=stealth]{>>}}}]
        \draw[postaction = {decorate}] (a.corner 2) to (a.corner 1);
        \draw[postaction = {decorate}] (a.corner 3) to (a.corner 2);
        \draw[postaction = {decorate}] (a.corner 4) to (a.corner 3);
        \draw[postaction = {decorate}] (a.corner 5) to (a.corner 4);
        \draw[postaction = {decorate}] (a.corner 1) to (a.corner 5);
    \end{scope}

    \begin{scope}[very thick, decoration={
        markings,
        mark=at position 0.52 with {\arrow[scale = 1.5, >=stealth]{>}}}]
        \draw[postaction = {decorate}] (a.corner 1) to (a.corner 3);
        \draw[postaction = {decorate}] (a.corner 2) to (a.corner 4);
        \draw[postaction = {decorate}] (a.corner 3) to (a.corner 5);
        \draw[postaction = {decorate}] (a.corner 4) to (a.corner 1);
        \draw[postaction = {decorate}] (a.corner 5) to (a.corner 2);
    \end{scope}

    \end{tikzpicture}
    \caption{Quiver associated to the orbifold geometry $\mathbb{C}^3/\mathbb{Z}_5(1,1,3)$.}
    \label{fig:C3Z5Example}
\end{figure}

\paragraph{Example} We begin by studying the following example of an isolated singularity:
\begin{equation}
X=\mathbb{C}^3/\Gamma, \qquad \Gamma=\mathbb{Z}_5(1,1,3).
\end{equation}
The adjacency matrix is obtained by the usual McKay prescription:\footnote{See Appendix \ref{app:QuiverAdjacency} for details on computing the adjacency matrix.}
\begin{equation}
A_{ij} =
\begin{pmatrix}
    0 & 2 & 0 & 1 & 0 \\
    0 & 0 & 2 & 0 & 1 \\
    1 & 0 & 0 & 2 & 0 \\
    0 & 1 & 0 & 0 & 2 \\
    2 & 0 & 1 & 0 & 0
\end{pmatrix}
.
\end{equation}
The associated quiver is shown in figure \ref{fig:C3Z5Example}.

From the quiver, we may compute the one-form symmetry group via the Dirac pairing / antisymmetrized adjacency matrix,
\begin{equation}
B=A-A^T.
\end{equation}
Taking the torsion part of the cokernel then determines the one-form symmetry group of the theory:
\begin{equation}
\mathrm{Tor}\,\mathrm{Coker}(B)\simeq \mathbb{Z}_5 \oplus \mathbb{Z}_5 .
\end{equation}
As in the general discussion, this captures the reduction of the electric one-form symmetry and magnetic two-form symmetry of the corresponding 5D SCFT.

Next, plugging the adjacency matrix into \eqref{eq:EagerMatrixHilbert}, we obtain the matrix-valued quiver Hilbert series
\begin{equation}
H(t)=\frac{1}{(-1+t)^3(1+t+t^2+t^3+t^4)^2}
\begin{pmatrix}
a(t) & b(t) & c(t) & d(t) & e(t) \\
e(t) & a(t) & b(t) & c(t) & d(t) \\
d(t) & e(t) & a(t) & b(t) & c(t) \\
c(t) & d(t) & e(t) & a(t) & b(t) \\
b(t) & c(t) & d(t) & e(t) & a(t)
\end{pmatrix},
\end{equation}
where
\begin{align}
a(t) &= -1+t-3t^3+t^4-3t^5+t^7-t^8,\\
b(t) &= t(-2+t+t^2-4t^3+t^4-2t^5),\\
c(t) &= t^2(-3+t+t^2-4t^3+t^4-t^5),\\
d(t) &= t(-1+t-4t^2+t^3+t^4-3t^5),\\
e(t) &= t^2(-2+t-4t^2+t^3+t^4-2t^5).
\end{align}
In particular, we extract the untwisted and twisted components. In this case the quantum symmetry associated to a flat bundle $L$ of charge $q$, acts as a cyclic permutation of the quiver, mapping the nodes according to $i \to i + j$ mod$\,5$. Once we fix node $0$, we can take $i = 0 + 1$ in \eqref{eq:AlphaFromQuiverGeneral}, corresponding to the index twisted by a flat bundle of charge $1$.
\begin{align}
    H_{00} &= \dfrac{-1+t-3t^3+t^4-3t^5+t^7-t^8}{(-1+t)^3(1+t+t^2+t^3+t^4)^2}\,,\\[1em]
    H_{01} &= \dfrac{t(-2+t+t^2-4t^3+t^4-2t^5)}{(-1+t)^3(1+t+t^2+t^3+t^4)^2}\,.
\end{align}
Using the prescription described in the previous section, we then compute
\begin{equation}
\alpha=\frac{1}{2}\eta^{\mathrm{ref}}
=
\operatorname*{Res}_{t=1}\left(\frac{H_{01}(t)-H_{00}(t)}{1-t}\right)
= \frac{1}{5} \qquad\textrm{mod}\,1 .
\end{equation}
In this particular example there are no refinements to the $\eta$-invariant due to 2-group data, in particular ${\operatorname*{Res}_{t=1}}^{(2)}\left(\frac{H_{01}(t)-H_{00}(t)}{1-t}\right)=0$. This is the expected result, as this is an example of an isolated singularity which means that there are no fixed loci in $\partial X$. We end by noting that this result is in agreement with Appendix A of \cite{Cvetic:2025lat}.

\paragraph{Example} We now move on to an example with additional non-isolated singularities. Consider
\begin{equation}
X=\mathbb{C}^3/\Gamma, \qquad \Gamma=\mathbb{Z}_6(1,1,4).
\end{equation}
This theory has a gauge theory phase, given by $SU(3)_3$. It has an emergent $\mathfrak{su}(2)$ flavor symmetry at the conformal fixed point. Indeed, unlike the previous example, this orbifold is not isolated: the element $\gamma_3 \in \mathbb{Z}_6$ action leaves a nontrivial fixed locus in $\partial X$, and so in addition to the higher-form symmetry data encoded by the full quiver, there is also additional flavor symmetry data associated to the non-isolated $\mathbb{C}^2/\mathbb{Z}_2$ singularity. As explained in the previous sections, this means that the refined $\eta$-invariant can now receive extra contributions reflecting the corresponding 2-group structure.

\begin{figure}
    \centering
    \begin{tikzpicture}

\node[draw=none, minimum size=7cm,regular polygon,regular polygon sides=6] (a) {};

  \foreach \corner/\labelnum/\pos in {
        1/1/above,
        2/0/above,
        3/5/left,
        4/4/below,
        5/3/below,
        6/2/right
    }
    \node[draw, circle, inner sep=2pt, fill=black, label=\pos:$\labelnum$] at (a.corner \corner) {};

\begin{scope}[very thick,decoration={
    markings,
    mark=at position 0.62 with {\arrow[scale = 1.5, >=stealth]{>>}}}
    ]
\draw[postaction = {decorate}] (a.corner 2) to (a.corner 1);
\draw[postaction = {decorate}] (a.corner 3) to (a.corner 2);
\draw[postaction = {decorate}] (a.corner 4) to (a.corner 3);
\draw[postaction = {decorate}] (a.corner 5) to (a.corner 4);
\draw[postaction = {decorate}] (a.corner 6) to (a.corner 5);
\draw[postaction = {decorate}] (a.corner 1) to (a.corner 6);
\end{scope}

\begin{scope}[very thick,decoration={
    markings,
    mark=at position 0.52 with {\arrow[scale = 1.5, >=stealth]{>}}}
    ]
\draw[postaction = {decorate}] (a.corner 1) to (a.corner 3);
\draw[postaction = {decorate}] (a.corner 2) to (a.corner 4);
\draw[postaction = {decorate}] (a.corner 3) to (a.corner 5);
\draw[postaction = {decorate}] (a.corner 4) to (a.corner 6);
\draw[postaction = {decorate}] (a.corner 5) to (a.corner 1);
\draw[postaction = {decorate}] (a.corner 6) to (a.corner 2);
\end{scope}
\end{tikzpicture}
    \caption{Quiver associated to the orbifold geometry $\mathbb{C}^3/\mathbb{Z}_6(1,1,4)$}.
    \label{fig:C3Z6Example}
\end{figure}
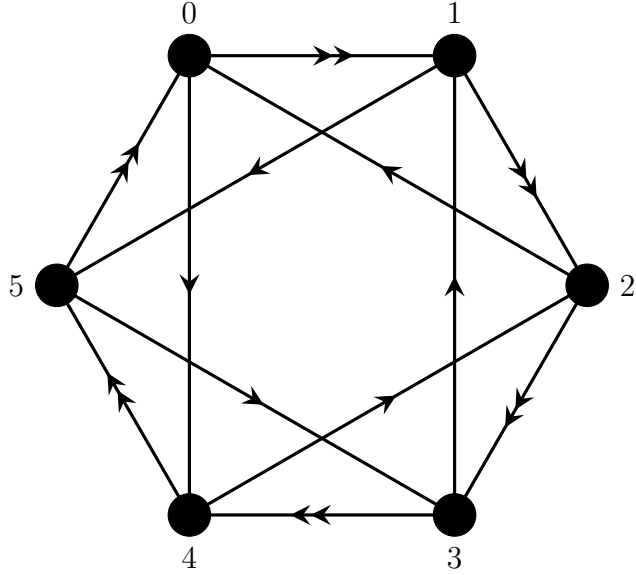

The adjacency matrix is again obtained from the generalized McKay correspondence (see Appendix \ref{app:QuiverAdjacency}):
\begin{equation}
A_{ij}=
\begin{pmatrix}
0 & 2 & 0 & 0 & 1 & 0 \\
0 & 0 & 2 & 0 & 0 & 1 \\
1 & 0 & 0 & 2 & 0 & 0 \\
0 & 1 & 0 & 0 & 2 & 0 \\
0 & 0 & 1 & 0 & 0 & 2 \\
2 & 0 & 0 & 1 & 0 & 0
\end{pmatrix}
.
\end{equation}
The associated quiver is shown in figure \ref{fig:C3Z6Example}.

From the quiver, we may compute the one-form symmetry group via the Dirac pairing / antisymmetrized adjacency matrix. Taking the torsion part of the cokernel then determines the one-form symmetry group of the theory:
\begin{equation}
\mathrm{Coker}(B)\simeq \mathbb{Z}_3 \oplus \mathbb{Z}_3 \oplus \mathbb{Z}^2.
\end{equation}
As in the isolated case, this captures the reduction of the electric one-form symmetry and magnetic two-form symmetry of the corresponding 5D SCFT.

Next, plugging the adjacency matrix into \eqref{eq:EagerMatrixHilbert}, we obtain the matrix-valued quiver Hilbert series
\begin{equation}
H(t)=\frac{1}{(-1+t^3)^3(1+t^3)^2}
\begin{pmatrix}
a(t) & b(t) & c(t) & d(t) & e(t) & f(t) \\
f(t) & a(t) & b(t) & c(t) & d(t) & e(t) \\
e(t) & f(t) & a(t) & b(t) & c(t) & d(t) \\
d(t) & e(t) & f(t) & a(t) & b(t) & c(t) \\
c(t) & d(t) & e(t) & f(t) & a(t) & b(t) \\
b(t) & c(t) & d(t) & e(t) & f(t) & a(t)
\end{pmatrix},
\end{equation}
where
\begin{align}
a(t) &= -(1+3t^3+10t^6+3t^9+t^{12}),\\
b(t) &= -2(t+2t^4+5t^7+t^{10}),\\
c(t) &= -t^2(4+5t^3+8t^6+t^9),\\
d(t) &= -6(t^3+t^6+t^9),\\
e(t) &= -t(1+8t^3+5t^6+4t^9),\\
f(t) &= -2(t^2+5t^5+2t^8+t^{11}).
\end{align}

In this case we have both a refined and unrefined $\eta$-invariant. The unrefined one is sensitive only to the $\mathbb{Z}_3$ one-form symmetry associated to the quantum subgroup which permutes the nodes as $i \to i + 2$ mod$\,6$, while the refined one is associated to the full quantum group, $i \to i + 1$ mod$\,6$. The former index is computed from
\begin{align}
    H_{00} &= -\dfrac{1+3t^3+10t^6+3t^9+t^{12}}{(-1+t^3)^3(1+t^3)^2},\\[1em]
    H_{02} &= -\dfrac{-t^2(4+5t^3+8t^6+t^9)}{(-1+t^3)^3(1+t^3)^2} .
\end{align}
and by applying the prescription \eqref{eq:AlphaFromQuiverGeneral}
\begin{equation}
\alpha=\frac{1}{2}\eta^{\mathrm{unref}}= \operatorname*{Res}_{t=1}\left(\frac{H_{02}(t)-H_{00}(t)}{1-t}\right)= \frac{1}{9}\qquad\textrm{mod}\,1 .
\end{equation}

While the refined invariant, containing information of the flavor group as well, is given in terms of
\begin{align}
    H_{00} &= -\dfrac{1+3t^3+10t^6+3t^9+t^{12}}{(-1+t^3)^3(1+t^3)^2},\\[1em]
    H_{01} &= -\dfrac{2(t+2t^4+5t^7+t^{10})}{(-1+t^3)^3(1+t^3)^2} .
\end{align}

Using again the prescription above we have
\begin{equation}
\alpha^{\mathrm{ref}}=\frac{1}{2}\eta^{\mathrm{ref}}= {\operatorname*{Res}}^{(2)}_{t=1}\left(\frac{H_{01}(t)-H_{00}(t)}{1-t}\right) + \operatorname*{Res}_{t=1}\left(\frac{H_{01}(t)-H_{00}(t)}{1-t}\right) =\frac{1}{12} + \frac{5}{36} = \frac{2}{9}\qquad\textrm{mod}\,1 .
\end{equation}
In contrast with the isolated example $\mathbb{C}^3/\mathbb{Z}_5(1,1,3)$, the additional term here reflects the presence of the non-isolated singular locus and the corresponding 2-group refinement. This provides a simple example in which the $\eta$-invariant is sensitive not only to the higher-form symmetry encoded by the full quiver, but also to the extra flavor symmetry sector. We end by noting that this result is in agreement with Appendix A of \cite{Cvetic:2025lat}.

\paragraph{Example} We now consider a second example with additional non-isolated singularities. Let
\begin{equation}
X=\mathbb{C}^3/\Gamma, \qquad \Gamma=\mathbb{Z}_9(1,2,6).
\end{equation}
As in the previous example, this orbifold is not isolated. Indeed, the elements $\gamma_3,\gamma_6\in \mathbb{Z}_9$ leaves a nontrivial fixed locus in $\partial X$, and so in addition to the higher-form symmetry data encoded by the full quiver, there is also additional flavor symmetry data associated to the non-isolated $\mathbb{C}^2/\mathbb{Z}_3$ singularities. Indeed, there is an emergent $\mathfrak{su}(3)$ flavor symmetry at the conformal fixed point. As explained in the previous sections, this means that the refined $\eta$-invariant can again receive extra contributions reflecting the corresponding 2-group structure.

\begin{figure}
    \centering
    \begin{tikzpicture}

\node[draw=none, minimum size=7cm,regular polygon,regular polygon sides=9] (a) {};

\foreach \corner/\labelnum/\pos in {
        1/0/above,
        2/8/above,
        3/7/left,
        4/6/left,
        5/5/left,
        6/4/right,
        7/3/right,
        8/2/right,
        9/1/above
    }
    \node[draw, circle, inner sep=2pt, fill=black, label=\pos:$\labelnum$] at (a.corner \corner) {};

\begin{scope}[very thick,decoration={
    markings,
    mark=at position 0.6 with {\arrow[scale = 1.5, >=stealth]{>}}}
    ]
\draw[postaction = {decorate}] (a.corner 2) to (a.corner 1);
\draw[postaction = {decorate}] (a.corner 3) to (a.corner 2);
\draw[postaction = {decorate}] (a.corner 4) to (a.corner 3);
\draw[postaction = {decorate}] (a.corner 5) to (a.corner 4);
\draw[postaction = {decorate}] (a.corner 6) to (a.corner 5);
\draw[postaction = {decorate}] (a.corner 7) to (a.corner 6);
\draw[postaction = {decorate}] (a.corner 8) to (a.corner 7);
\draw[postaction = {decorate}] (a.corner 9) to (a.corner 8);
\draw[postaction = {decorate}] (a.corner 1) to (a.corner 9);
\end{scope}

\begin{scope}[very thick,decoration={
    markings,
    mark=at position 0.52 with {\arrow[scale = 1.5, >=stealth]{>}}}
    ]
\draw[postaction = {decorate}] (a.corner 3) to (a.corner 1);
\draw[postaction = {decorate}] (a.corner 4) to (a.corner 2);
\draw[postaction = {decorate}] (a.corner 5) to (a.corner 3);
\draw[postaction = {decorate}] (a.corner 6) to (a.corner 4);
\draw[postaction = {decorate}] (a.corner 7) to (a.corner 5);
\draw[postaction = {decorate}] (a.corner 8) to (a.corner 6);
\draw[postaction = {decorate}] (a.corner 9) to (a.corner 7);
\draw[postaction = {decorate}] (a.corner 1) to (a.corner 8);
\draw[postaction = {decorate}] (a.corner 2) to (a.corner 9);
\end{scope}

\begin{scope}[very thick,decoration={
    markings,
    mark=at position 0.55 with {\arrow[scale = 1.5, >=stealth]{>}}}
    ]
\draw[postaction = {decorate}] (a.corner 7) to (a.corner 1);
\draw[postaction = {decorate}] (a.corner 8) to (a.corner 2);
\draw[postaction = {decorate}] (a.corner 9) to (a.corner 3);
\draw[postaction = {decorate}] (a.corner 1) to (a.corner 4);
\draw[postaction = {decorate}] (a.corner 2) to (a.corner 5);
\draw[postaction = {decorate}] (a.corner 3) to (a.corner 6);
\draw[postaction = {decorate}] (a.corner 4) to (a.corner 7);
\draw[postaction = {decorate}] (a.corner 5) to (a.corner 8);
\draw[postaction = {decorate}] (a.corner 6) to (a.corner 9);
\end{scope}
\end{tikzpicture}
    \caption{Quiver associated to the orbifold geometry $\mathbb{C}^3/\mathbb{Z}_9(1,2,6)$.}
    \label{fig:C3Z9Example}
\end{figure}
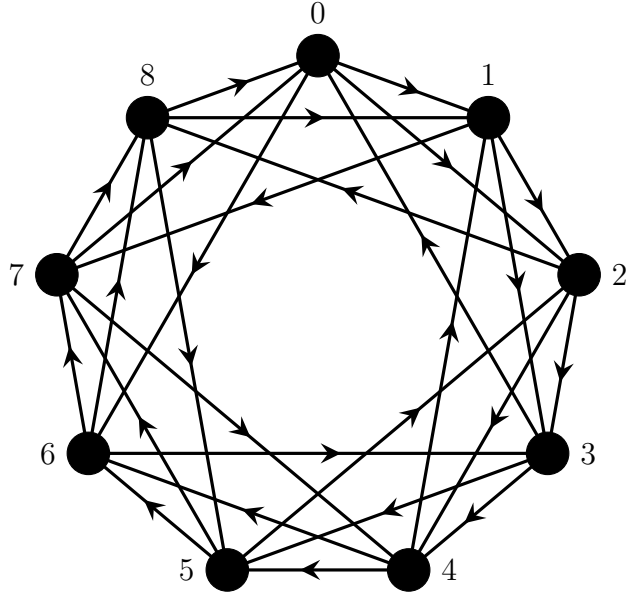

The adjacency matrix is again obtained from the generalized McKay correspondence:
\begin{equation}
A_{ij}=
\begin{pmatrix}
0 & 1 & 1 & 0 & 0 & 0 & 1 & 0 & 0 \\
0 & 0 & 1 & 1 & 0 & 0 & 0 & 1 & 0 \\
0 & 0 & 0 & 1 & 1 & 0 & 0 & 0 & 1 \\
1 & 0 & 0 & 0 & 1 & 1 & 0 & 0 & 0 \\
0 & 1 & 0 & 0 & 0 & 1 & 1 & 0 & 0 \\
0 & 0 & 1 & 0 & 0 & 0 & 1 & 1 & 0 \\
0 & 0 & 0 & 1 & 0 & 0 & 0 & 1 & 1 \\
1 & 0 & 0 & 0 & 1 & 0 & 0 & 0 & 1 \\
1 & 1 & 0 & 0 & 0 & 1 & 0 & 0 & 0
\end{pmatrix}
.
\end{equation}
The associated quiver is shown in figure \ref{fig:C3Z9Example}.

From the quiver, we may compute the one-form symmetry group via the Dirac pairing / antisymmetrized adjacency matrix. Taking the cokernel then determines the one-form symmetry group of the theory as well as the rank of the flavor algebra:
\begin{equation}
\mathrm{Coker}(B)\simeq \mathbb{Z}_3 \oplus \mathbb{Z}_3 \oplus \mathbb{Z}^3.
\end{equation}
As in the isolated case, this captures the reduction of the electric one-form symmetry and magnetic two-form symmetry of the corresponding 5D SCFT.

Next, plugging the adjacency matrix into \eqref{eq:EagerMatrixHilbert}, we obtain the matrix-valued quiver Hilbert series
\begin{equation}
H(t)=\frac{1}{(-1+t)^3(1+t+t^2)(1+t^3+t^6)}
\begin{pmatrix}
a(t) & b(t) & c(t) & d(t) & e(t) & f(t) & g(t) & h(t) & i(t) \\
i(t) & a(t) & b(t) & c(t) & d(t) & e(t) & f(t) & g(t) & h(t) \\
h(t) & i(t) & a(t) & b(t) & c(t) & d(t) & e(t) & f(t) & g(t) \\
g(t) & h(t) & i(t) & a(t) & b(t) & c(t) & d(t) & e(t) & f(t) \\
f(t) & g(t) & h(t) & i(t) & a(t) & b(t) & c(t) & d(t) & e(t) \\
e(t) & f(t) & g(t) & h(t) & i(t) & a(t) & b(t) & c(t) & d(t) \\
d(t) & e(t) & f(t) & g(t) & h(t) & i(t) & a(t) & b(t) & c(t) \\
c(t) & d(t) & e(t) & f(t) & g(t) & h(t) & i(t) & a(t) & b(t) \\
b(t) & c(t) & d(t) & e(t) & f(t) & g(t) & h(t) & i(t) & a(t)
\end{pmatrix},
\end{equation}
where
\begin{align}
a(t) &= -\Bigl(1+(-1+t)t(1+t)(2-t+2t^3-2t^4+t^5)\Bigr),\\
b(t) &= -t(1+(-1+t)t)(1-t+t^3),\\
c(t) &= t\Bigl(-1-(-1+t)t\bigl(1+t(2+(-1+t)t)\bigr)\Bigr),\\
d(t) &= -t^2\Bigl(2+t(-3+t+t^2(2+(-2+t)t))\Bigr),\\
e(t) &= -(t^2-2t^4+2t^5),\\
f(t) &= -t^3(2-2t+t^3),\\
g(t) &= -t\Bigl(1+(-1+t)t(2+t^2(-1+2t))\Bigr),\\
h(t) &= t^2\Bigl(-1-(-1+t)t(2-t+t^3)\Bigr),\\
i(t) &= t^2(-1+t-2t^3+2t^4-t^5).
\end{align}
In particular, we extract the untwisted and twisted components
\begin{align}
H_{00}(t) &= -\dfrac{1+(-1+t)t(1+t)(2-t+2t^3-2t^4+t^5)}{(-1+t)^3(1+t+t^2)(1+t^3+t^6)},\\[1em]
H_{01}(t) &= -\dfrac{t(1+(-1+t)t)(1-t+t^3)}{(-1+t)^3(1+t+t^2)(1+t^3+t^6)},
\\[1em]
H_{03}(t) &= -\dfrac{ -t^2\Bigl(2+t(-3+t+t^2(2+(-2+t)t))\Bigr)}{(-1+t)^3(1+t+t^2)(1+t^3+t^6)},
\end{align}
Where $H_{01}$ is used to compute the refined $\eta$, while $H_{03}$ the unrefined one.

Using the prescription described in the previous section, we get
\begin{align}
\alpha &=\frac{1}{2}\eta^{\mathrm{unref}}=\operatorname*{Res}_{t=1}\left(\frac{H_{03}(t)-H_{00}(t)}{1-t}\right) = 0 \qquad\textrm{mod}\,1
\\[1em]
\alpha^{\textrm{ref}} &=\frac{1}{2}\eta^{\mathrm{ref}}={\operatorname*{Res}}^{(2)}_{t=1}\left(\frac{H_{01}(t)-H_{00}(t)}{1-t}\right) + \operatorname*{Res}_{t=1}\left(\frac{H_{01}(t)-H_{00}(t)}{1-t}\right) =\frac{1}{9}+0 = \frac{1}{9}\qquad\textrm{mod}\,1 .
\end{align}
Note that in this case there is no one-form symmetry anomaly, the only contributions to the $\eta$-invariant is coming from the 2-group refinement. We end by noting that this result is in agreement with table 2 of \cite{Cvetic:2025lat}.

\subsection{\texorpdfstring{$X = \mathrm{Cone}(Y^{p,q})$}{Ypq}} \label{sec:Ypq}

We now turn to the $X = \mathrm{Cone}(Y^{p,q})$ class of Calabi-Yau geometries, where $Y^{p,q}$ refers to the Sasaki-Einstein 5-manifold of reference \cite{Gauntlett:2004zh, Gauntlett:2004yd}. In contrast with the orbifold examples discussed above, when $p \neq q$, these are not abelian orbifolds of $\mathbb{C}^3$, and so the computation of the corresponding $\eta$-invariants is considerably less direct from a purely geometric point of view. In fact, to the best of our knowledge, these $\eta$-invariants have not previously been computed by geometric methods.\footnote{See, however, reference \cite{savale2026etainvariantcirclebundle}.} Nevertheless, by applying the same quiver-theoretic framework developed in the previous sections, we obtain a physics-based prediction for the value of the $\eta$-invariant in these examples. In particular, the results of this section agree with that obtained via toric geometry (see Appendix \ref{app:ToricGeometry} as well \cite{Apruzzi:2021nmk}). We will only be computing the $\eta$-invariant for theories without a flavor symmetry, so we will only need the unrefined $\eta^\textrm{unref}$. In particular, the examples we consider are obtained as orbifolds of $Y^{2,1}$ via actions free of fixed loci, thus no new flavor symmetry is introduced. Before considering explicit examples, we remark that the one-form symmetry group for the $Y^{p,q}$ geometries was computed in \cite{Albertini:2020mdx} to be $\mathbb Z_{\textrm{gcd}(p,q)}$.

\paragraph{Example} We now consider the geometry $X=\textrm{Cone}(Y^{4,2}$). As explained above, unlike the orbifold examples, this space is not presented directly as a quotient of $\mathbb{C}^3$. Nevertheless, it admits a quiver description from which the relevant symmetry data may be extracted (see Appendix \ref{app:QuiverAdjacency}).

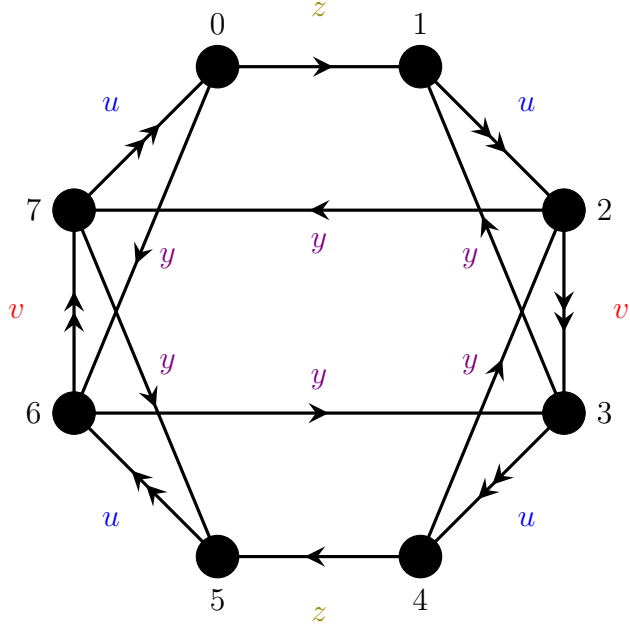
\begin{figure}
    \centering
    \begin{tikzpicture}
        \node[draw=none, minimum size=7cm,regular polygon,regular polygon sides=8] (a) {};

        \foreach \corner/\labelnum/\pos in {
        1/1/above,
        2/0/above,
        3/7/left,
        4/6/left,
        5/5/below,
        6/4/below,
        7/3/right,
        8/2/right
    }
    \node[draw, circle, inner sep=2pt, fill=black, label=\pos:$\labelnum$] at (a.corner \corner) {};

        \begin{scope}[very thick,decoration={markings, mark=at position 0.6 with {\arrow[scale = 1.5, >=stealth]{>>}}}]
        \draw[postaction = {decorate}] (a.corner 1) to (a.corner 8);
        \draw[postaction = {decorate}] (a.corner 8) to (a.corner 7);
        \draw[postaction = {decorate}] (a.corner 7) to (a.corner 6);
        \draw[postaction = {decorate}] (a.corner 5) to (a.corner 4);
        \draw[postaction = {decorate}] (a.corner 4) to (a.corner 3);
        \draw[postaction = {decorate}] (a.corner 3) to (a.corner 2);
        \end{scope}

        \begin{scope}[very thick,decoration={markings, mark=at position 0.57 with {\arrow[scale = 1.5, >=stealth]{>}}}]
        \draw[postaction = {decorate}] (a.corner 7) to (a.corner 1);
        \draw[postaction = {decorate}] (a.corner 6) to (a.corner 8);
        \draw[postaction = {decorate}] (a.corner 3) to (a.corner 5);
        \draw[postaction = {decorate}] (a.corner 2) to (a.corner 4);
        \end{scope}

        \begin{scope}[very thick,decoration={markings, mark=at position 0.57 with {\arrow[scale = 1.5, >=stealth]{>}}}]
        \draw[postaction = {decorate}] (a.corner 2) to (a.corner 1);
        \draw[postaction = {decorate}] (a.corner 6) to (a.corner 5);
        \end{scope}

        \begin{scope}[very thick,decoration={markings, mark=at position 0.52 with {\arrow[scale = 1.5, >=stealth]{>}}}]
        \draw[postaction = {decorate}] (a.corner 8) to (a.corner 3);
        \draw[postaction = {decorate}] (a.corner 4) to (a.corner 7);
        \end{scope}

        \node at (2.75,2.75) {\textcolor{blue}{$u$}};
        \node at (2.75,-2.75) {\textcolor{blue}{$u$}};
        \node at (4,0) {\textcolor{red}{$v$}};
        \node at (-2.75,2.75) {\textcolor{blue}{$u$}};
        \node at (-2.75,-2.75) {\textcolor{blue}{$u$}};
        \node at (-4,0) {\textcolor{red}{$v$}};

        \node at (0,0.9) {\textcolor{violet}{$y$}};
        \node at (0,-0.9) {\textcolor{violet}{$y$}};
        \node at (2.0,0.7) {\textcolor{violet}{$y$}};
        \node at (-2.0,0.7) {\textcolor{violet}{$y$}};
        \node at (-2.0,-0.7) {\textcolor{violet}{$y$}};
        \node at (2.0,-0.7) {\textcolor{violet}{$y$}};

        \node at (0,4) {\textcolor{olive}{$z$}};
        \node at (0,-4) {\textcolor{olive}{$z$}};

        \end{tikzpicture}
    \caption{Quiver for D-brane probe of $X=\textrm{Cone}(Y^{4,2})$.}
    \label{fig:Y21Quiver}
\end{figure}

The weighted adjacency matrix for this example is given by\footnote{Note that now the bifundamentals have different scaling dimensions.}
\begin{equation}
A_{ij}(t)=
\begin{pmatrix}
0 & t^{z(4,2)} & 0 & 0 & 0 & 0 & t^{y(4,2)} & 0 \\
0 & 0 & 2 t^{u(4,2)} & 0 & 0 & 0 & 0 & 0 \\
0 & 0 & 0 & 2 t^{v(4,2)} & 0 & 0 & 0 & t^{y(4,2)} \\
0 & t^{y(4,2)} & 0 & 0 & 2 t^{u(4,2)} & 0 & 0 & 0 \\
0 & 0 & t^{y(4,2)} & 0 & 0 & t^{z(4,2)} & 0 & 0 \\
0 & 0 & 0 & 0 & 0 & 0 & 2 t^{u(4,2)} & 0 \\
0 & 0 & 0 & t^{y(4,2)} & 0 & 0 & 0 & 2 t^{v(4,2)} \\
2 t^{u(4,2)} & 0 & 0 & 0 & 0 & t^{y(4,2)} & 0 & 0
\end{pmatrix}.
\end{equation}
See figure \ref{fig:Y21Quiver} for the quiver. Here, the scaling dimensions are fixed via the procedure of $a$-maximization \cite{Intriligator:2003jj}. This yields \cite{Benvenuti:2004wx, Benvenuti:2004dy}:
\begin{align}
    y(4,2)&=\frac{3}{2}\left(-3 + \sqrt{13} \right),\\[1em]
    z(4,2)&=\frac{1}{2}\left(-17 + 5 \sqrt{13}\right),\\[1em]
    u(4,2)&=2\left(4 - \sqrt{13}\right),\\[1em]
    v(4,2)&=\frac{1}{2}\left(-1 + \sqrt{13}\right).
\end{align}

In this case, the defect group is:
\begin{equation}
    \mathbb{D} = \textrm{Tor}\,\textrm{Coker}(B) = \mathbb{Z}_2 \oplus \mathbb{Z}_2,
\end{equation}
which agrees with \cite{Albertini:2020mdx}.\footnote{Note that the R-charges of the fields are not important in computing the one-form symmetry group. The only thing that contributes is the number of chiral fields connecting each node of the quiver. As such, we can simply set $t=1$ for this computation. We then reintroduce the R-charges when computing the Hilbert series.} With this in hand, we can extract the quiver matrix Hilbert series from \eqref{eq:GeneralQuiverHilbert}.\footnote{Recall that we must use the general matrix Hilbert series for the $Y^{p,q}$ case.} The exact form of the matrix is too long to be included, so we simply provide numerical evaluations of the relevant entries in the 
matrix Hilbert series:

\begin{align}
H_{00}(t) &= \frac{0.143321}{(1-1)^3}+\frac{0.0207026}{1-t} + 0.0207026 + O(1-t) \\
H_{04}(t) &= \frac{0.143321}{(1-t)^3}+\frac{0.0207026}{1-t} + 0.0207026 + O(1-t).
\end{align}
Note that we use $H_{04}$ instead of $H_{01}$ for $\eta^\textrm{unref}$, as explained in Appendix \ref{app:QuantumSymmetry}.

Applying the residue prescription from the previous sections we get:
\begin{equation}
    \alpha=\frac{1}{2}\eta=\operatorname*{Res}_{t=1}\left( \frac{H_{04}(t)-H_{00}(t)}{1-t} \right) =0\qquad\textrm{mod}\,1 ,
\end{equation}
which is in agreement with \cite{Apruzzi:2021nmk}. Since a direct geometric computation of the $\eta$-invariant is not presently available for this example, we regard this as a physics-based prediction arising from the quiver description of $Y^{4,2}$.

\paragraph{Example} We now consider an example that has a non-trivial one-form symmetry and anomaly: the geometry $X=\textrm{Cone}(Y^{6,3})$. As in the previous example, this space is not presented directly as a quotient of $\mathbb{C}^3$. Nevertheless, it admits a quiver description from which the relevant symmetry data may be extracted (see Appendix \ref{app:QuiverAdjacency}). This allows us to evaluate the matrix-valued quiver Hilbert series entering our proposal for the $\eta$-invariant.

\begin{figure}
    \usetikzlibrary{arrows}
    \centering
    \scalebox{1.2}{
    \begin{tikzpicture}
        \node[draw=none, minimum size=7cm,regular polygon,regular polygon sides=12] (a) {};

        \foreach \corner/\labelnum/\pos in {
        1/1/above,
        2/0/above,
        3/11/left,
        4/10/left,
        5/9/left,
        6/8/left,
        7/7/below,
        8/6/below,
        9/5/right,
        10/4/right,
        11/3/right,
        12/2/right
    }
    \node[draw, circle, inner sep=2pt, fill=black, label=\pos:$\labelnum$] at (a.corner \corner) {};

        \begin{scope}[very thick,decoration={markings, mark=at position 0.65 with {\arrow[scale = 1.5, >=stealth]{>>}}}]
        \draw[postaction = {decorate}] (a.corner 1) to (a.corner 12);
        \draw[postaction = {decorate}] (a.corner 12) to (a.corner 11);
        \draw[postaction = {decorate}] (a.corner 11) to (a.corner 10);
        \draw[postaction = {decorate}] (a.corner 9) to (a.corner 8);
        \draw[postaction = {decorate}] (a.corner 8) to (a.corner 7);
        \draw[postaction = {decorate}] (a.corner 7) to (a.corner 6);
        \draw[postaction = {decorate}] (a.corner 5) to (a.corner 4);
        \draw[postaction = {decorate}] (a.corner 4) to (a.corner 3);
        \draw[postaction = {decorate}] (a.corner 3) to (a.corner 2);
        \end{scope}

        \begin{scope}[very thick,decoration={markings, mark=at position 0.55 with {\arrow[scale = 1.5, >=stealth]{>}}}]
        \draw[postaction = {decorate}] (a.corner 11) to (a.corner 1);
        \draw[postaction = {decorate}] (a.corner 10) to (a.corner 12);
        \draw[postaction = {decorate}] (a.corner 7) to (a.corner 9);
        \draw[postaction = {decorate}] (a.corner 6) to (a.corner 8);
        \draw[postaction = {decorate}] (a.corner 3) to (a.corner 5);
        \draw[postaction = {decorate}] (a.corner 2) to (a.corner 4);
        \end{scope}

        \begin{scope}[very thick,decoration={markings, mark=at position 0.62 with {\arrow[scale = 1.5, >=stealth]{>}}}]
        \draw[postaction = {decorate}] (a.corner 2) to (a.corner 1);
        \draw[postaction = {decorate}] (a.corner 10) to (a.corner 9);
        \draw[postaction = {decorate}] (a.corner 6) to (a.corner 5);
        \end{scope}

        \begin{scope}[very thick,decoration={markings, mark=at position 0.53 with {\arrow[scale = 1.5, >=stealth]{>}}}]
        \draw[postaction = {decorate}] (a.corner 12) to (a.corner 3);
        \draw[postaction = {decorate}] (a.corner 8) to (a.corner 11);
        \draw[postaction = {decorate}] (a.corner 4) to (a.corner 7);
        \end{scope}

        \node at (2,3.5) {\textcolor{blue}{$u$}};
        \node at (3.4,2) {\textcolor{red}{$v$}};
        \node at (4,0) {\textcolor{blue}{$u$}};
        \node at (2,-3.5) {\textcolor{blue}{$u$}};
        \node at (0,-4) {\textcolor{red}{$v$}};
        \node at (-2,-3.5) {\textcolor{blue}{$u$}};
        \node at (-4,0) {\textcolor{blue}{$u$}};
        \node at (-3.4,2) {\textcolor{red}{$v$}};
        \node at (-2,3.5) {\textcolor{blue}{$u$}};

        \node at (1.85,1.9) {\textcolor{violet}{$y$}};
        \node at (-1.85,1.9) {\textcolor{violet}{$y$}};
        \node at (2.6,0.7) {\textcolor{violet}{$y$}};
        \node at (-2.6,0.7) {\textcolor{violet}{$y$}};
        \node at (0.7,-2.6) {\textcolor{violet}{$y$}};
        \node at (-0.7,-2.6) {\textcolor{violet}{$y$}};
        \node at (0,2) {\textcolor{violet}{$y$}};
        \node at (1.75,-1) {\textcolor{violet}{$y$}};
        \node at (-1.75,-1) {\textcolor{violet}{$y$}};

        \node at (0,4) {\textcolor{olive}{$z$}};
        \node at (3.4,-2) {\textcolor{olive}{$z$}};
        \node at (-3.4,-2) {\textcolor{olive}{$z$}};
    \end{tikzpicture}
    }
    \caption{Quiver for D-brane probe of $X=\textrm{Cone}(Y^{6,3})$.}
    \label{fig:Y63Quiver}
\end{figure}
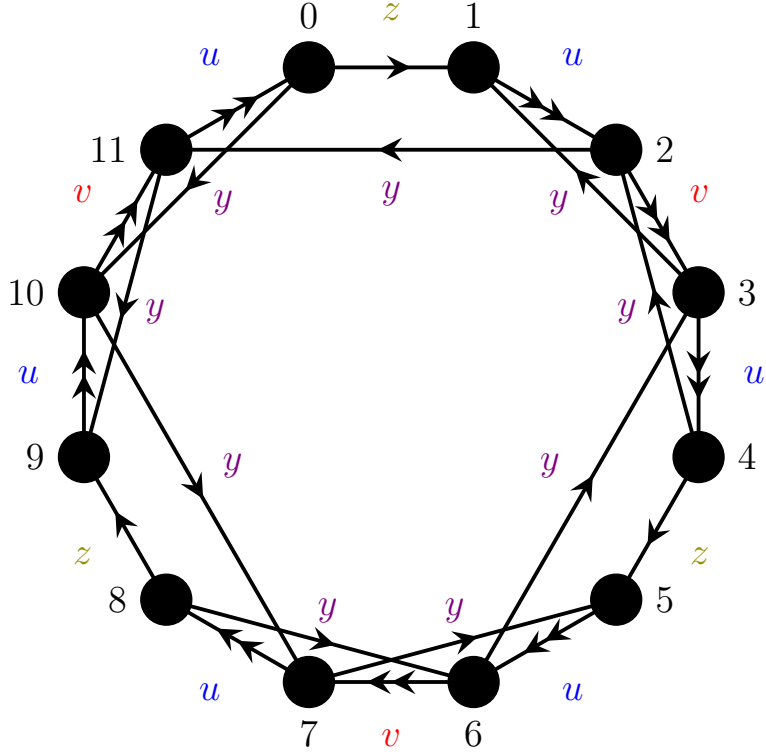

The weighted adjacency matrix for this example is given by
\begin{equation}
A_{ij}(t) =
\scalebox{0.8}{$
\begin{pmatrix}
0 & t^{z(6,3)} & 0 & 0 & 0 & 0 & 0 & 0 & 0 & 0 & t^{y(6,3)} & 0 \\
0 & 0 & 2t^{u(6,3)} & 0 & 0 & 0 & 0 & 0 & 0 & 0 & 0 & 0 \\
0 & 0 & 0 & 2t^{v(6,3)} & 0 & 0 & 0 & 0 & 0 & 0 & 0 & t^{y(6,3)} \\
0 & t^{y(6,3)} & 0 & 0 & 2t^{u(6,3)} & 0 & 0 & 0 & 0 & 0 & 0 & 0 \\
0 & 0 & t^{y(6,3)} & 0 & 0 & t^{z(6,3)} & 0 & 0 & 0 & 0 & 0 & 0 \\
0 & 0 & 0 & 0 & 0 & 0 & 2t^{u(6,3)} & 0 & 0 & 0 & 0 & 0 \\
0 & 0 & 0 & t^{y(6,3)} & 0 & 0 & 0 & 2t^{v(6,3)} & 0 & 0 & 0 & 0 \\
0 & 0 & 0 & 0 & 0 & t^{y(6,3)} & 0 & 0 & 2t^{u(6,3)} & 0 & 0 & 0 \\
0 & 0 & 0 & 0 & 0 & 0 & t^{y(6,3)} & 0 & 0 & t^{z(6,3)} & 0 & 0 \\
0 & 0 & 0 & 0 & 0 & 0 & 0 & 0 & 0 & 0 & 2t^{u(6,3)} & 0 \\
0 & 0 & 0 & 0 & 0 & 0 & 0 & t^{y(6,3)} & 0 & 0 & 0 & 2t^{v(6,3)} \\
2t^{u(6,3)} & 0 & 0 & 0 & 0 & 0 & 0 & 0 & 0 & t^{y(6,3)} & 0 & 0
\end{pmatrix}
$},
\end{equation}
See figure \ref{fig:Y63Quiver} for the quiver. Here
\begin{align}
    y(6, 3)&=\frac{1}{18}\left(-81 + 27 \sqrt{13}\right),\\[1em]
    z(6, 3)&=\frac{1}{18}\left(-153 + 45 \sqrt{13}\right),\\[1em]
    u(6, 3)&=\frac{2}{3}\left(12 - 3 \sqrt{13}\right),\\[1em]
    v(6, 3)&=\frac{1}{6}\left(-3 + 3 \sqrt{13}\right).
\end{align}

In this case, the defect group is:
\begin{equation}
    \mathbb{D} = \textrm{Tor}\,\textrm{Coker}(B) = \mathbb{Z}_3 \oplus \mathbb{Z}_3,
\end{equation}
which agrees with the result of \cite{Albertini:2020mdx}.\footnote{As in the previous example, we set $t=1$ when computing the symmetry group, and restore the scaling dimension dependence when computing the Hilbert matrix.}. With this in hand, we can extract the quiver matrix Hilbert series.
The exact form of the result is rather lengthy, so we omit this and instead focus on the essential points. The residue prescription from the previous section yields:
\begin{equation}
\alpha=\frac{1}{2}\eta^\textrm{unref} = \operatorname*{Res}_{t=1}\left( \frac{H_{04}(t)-H_{00}(t)}{1-t} \right)  =\frac{2}{9}\qquad\textrm{mod}\,1 .
\end{equation}
Since a direct geometric computation of the $\eta$-invariant is not presently available for this example, we regard this as a physics-based prediction.

\section{Conclusions} \label{sec:CONC}

In this paper we developed a quiver based method for extracting the SymTh / SymTFT of 5D SCFTs engineered by M-theory on a non-compact Calabi-Yau cone $X$. This provides a complementary approach to extracting this data in situations where more geometric approaches are computationally cumbersome / currently unavailable. One of the key ideas in our approach was to show how a residue of a matrix Hilbert series directly matches to the data of twisted $\eta$-invariants and their associated Molien series. We used this to provide a quiver-based computation of $\eta$-invariants and their refinements (in cases where $\partial X$ is singular) to extract the relevant anomaly terms in toric geometries, including both abelian orbifolds and Calabi-Yau cones of Sasaki-Einstein five-manifolds. More broadly, we anticipate that this procedure works in all situations where the quiver gauge theory of branes probing a singularity is known. In the remainder of this section we discuss some potential avenues for future investigation.

It would be useful to compute the 2-group symmetry directly from the quiver. In this paper we extracted anomaly data in situations where non-isolated singularities lead to 2-group refinements of the symmetry theory, but we did not give a purely quiver based derivation of the 2-group symmetry itself. Such a derivation would be useful because the quiver is sensitive not only to the flavor symmetries of the 5D SCFT, but also to the zero-form R-symmetry group, which should also contribute to the full 2-group symmetry structure. Thus, understanding how these ingredients are represented in the quiver would provide a route to also including R-symmetry contributions to higher-group structures.

Another natural direction is to extend this analysis to non-supersymmetric orbifolds (see e.g. \cite{Morrison:2004fr, Adams:2001sv, Harvey:2001wm, Martinec:2001cf, Dabholkar:2001wn, Vafa:2001ra, Narayan:2009uy, Braeger:2024jcj, Braeger:2025rov}) and to orbifolds of other complex dimensions, such as Calabi-Yau fourfold singularities (see e.g. \cite{Becker:1996gj, Sethi:1996es, Klemm:1996ts}). In these cases, the main new ingredient one would need is the appropriate replacement of the quiver Hilbert series used throughout this paper. For Calabi-Yau threefold orbifolds, the Hilbert series can be derived from the McKay correspondence by applying the orbifold projection to the Koszul resolution \cite{eisenbud1995commutative}
\begin{equation}
1 \longrightarrow \mathbf{3} \longrightarrow \overline{\mathbf{3}} \longrightarrow 1,
\end{equation}
or, more schematically,
\begin{equation}
\rho \otimes \left( 1 \longrightarrow \mathbf{3} \longrightarrow \overline{\mathbf{3}} \longrightarrow 1 \right) \quad \longrightarrow \quad H(t),
\end{equation}
where $\rho$ is an irreducible representation of the orbifold group.\footnote{We thank M. Hübner for discussion on this point.} For non-supersymmetric orbifolds, or for Calabi-Yau fourfolds, the corresponding Koszul complex and representation content are modified. One should therefore be able to define an analogous matrix-valued Hilbert series by replacing the above complex with the appropriate representation-theoretic resolution. It would be interesting to understand whether the resulting Hilbert series continues to capture relative $\eta$-invariants and anomaly data in these more general settings.

It would be interesting to apply our proposal to broader families of examples. To name a few, the $L^{p,q,r}$ case \cite{Martelli:2005wy, Cvetic:2005ft}, non-abelian orbifolds, and more general non-geometric compactifications, where the appropriate notion of a geometric internal space is absent or only partially available, would all be interesting to explore further. For these cases the geometric computation of the $\eta$-invariant is unknown for the most part, but we anticipate that the quiver provides a straightforward route to acces this information regardless.\footnote{Some results on non-abelian orbifolds can be found in \cite{Cvetic:2025lat}.}

Finally, we have presented a general quiver based prescription which directly computes the geometric twisted $\eta$-invariant computations on $\partial X$. That being said, it would be desirable to have a more complete physical understanding of why the matrix Hilbert series is the correct quantity to consider.

\section*{Acknowledgments}

We thank M. H\"ubner for comments on an earlier draft. 
We thank D. Brennan, D. Dramburg, M. H\"ubner, and W. Xing for helpful discussions. 
VC and JJH thank the 2025 Simons Summer workshop for hospitality during part of this work.
The work of VC is supported by a NSF Graduate Research Fellowship. The work of
VC, JJH and SM is supported by DOE (HEP) Award DE-SC0013528. The work of VC
is also supported by the Slovenian Research Agency (ARRS No. P1-0306) and Fay R. and
Eugene L. Langberg Endowed Chair funds. The work of JJH and SM is also supported by BSF grant
2022100 and a University Research Foundation grant at the University of Pennsylvania.

\newpage

\appendix

\section{Anomalies via Toric Geometry} \label{app:ToricGeometry}

In this Appendix we review the geometrical approach to the computation of anomalies discussed in \cite{Apruzzi:2021nmk}. We will not deal with the formal derivation, which can be found in the refence, but we will describe how this computation can be carried out in general, applying it then to the cases of interest in this paper.

In \cite{Apruzzi:2021nmk}, the computation of the anomaly for one-form symmetries and the mixed anomaly with gravity is done via the triple intersection of a specific compact divisor with itself and with the Pontrjagin class $p_1$. The  divisor of interest can than be mapped to the boundary cycle generating the higher form symmetry. The compact divisor is identified as the 4-cycle that has intersection $0$ mod$\, n$ with every curve, where $n$ is the order of the one-form symmetry, without being 0 mod$\, n$ itself.

The simplest example is local $\mathbb{P}^2$. In the resolved geometry there is only one compact divisor, $E_0$, and the theory has a $\mathbb{Z}_3$ one-form symmetry. It is easy to check that $C \cdot E_0 = 0$ mod$\, 3$ for every compact curve $C$ and the (mixed) anomaly coefficients can be computed via
\begin{align}
    \beta &= \left[\frac{1}{6} \frac{E_0 \cdot E_0 \cdot E_0}{3^3} \right]_{\mathrm{mod}\, 1} = \frac{1}{6 \cdot 3} \, \nonumber
    \\
    \gamma &= \left[-\frac{1}{24}\frac{E_0\cdot p_1}{3}\right]_{\mathrm{mod}\, 1} = -\frac{4}{24} \, .
\end{align}

These quantities can be combined to give the $\eta$ invariant as
\begin{align}
    \alpha = \left[\frac{1}{6} \left( \frac{E_0 \cdot E_0 \cdot E_0}{3^3} - \frac{E_0\cdot p_1}{4\cdot3} \right)\right]_{\mathrm{mod}\, 1} = - \frac{1}{9} \, .
\end{align}

This procedure can be applied to any local Calabi Yau geometry, however, the computation of self triple intersections is not always easy, since it relies on establishing linear relations among divisors. In what follow we will present examples of non-compact toric geometries for which the above computation can be carried out via combinatorial methods. For the interested reader, a \texttt{Sage} notebook can be found in the attachments for the \texttt{arXiv} submission that carries out the computation for any toric diagram.

\subsection{Abelian Orbifolds of \texorpdfstring{$\mathbb C^3$}{C3}} \label{sec:C3/Z3Toric}

The simplest example of local toric CY is given by abelian orbifolds of the form $\mathbb C^3/\Gamma$. Given the action of $\Gamma$ on the coordinates, $(a,b,c)$ such that $a+b+c=0$ mod$|\Gamma|$, the toric diagram of the orbifolded geometry can be obtained starting from the fan of $\mathbb C^3$, $v_i \in \{(1,0,0),(0,1,0),(0,0,1)\}$, and adding a vector $e = (a,b,c)/n$. The new lattice generated by $\{v_i, e \}$ can be casted in the usual 2d toric diagram presentation via a change of coordinates.

Once the toric diagram is computed, it is straightforward to compute the intersections between curves and divisors following standard methods, see \cite{Closset:2018bjz} for a easy reference. Once this is done, the divisor associated via relative homology to the boundary cycle generating the one form symmetry is given by solving
\begin{align}
    C_i \cdot \left( \sum_k a_k E_k \right) = 0 \quad \mathrm{mod}\, n
\end{align}
for all $a_k$ not zero modulo $n$, the order of the one-form symmetry.

\subsubsection{\texorpdfstring{$\mathbb{C}^3/\mathbb{Z}_5$}{}}

The toric diagram for $\mathbb{C}^3/\mathbb{Z}_5$ is the triangle with vertices $\{a,b,c\}=\{(0,0),(2,3),(3,2)\}$. We will label the non-compact divisors associated to the points on the perimeter as $D_{a,b,c}$, while the compact ones $E_{0,1}$ are associated to the internal points $\{0,1\}=\{(1,1),(2,2)\}$. Finally, the compact curves arising by the intersection of compact and (non-)compact divisors are labeled according to the intersecting divisors, e.g. $C_{a,0}=D_a \cdot E_0$.

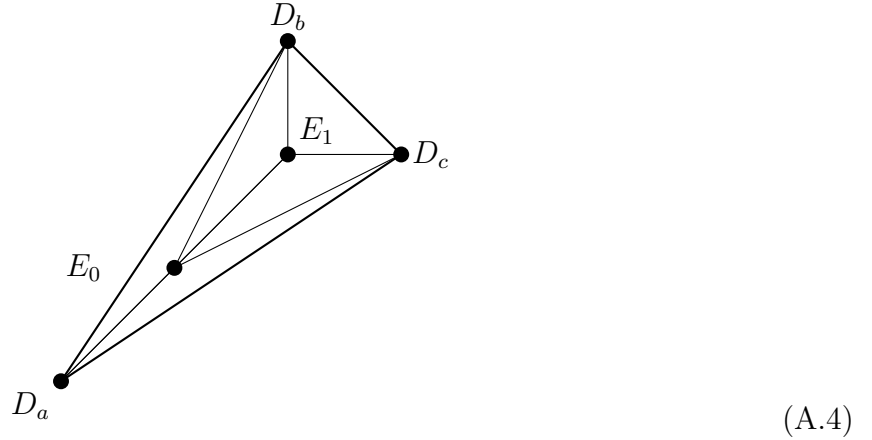
\begin{equation}
\begin{tikzpicture}[scale=1.5]

\coordinate (A) at (0,0);
\coordinate (B) at (2,3);
\coordinate (C) at (3,2);

\coordinate (E0) at (1,1);
\coordinate (E1) at (2,2);

\draw[thick] (A) -- (B) -- (C) -- cycle;

\draw (A) -- (E0);
\draw (A) -- (E1);
\draw (B) -- (E0);
\draw (B) -- (E1);
\draw (C) -- (E0);
\draw (C) -- (E1);
\draw (E0) -- (E1);

\fill (A) circle (2pt);
\fill (B) circle (2pt);
\fill (C) circle (2pt);
\fill (E0) circle (2pt);
\fill (E1) circle (2pt);

\node[below left] at (A) {$D_a$};
\node[above] at (B) {$D_b$};
\node[right] at (C) {$D_c$};

\node at ($(E0)+(-0.8,0)$) {$E_0$};
\node[above right] at (E1) {$E_1$};

\end{tikzpicture}
\end{equation}

The intersection of curves and divisors can be readily computed following \cite{Closset:2018bjz, Tian:2021cif}.
\begin{equation}
\begin{tabular}{c|ccccc}
 & $D_a$ & $D_b$ & $D_c$ & $E_0$ & $E_1$ \\
\hline
$C_{a,0}$ & 3 & 1 & 1 & -5 & 0 \\
$C_{b,0}$ & 1 & 0 & 0 & -2 & 1 \\
$C_{c,0}$ & 1 & 0 & 0 & -2 & 1 \\
$C_{b,1}$ & 0 & 1 & 1 & 1 & -3 \\
$C_{c,1}$ & 0 & 1 & 1 & 1 & -3 \\
$C_{0,1}$ & 0 & 1 & 1 & 1 & -3 \\
\end{tabular}
\end{equation}

The theory admits a $\mathbb{Z}_5$ one-form symmetry and from the table, we can find the divisor associated to the one-form symmetry to be $\mathcal{D}= - E_0 - 2 E_1$, since indeed
\begin{align}
    C_{(a,b,c),(0,1)} \cdot \mathcal{D} = 0 \text{ mod} \,5 \, .
\end{align}

The anomaly for this symmetry is therefore:
\begin{align}
    \alpha = \frac{1}{6} \left[ \frac{\mathcal{D} \cdot \mathcal{D} \cdot \mathcal{D}}{5^3} - \frac{\mathcal{D}\cdot p_1}{4\cdot5} \right]_{\mathrm{mod} \, 1} = \frac{1}{5} \, ,
\end{align}
corresponding to the $\eta$-invariant computed in section \ref{sec:EXAMPLES}.


\subsubsection{\texorpdfstring{$\mathbb{C}^3/\mathbb{Z}_6$}{C3Z6}}

We now repeat the above computation in the case of a non-isolated singularity. In this case the effective one-form symmetry is reduced and, as we will see, this needs to be take into account when picking up the divisor associated to the center symmetry.

The intersection matrix is given by
\begin{equation}
\begin{tabular}{c|cccccc}
 & $D_a$ & $D_b$ & $D_c$ & $D_e$ & $E_0$ & $E_1$ \\
\hline
$C_{a,0}$ & 0 & 0 & 0 & 1 & -2 & 1 \\
$C_{b,0}$ & 0 & 0 & 0 & 1 & -2 & 1 \\
$C_{e,0}$ & 1 & 1 & 0 & -2 & 0 & 0 \\
$C_{a,1}$ & 0 & 0 & 1 & 0 & 1 & -2 \\
$C_{b,1}$ & 0 & 0 & 1 & 0 & 1 & -2 \\
$C_{c,1}$ & 1 & 1 & 4 & 0 & 0 & -6 \\
$C_{0,1}$ & 1 & 1 & 0 & -2 & 0 & 0 \\
\end{tabular}
\end{equation}
And the divisor associated to the one-form symmetry is $\mathcal{D}=E_0+2E_1$.

\begin{equation}
\begin{tikzpicture}[scale=1.2]

\coordinate (A) at (0,0);
\coordinate (B) at (2,0);
\coordinate (C) at (1,3);

\coordinate (E0) at (1,2);
\coordinate (E1) at (1,1);
\coordinate (E2) at (1,0);

\draw[thick] (A) -- (B) -- (C) -- cycle;

\draw (A) -- (E0);
\draw (A) -- (E1);
\draw (A) -- (E2);

\draw (B) -- (E0);
\draw (B) -- (E1);
\draw (B) -- (E2);

\draw (C) -- (E0);
\draw (C) -- (E1);
\draw (C) -- (E2);

\draw (E0) -- (E1);
\draw (E1) -- (E2);

\fill (A) circle (2pt);
\fill (B) circle (2pt);
\fill (C) circle (2pt);
\fill (E0) circle (2pt);
\fill (E1) circle (2pt);
\fill (E2) circle (2pt);

\node[below left] at (A) {$D_a$};
\node[below] at (B) {$D_b$};
\node[right] at (C) {$D_c$};

\node[left] at ($(E0)+(-0.3,0)$) {$E_1$};
\node[below] at ($(E1)+(-0.3,-0.4)$) {$E_0$};
\node[below] at (E2) {$D_e$};

\end{tikzpicture}
\end{equation}

Once again, the anomaly can be computed to be
\begin{align}
    \alpha = \frac{1}{6} \left[ \frac{\mathcal{D} \cdot \mathcal{D} \cdot \mathcal{D}}{3^3} - \frac{\mathcal{D}\cdot p_1}{4\cdot3} \right]_{\mathrm{mod}\,1} = \frac{1}{9}   ,
\end{align}
again in agreement with the results in section \ref{sec:EXAMPLES}.


\subsection{\texorpdfstring{$\mathrm{Cone}(Y^{p,q})$}{Ypq}} \label{sec:YpqToric}

We now consider some other well known toric geometries, the $Y^{p,q}$. Here we will reproduce the detailed geometric computation of the examples proposed in the main text.

\paragraph{Example $\mathrm{Cone}(Y^{6,3})$:} This theory can be obtained as an orbifold of $\mathrm{Cone}(Y^{2,1})$ by a suitable action of $\mathbb{Z}_3$, thus we have a $\mathbb{Z}_3$ one-form symmetry.

The intersection matrix is given by
\begin{equation}
\begin{tabular}{c|ccccccccc}
 & $D_a$ & $D_b$ & $D_c$ & $D_d$ & $E_0$ & $E_1$ & $E_2$ & $E_3$ & $E_4$\\
\hline
$C_{a,0}$ & 0 & 1 & 0 & 0 & -2 & 1 & 0 & 0 & 0 \\
$C_{a,1}$ & 0 & 0 & 0 & 0 & 1 & -2 & 1 & 0 & 0 \\
$C_{a,2}$ & 0 & 0 & 0 & 0 & 0 & 1 & -2 & 1 & 0 \\
$C_{a,3}$ & 0 & 0 & 0 & 0 & 0 & 0 & 1 & -2 & 1 \\
$C_{a,4}$ & 0 & 0 & 0 & 1 & 0 & 0 & 0 & 1 & -2 \\
$C_{c,0}$ & 0 & 1 & 0 & 0 & -2 & 1 & 0 & 0 & 0 \\
$C_{c,1}$ & 0 & 0 & 0 & 0 & 1 & -2 & 1 & 0 & 0 \\
$C_{c,2}$ & 0 & 0 & 0 & 0 & 0 & 1 & -2 & 1 & 0 \\
$C_{c,3}$ & 0 & 0 & 0 & 0 & 0 & 0 & 1 & -2 & 1 \\
$C_{c,4}$ & 0 & 0 & 0 & 1 & 1 & 0 & 0 & 1 & -2 \\
$C_{b,0}$ & 1 & 1 & 1 & 0 & -3 & 0 & 0 & 0 & 0 \\
$C_{d,4}$ & 1 & 0 & 1 & 7 & 0 & 0 & 0 & 0 & -9 \\
$C_{0,1}$ & 1 & 0 & 1 & 0 & -1 & -1 & 0 & 0 & 0 \\
$C_{1,2}$ & 1 & 0 & 1 & 0 & 0 & -3 & 1 & 0 & 0 \\
$C_{2,3}$ & 1 & 0 & 1 & 0 & 0 & 0 & -5 & 3 & 0 \\
$C_{3,4}$ & 1 & 0 & 1 & 0 & 0 & 0 & 0 & -7 & 5 \\
\end{tabular}
\end{equation}
And the divisor associated to the one-form symmetry is $\mathcal{D}=E_0+2E_1+E_3+2E_4$, with anomaly

\begin{align}
    \alpha = \frac{1}{6} \left[ \frac{\mathcal{D} \cdot \mathcal{D} \cdot \mathcal{D}}{3^3} - \frac{\mathcal{D}\cdot p_1}{4\cdot3}\right]_{\mathrm{mod}\,1} = \frac{2}{9} ,
\end{align}
once again in agreement with the matrix Hilbert series computation.

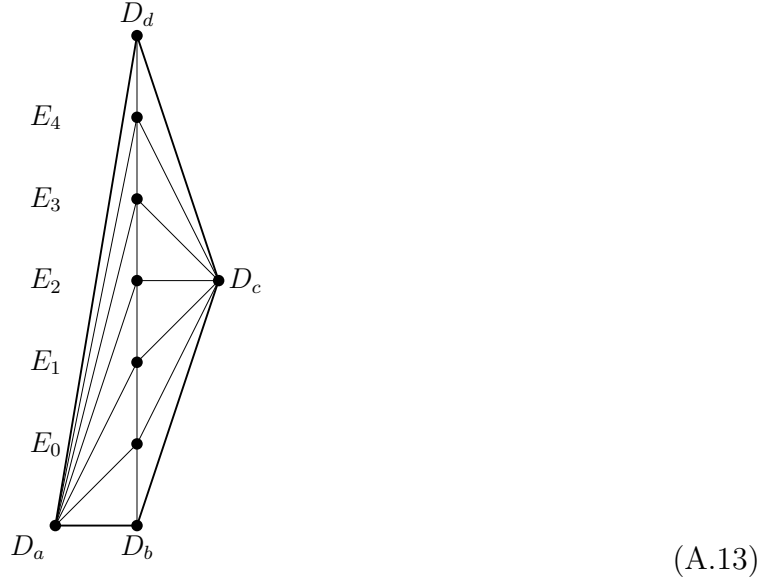
\begin{equation}
\scalebox{0.9}{
\begin{tikzpicture}[scale=1.2]

\coordinate (A) at (-1,0);
\coordinate (B) at (0,0);
\coordinate (C) at (1,3);
\coordinate (D) at (0,6);

\coordinate (E0) at (0,1);
\coordinate (E1) at (0,2);
\coordinate (E2) at (0,3);
\coordinate (E3) at (0,4);
\coordinate (E4) at (0,5);

\draw[thick] (A) -- (B) -- (C) -- (D) -- cycle;

\draw (A) -- (E0);
\draw (A) -- (E1);
\draw (A) -- (E2);
\draw (A) -- (E3);
\draw (A) -- (E4);

\draw (B) -- (E0);

\draw (C) -- (E0);
\draw (C) -- (E1);
\draw (C) -- (E2);
\draw (C) -- (E3);
\draw (C) -- (E4);

\draw (D) -- (E4);

\draw (E0) -- (E1);
\draw (E1) -- (E2);
\draw (E2) -- (E3);
\draw (E3) -- (E4);

\fill (A) circle (2pt);
\fill (B) circle (2pt);
\fill (C) circle (2pt);
\fill (D) circle (2pt);
\fill (E0) circle (2pt);
\fill (E1) circle (2pt);
\fill (E2) circle (2pt);
\fill (E3) circle (2pt);
\fill (E4) circle (2pt);

\node[below left] at (A) {$D_a$};
\node[below] at (B) {$D_b$};
\node[right] at (C) {$D_c$};
\node[above] at (D) {$D_d$};

\node[left] at ($(E0)+(-0.8,0)$) {$E_0$};
\node[left] at ($(E1)+(-0.8,0)$) {$E_1$};
\node[left] at ($(E2)+(-0.8,0)$) {$E_2$};
\node[left] at ($(E3)+(-0.8,0)$) {$E_3$};
\node[left] at ($(E4)+(-0.8,0)$) {$E_4$};

\end{tikzpicture}
}
\end{equation}

\section{Quiver Adjacency Matrices} \label{app:QuiverAdjacency}

In this Appendix, we review how to compute the adjacency matrices of the quivers used throughout this paper. We focus on the two classes of examples relevant for our analysis: orbifolds of $\mathbb C^3$ and the $\mathrm{Cone}(Y^{p,q})$ geometries.

We begin with orbifolds of the form
\begin{equation}
X = \mathbb C^3 / \mathbb Z_N(m_1,m_2,m_3),
\end{equation}
where the generator $g \in \mathbb Z_N$ acts as
\begin{equation}
(z_1,z_2,z_3) \mapsto (\omega^{m_1} z_1,\, \omega^{m_2} z_2,\, \omega^{m_3} z_3), \qquad \omega = e^{2\pi i/N}.
\end{equation}
The Calabi–Yau condition requires
\begin{equation}
m_1 + m_2 + m_3 \equiv 0 \quad \mathrm{mod}\, N.
\end{equation}
The quiver is obtained via the generalized McKay correspondence.\footnote{See \cite{Douglas:1996sw, Kachru:1998ys, Lawrence:1998ja, McKay, ito96, PeterBKronheimer:1990zmj} for foundational work on the subject.} Nodes correspond to irreducible representations $R_i$ of $\mathbb Z_N$. Since $\mathbb Z_N$ is abelian, all irreducible representations are one-dimensional and labeled by $i=0,\dots,N-1$. Arrows are determined by the decomposition of the defining representation $R$ acting on $\mathbb C^3$. Writing
\begin{equation}
 R \simeq R_{m_1} \oplus R_{m_2} \oplus R_{m_3},
\end{equation}
the tensor product rule gives
\begin{equation}
R_i \otimes R = R_{i+m_1} \oplus R_{i+m_2} \oplus R_{i+m_3}.
\end{equation}
Thus, for each node $i$, there are arrows
\begin{equation}
i \longrightarrow i+m_1,\qquad i \longrightarrow i+m_2,\qquad i \longrightarrow i+m_3,
\end{equation}
with all indices understood modulo $N$.

Equivalently, the adjacency matrix $A_{ij}$ is given by
\begin{equation}
A_{ij} = \# \{ \alpha \in \{m_1,m_2,m_3\} \mid j \equiv i+\alpha \ (\mathrm{mod}\, N)\}.
\end{equation}
This construction follows directly from the McKay graph associated to the defining representation of the orbifold group, where arrows encode how irreducible representations appear in tensor products with the defining representation.

\paragraph{Example: $\mathbb C^3/\mathbb Z_3(1,1,1)$}

In this case, we have
\begin{equation}
R \simeq R_1 \oplus R_1 \oplus R_1,
\end{equation}
so each node $i$ has three arrows to $i+1$. This reproduces the cyclic quiver with three arrows between consecutive nodes used in section \ref{sec:IllustrativeExamplePart2}.

We now turn to the $\mathrm{Cone}(Y^{p,q})$ geometries. Unlike orbifolds, these do not arise directly from a group action on $\mathbb C^3$, but their quivers can be obtained from orbifold quivers via a simple and powerful procedure explained in \cite{Benvenuti:2004dy}.

The starting point is the observation that the $\mathrm{Cone}(Y^{p,p})$ geometry corresponds to the orbifold
\begin{equation}
\mathrm{Cone}(Y^{p,p}) \simeq \mathbb C^3/\mathbb Z_{2p}(1,1,-2).
\end{equation}
The corresponding quiver can therefore be constructed using the McKay prescription described above. The key idea, explained in \cite{Benvenuti:2004dy}, is that the quivers for general $\mathrm{Cone}(Y^{p,q})$ can be obtained from this orbifold quiver by removing arrows in a systematic way. Physically, this corresponds to turning on relevant deformations that partially resolve the singularity. The quiver for $\mathrm{Cone}(Y^{p,q})$ has $2p$ nodes, four types of bifundamental fields, conventionally denoted $Y$, $Z$, $U$, and $V$, and a superpotential determined by a periodic quiver / brane tiling construction. The fields carry different scaling dimensions / $R$-charges, which in the superconformal theory are given by
\begin{align}
\Delta_Y &= y(p,q)= \frac{-4 p^2 + 3 q^2 + 2 p q + (2 p - q) \sqrt{4 p^2 - 3 q^2}}{2 q^2}, \\
\Delta_Z &= z(p,q)= \frac{-4 p^2 + 3 q^2 - 2 p q + (2 p + q) \sqrt{4 p^2 - 3 q^2}}{2 q^2}, \\
\Delta_U &= u(p,q)= \frac{2 p \left( 2 p - \sqrt{4 p^2 - 3 q^2} \right)}{2 q^2}, \\
\Delta_V &= v(p,q)= \frac{3 q - 2 p + \sqrt{4 p^2 - 3 q^2}}{2 q}.
\end{align}
For $p=q$ these reduce to $\Delta_Y = \Delta_U = \Delta_V = 1$ and there are no $Z$ fields, so we can formally set ``$\Delta_Z=0$.'' This recovers the $Y^{p,p}$ orbifold theory. The adjacency matrix is determined by counting how many fields connect each pair of nodes. Unlike the orbifold case, different arrows now carry different weights.

\begin{figure}
    \usetikzlibrary{arrows}
    \centering
    \scalebox{1.2}{
    \begin{tikzpicture}
        \node[draw=none, minimum size=7cm,regular polygon,regular polygon sides=12] (a) {};

        \foreach \corner/\labelnum/\pos in {
        1/1/above,
        2/0/above,
        3/11/left,
        4/10/left,
        5/9/left,
        6/8/left,
        7/7/below,
        8/6/below,
        9/5/right,
        10/4/right,
        11/3/right,
        12/2/right
        }
        \node[draw, circle, inner sep=2pt, fill=black, label=\pos:$\labelnum$] at (a.corner \corner) {};

        \begin{scope}[very thick,decoration={markings, mark=at position 0.65 with {\arrow[scale = 1.5, >=stealth]{>>}}}]
        \draw[postaction = {decorate}] (a.corner 1) to (a.corner 12);
        \draw[postaction = {decorate}] (a.corner 12) to (a.corner 11);
        \draw[postaction = {decorate}] (a.corner 11) to (a.corner 10);
        \draw[postaction = {decorate}] (a.corner 10) to (a.corner 9);
        \draw[postaction = {decorate}] (a.corner 9) to (a.corner 8);
        \draw[postaction = {decorate}] (a.corner 8) to (a.corner 7);
        \draw[postaction = {decorate}] (a.corner 7) to (a.corner 6);
        \draw[postaction = {decorate}] (a.corner 6) to (a.corner 5);
        \draw[postaction = {decorate}] (a.corner 5) to (a.corner 4);
        \draw[postaction = {decorate}] (a.corner 4) to (a.corner 3);
        \draw[postaction = {decorate}] (a.corner 3) to (a.corner 2);
        \draw[postaction = {decorate}] (a.corner 2) to (a.corner 1);
        \end{scope}

        \begin{scope}[very thick,decoration={markings, mark=at position 0.55 with {\arrow[scale = 1.5, >=stealth]{>}}}]
        \draw[postaction = {decorate}] (a.corner 1) to (a.corner 3);
        \draw[postaction = {decorate}] (a.corner 12) to (a.corner 2);
        \draw[postaction = {decorate}] (a.corner 11) to (a.corner 1);
        \draw[postaction = {decorate}] (a.corner 10) to (a.corner 12);
        \draw[postaction = {decorate}] (a.corner 9) to (a.corner 11);
        \draw[postaction = {decorate}] (a.corner 8) to (a.corner 10);
        \draw[postaction = {decorate}] (a.corner 7) to (a.corner 9);
        \draw[postaction = {decorate}] (a.corner 6) to (a.corner 8);
        \draw[postaction = {decorate}] (a.corner 5) to (a.corner 7);
        \draw[postaction = {decorate}] (a.corner 4) to (a.corner 6);
        \draw[postaction = {decorate}] (a.corner 3) to (a.corner 5);
        \draw[postaction = {decorate}] (a.corner 2) to (a.corner 4);
        \end{scope}

        \node at (0,4) {\textcolor{red}{$v$}};
        \node at (2,3.5) {\textcolor{blue}{$u$}};
        \node at (3.4,2) {\textcolor{red}{$v$}};
        \node at (4,0) {\textcolor{blue}{$u$}};
        \node at (3.4,-2) {\textcolor{red}{$v$}};
        \node at (2,-3.5) {\textcolor{blue}{$u$}};
        \node at (0,-4) {\textcolor{red}{$v$}};
        \node at (-2,-3.5) {\textcolor{blue}{$u$}};
        \node at (-3.4,-2) {\textcolor{red}{$v$}};
        \node at (-4,0) {\textcolor{blue}{$u$}};
        \node at (-3.4,2) {\textcolor{red}{$v$}};
        \node at (-2,3.5) {\textcolor{blue}{$u$}};

        \node at (0.7,2.6) {\textcolor{violet}{$y$}};
        \node at (-0.7,2.6) {\textcolor{violet}{$y$}};
        \node at (1.85,1.9) {\textcolor{violet}{$y$}};
        \node at (-1.85,1.9) {\textcolor{violet}{$y$}};
        \node at (2.6,0.7) {\textcolor{violet}{$y$}};
        \node at (2.6,-0.7) {\textcolor{violet}{$y$}};
        \node at (-2.6,0.7) {\textcolor{violet}{$y$}};
        \node at (-2.6,-0.7) {\textcolor{violet}{$y$}};
        \node at (1.85,-1.9) {\textcolor{violet}{$y$}};
        \node at (-1.85,-1.9) {\textcolor{violet}{$y$}};
        \node at (0.7,-2.6) {\textcolor{violet}{$y$}};
        \node at (-0.7,-2.6) {\textcolor{violet}{$y$}};
    \end{tikzpicture}
    }
    \caption{Quiver for D-brane probe of $X=\textrm{Cone}(Y^{6,6})=\mathbb{C}^3/\mathbb{Z}_{12}$. }
    \label{fig:Z12Quiver}
\end{figure}
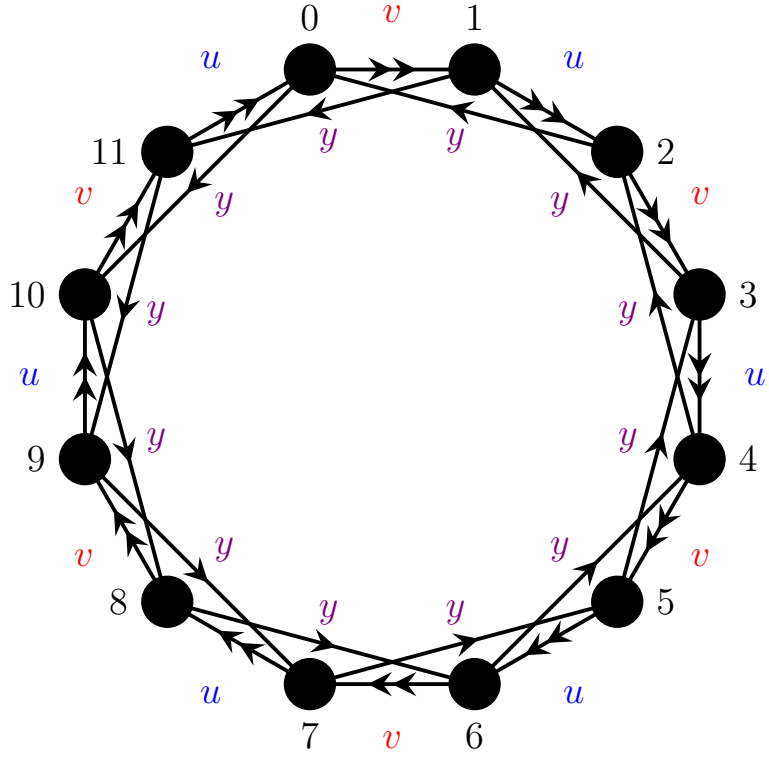

To illustrate the general procedure, we will now consider the explicit example of the $\mathrm{Cone}(Y^{6,5})$ geometry. The starting point is the $Y^{6,6}=\mathbb C^3/\mathbb Z_{12}(1,1,-2)$ geometry. See figure \ref{fig:Z12Quiver} for the corresponding quiver; the labeling of arrows in this quiver matches that of \cite{Benvenuti:2004dy} so as to make the erasing procedure explicit. In particular, we label each field's scaling dimension assignment as $u$, $v$, $y$, and $z$.
The weighted adjacency matrix can be written as
\begin{equation}
A^{Y^{6,6}}_{ij}(t)
    \begin{pmatrix}
        0 & 2t^{v} & 0 & 0 & 0 & 0 & 0 & 0 & 0 & 0 & t^y & 0 \\
        0 & 0 & 2t^{u} & 0 & 0 & 0 & 0 & 0 & 0 & 0 & 0 & t^y \\
        t^y & 0 & 0 & 2t^{v} & 0 & 0 & 0 & 0 & 0 & 0 & 0 & 0 \\
        0 & t^y & 0 & 0 & 2t^{u} & 0 & 0 & 0 & 0 & 0 & 0 & 0 \\
        0 & 0 & t^y & 0 & 0 & 2t^{v} & 0 & 0 & 0 & 0 & 0 & 0 \\
        0 & 0 & 0 & t^y & 0 & 0 & 2t^{u} & 0 & 0 & 0 & 0 & 0 \\
        0 & 0 & 0 & 0 & t^y & 0 & 0 & 2t^{v} & 0 & 0 & 0 & 0 \\
        0 & 0 & 0 & 0 & 0 & t^y & 0 & 0 & 2t^{u} & 0 & 0 & 0 \\
        0 & 0 & 0 & 0 & 0 & 0 & t^y & 0 & 0 & 2t^{v} & 0 & 0 \\
        0 & 0 & 0 & 0 & 0 & 0 & 0 & t^y & 0 & 0 & 2t^{u} & 0 \\
        0 & 0 & 0 & 0 & 0 & 0 & 0 & 0 & t^y & 0 & 0 & 2t^{v} \\
        2t^{u} & 0 & 0 & 0 & 0 & 0 & 0 & 0 & 0 & t^y & 0 & 0
    \end{pmatrix}.
\end{equation}

\begin{figure}
    \usetikzlibrary{arrows}
    \centering
    \scalebox{1.2}{
    \begin{tikzpicture}
        \node[draw=none, minimum size=7cm,regular polygon,regular polygon sides=12] (a) {};
        \foreach \corner/\labelnum/\pos in {
        1/1/above,
        2/0/above,
        3/11/left,
        4/10/left,
        5/9/left,
        6/8/left,
        7/7/below,
        8/6/below,
        9/5/right,
        10/4/right,
        11/3/right,
        12/2/right
        }
        \node[draw, circle, inner sep=2pt, fill=black, label=\pos:$\labelnum$] at (a.corner \corner) {};

        \begin{scope}[very thick,decoration={markings, mark=at position 0.65 with {\arrow[scale = 1.5, >=stealth]{>>}}}]
        \draw[postaction = {decorate}] (a.corner 1) to (a.corner 12);
        \draw[postaction = {decorate}] (a.corner 12) to (a.corner 11);
        \draw[postaction = {decorate}] (a.corner 11) to (a.corner 10);
        \draw[postaction = {decorate}] (a.corner 10) to (a.corner 9);
        \draw[postaction = {decorate}] (a.corner 9) to (a.corner 8);
        \draw[postaction = {decorate}] (a.corner 8) to (a.corner 7);
        \draw[postaction = {decorate}] (a.corner 7) to (a.corner 6);
        \draw[postaction = {decorate}] (a.corner 6) to (a.corner 5);
        \draw[postaction = {decorate}] (a.corner 5) to (a.corner 4);
        \draw[postaction = {decorate}] (a.corner 4) to (a.corner 3);
        \draw[postaction = {decorate}] (a.corner 3) to (a.corner 2);
        \end{scope}

        \begin{scope}[very thick,decoration={markings, mark=at position 0.55 with {\arrow[scale = 1.5, >=stealth]{>}}}]
        \draw[postaction = {decorate}] (a.corner 11) to (a.corner 1);
        \draw[postaction = {decorate}] (a.corner 10) to (a.corner 12);
        \draw[postaction = {decorate}] (a.corner 9) to (a.corner 11);
        \draw[postaction = {decorate}] (a.corner 8) to (a.corner 10);
        \draw[postaction = {decorate}] (a.corner 7) to (a.corner 9);
        \draw[postaction = {decorate}] (a.corner 6) to (a.corner 8);
        \draw[postaction = {decorate}] (a.corner 5) to (a.corner 7);
        \draw[postaction = {decorate}] (a.corner 4) to (a.corner 6);
        \draw[postaction = {decorate}] (a.corner 3) to (a.corner 5);
        \draw[postaction = {decorate}] (a.corner 2) to (a.corner 4);
        \end{scope}

        \begin{scope}[very thick,decoration={markings, mark=at position 0.62 with {\arrow[scale = 1.5, >=stealth]{>}}}]
        \draw[postaction = {decorate}] (a.corner 2) to (a.corner 1);
        \end{scope}

        \begin{scope}[very thick,decoration={markings, mark=at position 0.53 with {\arrow[scale = 1.5, >=stealth]{>}}}]
        \draw[postaction = {decorate}] (a.corner 12) to (a.corner 3);
        \end{scope}

        \node at (2,3.5) {\textcolor{blue}{$u$}};
        \node at (3.4,2) {\textcolor{red}{$v$}};
        \node at (4,0) {\textcolor{blue}{$u$}};
        \node at (3.4,-2) {\textcolor{red}{$v$}};
        \node at (2,-3.5) {\textcolor{blue}{$u$}};
        \node at (0,-4) {\textcolor{red}{$v$}};
        \node at (-2,-3.5) {\textcolor{blue}{$u$}};
        \node at (-3.4,-2) {\textcolor{red}{$v$}};
        \node at (-4,0) {\textcolor{blue}{$u$}};
        \node at (-3.4,2) {\textcolor{red}{$v$}};
        \node at (-2,3.5) {\textcolor{blue}{$u$}};

        \node at (1.85,1.9) {\textcolor{violet}{$y$}};
        \node at (-1.85,1.9) {\textcolor{violet}{$y$}};
        \node at (2.6,0.7) {\textcolor{violet}{$y$}};
        \node at (2.6,-0.7) {\textcolor{violet}{$y$}};
        \node at (-2.6,0.7) {\textcolor{violet}{$y$}};
        \node at (-2.6,-0.7) {\textcolor{violet}{$y$}};
        \node at (1.85,-1.9) {\textcolor{violet}{$y$}};
        \node at (-1.85,-1.9) {\textcolor{violet}{$y$}};
        \node at (0.7,-2.6) {\textcolor{violet}{$y$}};
        \node at (-0.7,-2.6) {\textcolor{violet}{$y$}};
        \node at (0,2) {\textcolor{violet}{$y$}};

        \node at (0,4) {\textcolor{olive}{$z$}};
    \end{tikzpicture}
    }
    \caption{Quiver for D-brane probe of $X=\textrm{Cone}(Y^{6,5})$.}
    \label{fig:Y65Quiver}
\end{figure}
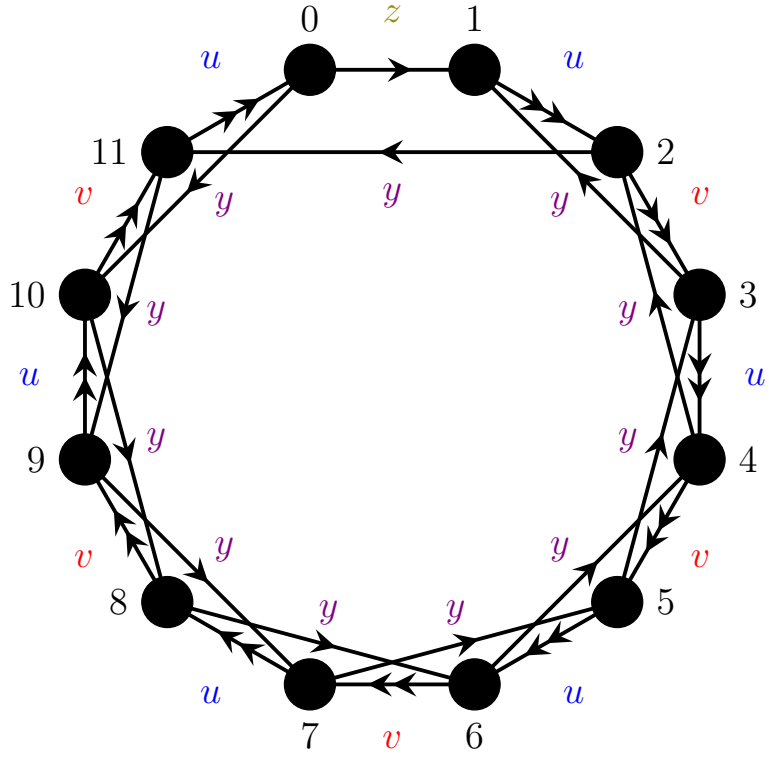

The erasing procedure is as follows. First, identify a field $v$ connecting node $i$ to node $i+1$ and erase this field. Then, replace the remaining $v$ field connecting these nodes with a $z$ field, which we denote as $t^z$ and color in olive. Also erase the $y$ fields connecting nodes $i+2$ to $i$ and $i+1$ to $i-1$. Finally, add a $y$ field connecting node $i+2$ to nodes $i-1$. See figure \ref{fig:Y65Quiver} for the $\mathrm{Cone}(Y^{6,5})$ example.
The weighted adjacency matrix can be written as
\begin{equation}
A^{Y^{6,5}}_{ij}(t) = 
    \begin{pmatrix}
        0 & t^{z} & 0 & 0 & 0 & 0 & 0 & 0 & 0 & 0 & t^y & 0 \\
        0 & 0 & 2t^{u} & 0 & 0 & 0 & 0 & 0 & 0 & 0 & 0 & 0 \\
        0 & 0 & 0 & 2t^{v} & 0 & 0 & 0 & 0 & 0 & 0 & 0 & t^y \\
        0 & t^y & 0 & 0 & 2t^{u} & 0 & 0 & 0 & 0 & 0 & 0 & 0 \\
        0 & 0 & t^y & 0 & 0 & 2t^{v} & 0 & 0 & 0 & 0 & 0 & 0 \\
        0 & 0 & 0 & t^y & 0 & 0 & 2t^{u} & 0 & 0 & 0 & 0 & 0 \\
        0 & 0 & 0 & 0 & t^y & 0 & 0 & 2t^{v} & 0 & 0 & 0 & 0 \\
        0 & 0 & 0 & 0 & 0 & t^y & 0 & 0 & 2t^{u} & 0 & 0 & 0 \\
        0 & 0 & 0 & 0 & 0 & 0 & t^y & 0 & 0 & 2t^{v} & 0 & 0 \\
        0 & 0 & 0 & 0 & 0 & 0 & 0 & t^y & 0 & 0 & 2t^{u} & 0 \\
        0 & 0 & 0 & 0 & 0 & 0 & 0 & 0 & t^y & 0 & 0 & 2t^{v} \\
        2t^{u} & 0 & 0 & 0 & 0 & 0 & 0 & 0 & 0 & t^y & 0 & 0
    \end{pmatrix},
\end{equation}
and the scaling dimensions of the fields are now \cite{Benvenuti:2004dy, Benvenuti:2004wx}:
\begin{align}
\Delta_Y &= y(6,5)= \frac{-9+7\sqrt{69}}{50}, \\
\Delta_Z &= z(6,5)= \frac{-129+17\sqrt{69}}{50}, \\
\Delta_U &= u(6,5)= \frac{144-12\sqrt{69}}{50}, \\
\Delta_V &= v(6,5)= \frac{3+\sqrt{69}}{10}.
\end{align}

The above prescription can be performed repeatedly to get the rest of the $Y^{6,q}$. In particular, after two more repetitions we arrive at the $Y^{6,3}$ quiver studied in the main text.

\section{Quantum Symmetry and the Twist} \label{app:QuantumSymmetry}

When a quiver is obtained via an orbifolding, there is a natural symmetry acting on its nodes. Following the discussion in \cite{Dramburg:2025tlb}, each node of the original theory is split into multiple ones labeled by irreducible representations of the orbifold group, a generalization of the McKay correspondence. One can then tensor the irreducible representations of each node by 1D irreducible representations and track where the nodes are mapped. Given a group $G$, its 1D irreducible representations form an abelian group isomorphic to the abelianization of $G$, this is know as the quantum symmetry of the quiver.

In the main text we discussed how the entries of matrix Hilbert series $H_{0i}$ can be interpreted as twisted Hilbert series. Using the relation between the quantum symmetry and the higher form symmetry discussed in \cite{Dramburg:2025tlb}, we can be more precise about which index $i$ encodes the series twisted by the flat bundle associated with the higher form symmetry generator.

First of all, if the orbifold group $\Gamma$ acts without fixed loci (other than the identity), the whole group can be put into one to one correspondence with the one-form symmetry of the theory. In this case, the entry $H_{0i}$ can be take to be the image of node $0$ tensored by any 1D irreducible representation $\rho_i$. This will lead to all possible $\eta$-invariants in \eqref{eq:AlphaFromQuiverGeneral}.

If the action is not free, there are fixed loci, and the quantum symmetry encodes the whole 2-group extension. The one-form symmetry part can be extracted by considering the subgroup of 1D irreducible representations that trivialize the action of $\Gamma$ elements that acts with fixed loci on the boundary. One can then use the either the full quantum symmetry or the subgroup associated to the one-form symmetry to get the refined or unrefined $\eta$-invariant, respectively.

As an example, let us consider $\mathbb{C}^3/\mathbb{Z}_6$ with action $(1,1,4)$ as in section \ref{sec:EXAMPLES}.
The character table is given by
\begin{align}\label{eq:chatbl}
 \begin{matrix}
    \rho_0 \\
    \rho_1 \\
    \rho_2 \\
    \rho_3 \\
    \rho_4 \\
    \rho_5 \\
\end{matrix}
\begin{pmatrix}
1 & 1 & 1 & 1 & 1 & 1   \\
1 & -\omega & \omega^2 & -1 & \omega & -\omega^2 \\
1 & \omega^2 & \omega & 1 & \omega^2 & \omega \\
1 & -1 & 1 & -1 & 1 & -1 \\
1 & \omega & \omega^2 & 1 & \omega & \omega^2 \\
1 & -\omega^2 & \omega & -1 & \omega^2 & -\omega \\
\end{pmatrix} \, ,
\end{align}
with $\omega^3=1$.

The desired geometry is obtained by orbifolding via the representation $R=\rho_1\oplus\rho_1\oplus\rho_4$, which is generated by the element $g = \textrm{diag}(-\omega,-\omega,\omega)$ acting on $\mathbb{C}^3$. In this case there is a $\mathbb{Z}_6$ quantum symmetry that rotates the whole quiver, but since the orbifold action is not free on the boundary, only a subset is identified with the one-form symmetry. It is easy to see that $g^3$ leaves the one direction invariant, thus acting with fixed loci on the boundary. Following the prescription of \cite{Dramburg:2025tlb}, the subgroup of the quantum symmetry associated to the $\mathbb{Z}_3$ one-form symmetry is generated by $\{ \rho_0, \rho_2,\rho_4 \}$, since these are the ones that trivialize the action of $g^3$, as can be seen in table \ref{eq:chatbl}.

\begin{figure}
\begin{tikzpicture}

    \newcommand{\drawHexagonOne}[1]{
        \begin{scope}[shift={#1}]
            \node[draw=none, minimum size=4cm, regular polygon, regular polygon sides=6] (h1) {};

            \foreach \corner/\labelnum/\pos in {
                1/\rho_1/above, 2/\rho_0/above, 3/\rho_5/left, 4/\rho_4/below, 5/\rho_3/below, 6/\rho_2/right
            }
            \node[draw, circle, inner sep=2pt, fill=black, label=\pos:$\labelnum$] at (h1.corner \corner) {};

            \begin{scope}[very thick, decoration={markings, mark=at position 0.65 with {\arrow[scale = 1.5, >=stealth]{>>}}}]
                \draw[postaction = {decorate}] (h1.corner 2) to (h1.corner 1);
                \draw[postaction = {decorate}] (h1.corner 3) to (h1.corner 2);
                \draw[postaction = {decorate}] (h1.corner 4) to (h1.corner 3);
                \draw[postaction = {decorate}] (h1.corner 5) to (h1.corner 4);
                \draw[postaction = {decorate}] (h1.corner 6) to (h1.corner 5);
                \draw[postaction = {decorate}] (h1.corner 1) to (h1.corner 6);
            \end{scope}

            \begin{scope}[very thick, decoration={markings, mark=at position 0.52 with {\arrow[scale = 1.5, >=stealth]{>}}}]
                \draw[postaction = {decorate}] (h1.corner 1) to (h1.corner 3);
                \draw[postaction = {decorate}] (h1.corner 2) to (h1.corner 4);
                \draw[postaction = {decorate}] (h1.corner 3) to (h1.corner 5);
                \draw[postaction = {decorate}] (h1.corner 4) to (h1.corner 6);
                \draw[postaction = {decorate}] (h1.corner 5) to (h1.corner 1);
                \draw[postaction = {decorate}] (h1.corner 6) to (h1.corner 2);
            \end{scope}
        \end{scope}
    }

    \newcommand{\drawHexagonTwo}[1]{
        \begin{scope}[shift={#1}]
            \node[draw=none, minimum size=4cm, regular polygon, regular polygon sides=6] (h2) {};

            \foreach \corner/\labelnum/\pos in {
                1/\rho_4/above, 2/\rho_3/above, 3/\rho_2/left, 4/\rho_1/below, 5/\rho_0/below, 6/\rho_5/right
            }
            \node[draw, circle, inner sep=2pt, fill=black, label=\pos:$\labelnum$] at (h2.corner \corner) {};

            \begin{scope}[very thick, decoration={markings, mark=at position 0.65 with {\arrow[scale = 1.5, >=stealth]{>>}}}]
                \draw[postaction = {decorate}] (h2.corner 2) to (h2.corner 1);
                \draw[postaction = {decorate}] (h2.corner 3) to (h2.corner 2);
                \draw[postaction = {decorate}] (h2.corner 4) to (h2.corner 3);
                \draw[postaction = {decorate}] (h2.corner 5) to (h2.corner 4);
                \draw[postaction = {decorate}] (h2.corner 6) to (h2.corner 5);
                \draw[postaction = {decorate}] (h2.corner 1) to (h2.corner 6);
            \end{scope}

            \begin{scope}[very thick, decoration={markings, mark=at position 0.52 with {\arrow[scale = 1.5, >=stealth]{>}}}]
                \draw[postaction = {decorate}] (h2.corner 1) to (h2.corner 3);
                \draw[postaction = {decorate}] (h2.corner 2) to (h2.corner 4);
                \draw[postaction = {decorate}] (h2.corner 3) to (h2.corner 5);
                \draw[postaction = {decorate}] (h2.corner 4) to (h2.corner 6);
                \draw[postaction = {decorate}] (h2.corner 5) to (h2.corner 1);
                \draw[postaction = {decorate}] (h2.corner 6) to (h2.corner 2);
            \end{scope}
        \end{scope}
    }

    \drawHexagonOne{(0,0)}
    \drawHexagonTwo{(9,0)}

    \draw[line width=1pt, -{Stealth[scale=2]}] (3,0) -- (6,0) node[midway, above] {$\rho_i\otimes \rho_3$};

\end{tikzpicture}
 \caption{Action of the quantum symmetry on $\mathbb{C}^3/\mathbb{Z}_6$. The nodes are labeles as the irreducible representations of $\mathbb{Z}_6$ and the symmetry is implemented by tensoring each node with the the desired irrep.}
    \label{fig:quantum_action}
\end{figure}
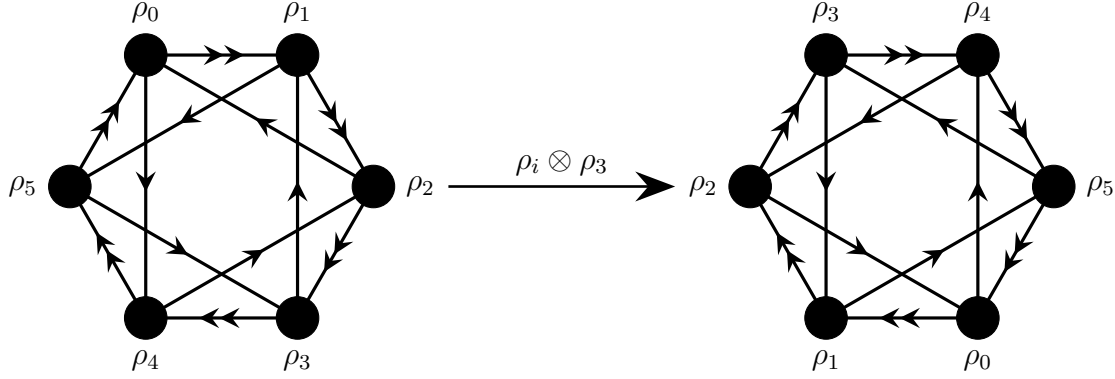

\section{Discrete Torsion} \label{app:DISCTORSION}

In this Appendix we discuss a related class of orbifold geometries where the target space is an orbifold with discrete torsion. Strictly speaking, the backgrounds we consider are obtained from type IIA rather than an M-theory background. Alternatively, one can view it as a circle compactification of the M-theory background $S^1 \times X$ in which we allow the $C_3$ potential to have one leg on the circle, i.e., we interpret it as a choice of (torsional) NSNS $B_2$-field on $X$.

We now generalize the class of orbifold backgrounds discussed above to include discrete torsion.\footnote{See for example \cite{Vafa:1986wx, Vafa:1994rv, Feng:2000af, Feng:2000mw, Sharpe:2003cs, Douglas:1998xa, Douglas:1999hq, Sharpe:1999pv, Sharpe:1999xw}.} For an orbifold of the form
\begin{equation}
X=\mathbb{C}^3/\Gamma,\qquad\Gamma=\mathbb{Z}_N (n_1,n_2,n_3) \times \mathbb{Z}_M(m_1,m_2,m_3),
\end{equation}
discrete torsion is specified by a choice of class
\begin{equation}
\delta \in H^2(\Gamma,U(1)).
\end{equation}
Physically, this may be viewed as turning on a flat NSNS $B$-field background on the orbifold, or equivalently as introducing phases in the twisted sectors of the orbifold theory.

More explicitly, the worldsheet orbifold partition function decomposes into twisted sectors labelled by commuting pairs of group elements. Turning on discrete torsion corresponds to the ambiguity of assigning phases to these sectors. Thus, for a choice $\delta\in H^2(\Gamma,U(1))$, the partition function takes the form
\begin{equation}
Z_{\delta} = \sum_{[g]\in \mathrm{Conj}(\Gamma)} \frac{1}{|C(g)|} \sum_{\sigma\in C(g)} \delta(g,\sigma)\,Z_{g,\sigma},
\end{equation}
where $C(g)$ is the centralizer of $g$, and $Z_{g,\sigma}$ denotes the contribution from the $g$-twisted sector with an $\sigma$-insertion. For the abelian orbifolds considered below, all conjugacy classes are singletons and $C(g)=\Gamma$.

From the D-brane point of view, this means that branes probing the singularity are no longer described by ordinary linear representations of $\Gamma$, but rather by projective representations, with multiplication twisted by the cocycle $\delta$. In particular, if $\rho$ denotes such a projective representation, then
\begin{equation}
\rho(g)\rho(h)=\delta(g,h)\rho(gh),
\qquad g,h\in \Gamma.
\end{equation}

For the abelian groups of interest here, the set of possible choices is especially simple:
\begin{equation}
H^2(\mathbb{Z}_N\times \mathbb{Z}_M,U(1))\cong\mathbb{Z}_{\gcd(N,M)}.
\end{equation}
A useful point emphasized in \cite{Braeger:2025rov} is that, in these examples, the effect of discrete torsion can be geometrized. More precisely, if $\delta$ has order
\begin{equation}
\mathrm{ord}(\delta)=\frac{\gcd(N,M)}{\gcd(\delta,\gcd(N,M))},
\end{equation}
then one can associate to it a subgroup
\begin{equation}
\Gamma_\delta=\Gamma / \bigl(\mathbb{Z}_{\mathrm{ord}(\delta)}\times\mathbb{Z}_{\mathrm{ord}(\delta)}\bigr) \cong \mathbb{Z}_{N/\mathrm{ord}(\delta)}\times \mathbb{Z}_{M/\mathrm{ord}(\delta)},
\end{equation}
and hence a covering orbifold
\begin{equation}
X_\delta=\mathbb{C}^3/\Gamma_\delta \longrightarrow X=\mathbb{C}^3/\Gamma.
\end{equation}
The key point is that $X_\delta$ is a less singular geometry with no discrete torsion turned on. In this sense, a background with discrete torsion can be reinterpreted as an ordinary orbifold by a smaller group. The weights of the orbifold group $\Gamma_\alpha$ are also determined by those of $\Gamma$ as $n_i\; \textrm{mod}\;\textrm{ord}(\delta)$ and $m_i\; \textrm{mod}\;\textrm{ord}(\delta)$.

This statement is particularly powerful from the quiver perspective. Rather than working directly with a projective McKay quiver, which is obtained via projective representations, for $(\Gamma,\delta)$, one can instead pass to the ordinary McKay quiver, which is obtained via linear representations, of the covering geometry $X_\delta$. Denoting by $Q_{X,\delta}$ the quiver of the orbifold $X=\mathbb{C}^3/\Gamma$ with discrete torsion $\delta$, and by $Q_{X_\delta}$ the usual quiver of the torsion-free orbifold $X_\delta$, the result of \cite{Braeger:2025rov} may be summarized as
\begin{equation}
Q_{X,\delta}\cong Q_{X_\delta}.
\end{equation}

For our purposes, this gives a simple and practical way to analyze discrete torsion examples. We may begin with
\begin{equation}
X=\mathbb{C}^3/\bigl(\mathbb{Z}_N\times \mathbb{Z}_M\bigr)
\end{equation}
together with a chosen class $\delta \in H^2(\Gamma,U(1))$, determine $\mathrm{ord}(\delta)$, and then replace the original problem by the torsion-free orbifold
\begin{equation}
X_\delta=\mathbb{C}^3/\Bigl(\mathbb{Z}_{N/\mathrm{ord}(\delta)}\times \mathbb{Z}_{M/\mathrm{ord}(\delta)}\Bigr).
\end{equation}

\paragraph{Example} We now illustrate this with a simple example. Consider
\begin{equation}
X=\mathbb{C}^3/\Gamma, \qquad \Gamma=\mathbb{Z}_8(2,1,5)\times \mathbb{Z}_2(1,0,1).
\end{equation}
In this case,
\begin{equation}
H^2(\Gamma,U(1))\cong \mathbb{Z}_2,
\end{equation}
so there are precisely two choices of discrete torsion: the trivial class $\delta=0$, corresponding to the ordinary orbifold with no discrete torsion, and the unique non-trivial class $\delta=1$. For this example, the master quiver therefore decomposes into two disconnected components, one for each choice of discrete torsion. The corresponding geometries are
\begin{equation}
X_{\delta=0}=\mathbb{C}^3/\mathbb{Z}_8(2,1,5)\times \mathbb{Z}_2(1,0,1),\qquad X_{\delta=1} = \mathbb{C}^3/\mathbb{Z}_4(2,1,1).
\end{equation}
Thus, turning on the unique non-trivial discrete torsion class maps the original orbifold to a simpler one with no discrete torsion.

\begin{figure}
    \centering
    \begin{tikzpicture}
        \node[draw=none, minimum size=7cm,regular polygon,regular polygon sides=4] (a) {};

        \foreach \corner/\labelnum/\pos in {
        1/1/above,
        2/0/above,
        3/3/below,
        4/2/below
    }
    \node[draw, circle, inner sep=2pt, fill=black, label=\pos:$\labelnum$] at (a.corner \corner) {};

        \begin{scope}[very thick,decoration={markings, mark=at position 0.57 with {\arrow[scale = 1.5, >=stealth]{>>}}}]
        \draw[postaction = {decorate}] (a.corner 2) to (a.corner 1);
        \draw[postaction = {decorate}] (a.corner 3) to (a.corner 2);
        \draw[postaction = {decorate}] (a.corner 4) to (a.corner 3);
        \draw[postaction = {decorate}] (a.corner 1) to (a.corner 4);
        \end{scope}

        \begin{scope}[very thick,decoration={markings, mark=at position 0.75 with {\arrow[scale = 1.5, >=stealth]{>}}}]
        \draw[postaction = {decorate}] (a.corner 1) to (a.corner 3);
        \draw[postaction = {decorate}] (a.corner 2) to (a.corner 4);
        \draw[postaction = {decorate}] (a.corner 3) to (a.corner 1);
        \draw[postaction = {decorate}] (a.corner 4) to (a.corner 2);
        \end{scope}
        \end{tikzpicture}
    \caption{Quiver associated to the geometry $\mathbb{C}^3/\mathbb{Z}_8(2,1,5)\times \mathbb{Z}_2(1,0,1)$ with discrete torsion or equivalently $\mathbb{C}^3/\mathbb{Z}_4(2,1,1)$ without discrete torsion.}
    \label{fig:DTExample}
\end{figure}
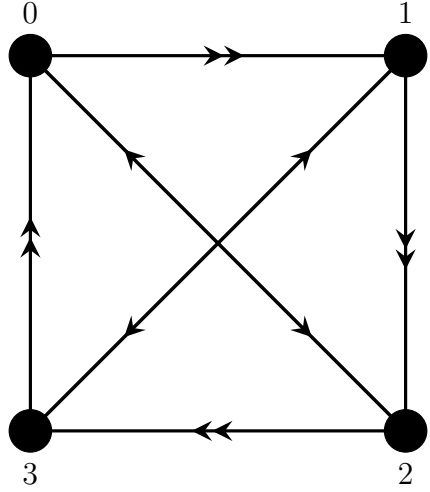

From the viewpoint of the previous discussion, this means that the brane-probe theory for
\begin{equation}
\mathbb{C}^3/\mathbb{Z}_8(2,1,5)\times \mathbb{Z}_2(1,0,1)
\quad \text{with } \delta=1
\end{equation}
may be analyzed instead using the ordinary McKay quiver (see figure \ref{fig:DTExample}) of
\begin{equation}
\mathbb{C}^3/\mathbb{Z}_4(2,1,1).
\end{equation}
The adjacency matrix for this example is given by
\begin{equation}
A_{ij}=
    \begin{pmatrix}
        0 & 2 & 1 & 0 \\
        0 & 0 & 2 & 1 \\
        1 & 0 & 0 & 2 \\
        2 & 1 & 0 & 0
    \end{pmatrix}.
\end{equation}

The one-form symmetry group for this example can be computed from the Dirac pairing / antisymmetrized adjacency matrix:
\begin{equation}
\mathrm{Tor}\,\mathrm{Coker}(B)\simeq \mathbb Z_2\oplus \mathbb Z_2.
\end{equation}
Recall that this is the reduction of the electric one-form symmetry and magnetic two-form symmetry in the 5D SCFT. Plugging the adjacency matrix into \eqref{eq:EagerMatrixHilbert}, we get the following quiver Hilbert series:
\begin{equation}
H(t)=
\begin{pmatrix}
-\frac{t^6-t^5+t^4+2 t^3+t^2-t+1}{(t-1)^3 \left(t^3+t^2+t+1\right)^2}
&
-\frac{2 t \left(t \left(t^2+t-1\right)+1\right)}{(t-1)^3 \left(t^3+t^2+t+1\right)^2}
&
-\frac{t \left(t \left(t \left(t (t+2)-2\right)+2\right)+1\right)}{(t-1)^3 \left(t^3+t^2+t+1\right)^2}
&
-\frac{2 t^2 \left(t^3-t^2+t+1\right)}{(t-1)^3 \left(t^3+t^2+t+1\right)^2}
\\[1em]
-\frac{2 t^2 \left(t^3-t^2+t+1\right)}{(t-1)^3 \left(t^3+t^2+t+1\right)^2}
&
-\frac{t^6-t^5+t^4+2 t^3+t^2-t+1}{(t-1)^3 \left(t^3+t^2+t+1\right)^2}
&
-\frac{2 t \left(t \left(t^2+t-1\right)+1\right)}{(t-1)^3 \left(t^3+t^2+t+1\right)^2}
&
-\frac{t \left(t \left(t \left(t (t+2)-2\right)+2\right)+1\right)}{(t-1)^3 \left(t^3+t^2+t+1\right)^2}
\\[1em]
-\frac{t \left(t \left(t \left(t (t+2)-2\right)+2\right)+1\right)}{(t-1)^3 \left(t^3+t^2+t+1\right)^2}
&
-\frac{2 t^2 \left(t^3-t^2+t+1\right)}{(t-1)^3 \left(t^3+t^2+t+1\right)^2}
&
-\frac{t^6-t^5+t^4+2 t^3+t^2-t+1}{(t-1)^3 \left(t^3+t^2+t+1\right)^2}
&
-\frac{2 t \left(t \left(t^2+t-1\right)+1\right)}{(t-1)^3 \left(t^3+t^2+t+1\right)^2}
\\[1em]
-\frac{2 t \left(t \left(t^2+t-1\right)+1\right)}{(t-1)^3 \left(t^3+t^2+t+1\right)^2}
&
-\frac{t \left(t \left(t \left(t (t+2)-2\right)+2\right)+1\right)}{(t-1)^3 \left(t^3+t^2+t+1\right)^2}
&
-\frac{2 t^2 \left(t^3-t^2+t+1\right)}{(t-1)^3 \left(t^3+t^2+t+1\right)^2}
&
-\frac{t^6-t^5+t^4+2 t^3+t^2-t+1}{(t-1)^3 \left(t^3+t^2+t+1\right)^2}
\end{pmatrix}.
\end{equation}
In particular, we get that
\begin{align}
    H_{00} &= -\frac{t^6-t^5+t^4+2 t^3+t^2-t+1}{(t-1)^3 \left(t^3+t^2+t+1\right)^2}\,,\\[1em]
    H_{01} &= -\frac{2 t \left(t \left(t^2+t-1\right)+1\right)}{(t-1)^3 \left(t^3+t^2+t+1\right)^2}\,.
\end{align}
From this we compute the refined $\eta$-invariant:
\begin{equation}
    \alpha=\frac{1}{2}\eta^{\mathrm{ref}} = \operatorname*{Res}^{(2)}_{t=1} \left( \frac{H_{01}(t)-H_{00}(t)}{1-t} \right) + \operatorname*{Res}_{t=1}\left( \frac{H_{01}(t)-H_{00}(t)}{1-t} \right) = \frac{1}{8} + 0 = \frac{1}{8}\qquad\textrm{mod}\,1\,.
\end{equation}
For this particular example, all the contributions to the $\eta$-invariant are coming from refinements due to the 2-group structure discussed in section \ref{sec:Refinements}. We end by noting that this result agrees with Appendix A of \cite{Cvetic:2025lat}.

\bibliographystyle{utphys}
\bibliography{SymTFTviaQuivers}

\end{document}